\documentclass[secnumarabic,floatfix,tightenlines,pra,pra,reprint,showpacs,preprintnumbers,amsmath,amssymb,superscriptaddress]{revtex4-1}

\usepackage{graphicx}
\usepackage{dcolumn}
\usepackage{bm}
\usepackage[utf8]{inputenc}
\usepackage{hyperref}
\usepackage[all]{hypcap}
\usepackage[mathlines]{lineno}

\begin{document}

\title{Improved nonlinear plasmonic slot waveguide: a full study}

\author{Mahmoud M. R. Elsawy}
\affiliation{Aix Marseille Université, CNRS, Centrale Marseille, Institut Fresnel UMR 7249, 13013 Marseille, France}
\author{Virginie Nazabal}
\affiliation{CNRS UMR 6226, Institut Sciences Chimiques de Rennes, Campus de Beaulieu,  Universit{é} de Rennes 1, 35042 Rennes Cedex, France}
\author{Mathieu Chauvet}
\affiliation{University of Franche-Comt{é}, CNRS UMR 6174, FEMTO-ST Institute, 15B Avenue des Montboucons, 25000 Besan{ç}on, France}%
\author{Gilles Renversez}%
 \email{gilles.renversez@fresnel.fr}
\affiliation{Aix Marseille Université, CNRS, Centrale Marseille, Institut Fresnel UMR 7249, 13013 Marseille, France} 


\date{\today}

\begin{abstract}
We present a full study of an improved nonlinear plasmonic slot waveguides (NPSWs) in which buffer linear dielectric layers are added between the Kerr type nonlinear dielectric core and the two semi-infinite metal regions. Our approach computes the stationary solutions using the fixed power algorithm, in which for a given structure the wave power is an input parameter and the outputs are the propagation constant and the corresponding field components. For TM polarized waves, the inclusion of these supplementary layers have two consequences. First, they reduced the overall losses. Secondly, they modify the types of solutions that propagate in the NPSWs adding new profiles enlarging the possibilities offered by these nonlinear waveguides. In addition to the symmetric linear plasmonic profile obtained in the simple plasmonic structure with linear core such that its effective index is above the linear core refractive index, we obtained a new field profile which is more localized in the core with an effective index below the core linear refractive index. In the nonlinear case, if the effective index of the symmetric linear mode is above the core linear refractive index, the mode field profiles now exhibit a spatial transition from a plasmonic type profile to a solitonic type one. Our structure also provides longer propagation length due to the decrease of the losses compared to the simple nonlinear slot waveguide and exhibits, for well-chosen refractive index or thickness of the buffer layer, a spatial transition of its main modes that can be controlled by the power. We provide a full phase diagram of the TM wave operating regimes of these improved NPSWs. The stability of the main TM modes is then demonstrated numerically using the FDTD. We also demonstrate the existence of TE waves for both linear and nonlinear cases (for some configurations) in which the maximum intensity is located in the middle of the waveguide. We indicate the bifurcation of the nonlinear asymmetric TE mode from the symmetric nonlinear one through the Hopf bifurcation. This kind of bifurcation is similar to the ones already obtained in TM case for our improved structure, and also for the simple NPSWs. At high power, above the bifurcation threshold, the fundamental symmetric nonlinear TE mode moves gradually to new nonlinear mode in which the soliton peak displays two peaks in the core. The losses of the TE modes decrease with the power for all the cases. This kind of structures could be fabricated and characterized experimentally due to the realistic parameters chosen to model them.  
\begin{description}
\item[PACS numbers]
42.65.Wi, 42.65.Tg, 42.65.Hw, 73.20.Mf
\end{description}
\end{abstract}


\keywords{surface plasmons, waveguide slab, Kerr effect, spatial solitons, nonlinear plasmonics, TE waves, TM waves, bifurcation, stability}
\maketitle


\section{Introduction}
\label{sec:intro}
In recent years, plasmonics has been receiving a great attention from scientists because of its ability to achieve highly confined electromagnetic field and to manipulate light at nano-scale. Nonlinear plasmonics might be of interest because the nonlinearity can be controlled by the power allowing a management of the solution unlike the linear case. Nonlinear plasmonic slot waveguides (NPSWs) have been a flourishing research topic, at least since 2007,~\cite{Feigenbaum07} due to the strong light confinement in the nonlinear dielectric core ensured by the surrounding metal regions and due to its peculiar nonlinear effects~\cite{Davoyan08,Davoyan09,Rukhlenko11a,Ferrando13,Walasik14a}. One of the key property of symmetric NPSWs is the existence of asymmetric modes that emerge at critical powers through  Hopf bifurcations from the first symmetric mode and also from some higher order symmetric modes~\cite{Walasik15a,Walasik15b}. Several applications have already been proposed for NPSWs including phase matching in higher harmonic generation processes~\cite{Davoyan09a},  nonlinear plasmonic couplers~\cite{Salgueiro10} or switching~\cite{Nozhat12}, and also for biological and chemical sensors~\cite{Charbonneau08,Wijesinghe-sensing15}.
Nevertheless, the experimental observation of plasmon-soliton waves in these NPSWs is still lacking even if linear slot waveguides have already been fabricated~\cite{Han10}. Similarly to the case of the single nonlinear dielectric/metal interface structures~\cite{Ariyasu85,stegeman85,stegeman84,kushwaha87,Walasik12}, the modes already studied in the simple NPSWs suffer from high propagation losses that seriously limit the propagation length of the waves.
In the present study, we propose and study an improved structure in which buffer linear dielectric layers are added between the nonlinear dielectric core and the two semi-infinite metal regions. For the TM case, these supplementary layers have two main consequences. First, they modify the types of solutions that propagate in the NPSWs adding new profiles, enlarging the possibilities offered by theses nonlinear waveguides. Second, they reduced  overall losses and allow the losses to decrease with the power in most of the cases. For the TE case, they reduce the cut-off core thickness for the fundamental linear mode,  and increase its propagation length. For the nonlinear case, the losses decrease with the increase of the power for all the TE modes, and the nonlinear symmetric mode moves gradually to new nonlinear mode in which the soliton intensity profile displays two peaks in the middle of the waveguide. We also demonstrate the bifurcation of the asymmetric nonlinear mode from the symmetric one for the TE polarized waves.   

The article is organized as follows. In Sec.~\ref{sec:models}, we describe the model and numerical method we use to study the stationary nonlinear waves in the improved NPSW we propose. In Sec.~\ref{sec:results-TM}, we study the properties of the stationary transverse magnetic (TM) waves. In this case, first, we describe the linear solutions of the new structures in order to be able to classify the solutions that will be obtained in the nonlinear case. Second, we give the properties of TM nonlinear stationary solutions and show that, for some linear parameter configurations, new modal spatial transitions as a function of power occur compared to the simple NPSW case. Third, we provide the full nonlinear phase diagram of the main TM modes as a function of the total power and the buffer layer. In Sec.~\ref{sec:loss_study}, we prove that the added buffer dielectric layers are able to reduce propagation losses and allow them to decrease with power in most of the cases for the TM case. In Sec.~\ref{sec:stability} using the FDTD method we study the stability properties of the main nonlinear TM solutions for the most significant cases. Finally, in Sec.~\ref{sec:TE_modes}, we study the existence and the properties of TE modes for both linear and nonlinear core.    
\section{Model and method}
\label{sec:models}
\begin{figure}[htbp]
\centerline{\includegraphics[width=0.99\columnwidth,angle=-0,clip=true,trim= 0 0 0 0]{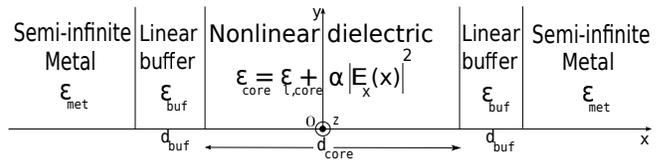}}
\caption{Symmetric improved NPSW  geometry with the supplementary linear dielectric buffer layers between the nonlinear dielectric core and the two semi-infinite metal regions.}
\label{fig:geom-5layers-NPSW}
\end{figure}

The improved NPSW we study is depicted in Fig.~\ref{fig:geom-5layers-NPSW}, compared to already studied simple NPSW made of a nonlinear dielectric core surrounded by two semi-infinite metal regions~\cite{Feigenbaum07,Davoyan08,Walasik14a}, the improved structure contains additional linear dielectric layers between the core and the metal regions. In this study, we will consider only symmetric structures even if asymmetric simple NPSWs have already been considered~\cite{Walasik15a}.  $d_{core}$ is the core thickness, and   $d_{buf}$ is the dielectric buffer thickness  of each of the supplementary layers with their permittivities denoted by $\varepsilon_{buf}$. We will only consider either monochromatic transverse magnetic (TM) waves or transverse electric (TE) waves propagating along the $z$ direction in a symmetric one-dimensional NPSW.  All field components evolve proportionally to $\exp[i( k_0 n_{eff} z - \omega t)]$ with $k_0 = \omega/c$, where $c$ denotes the speed of light in vacuum, $\omega$ is the angular frequency,  and $n_{eff}$ denotes the effective mode index. The nonlinear Kerr-type dielectric is isotropic with focusing nonlinearity. 

For the TM polarized waves, the permittivity in the nonlinear core is defined as $\varepsilon_{core} = \varepsilon_{l,core} + \alpha |E_x|^2$ ($\varepsilon_{l,core}$ being the linear part of the core permittivity, $E_x$ being the $x$ component of the electric field), and with $\alpha > 0$  (focusing nonlinearity).
This approximation of the Kerr nonlinearity, taking into account only the transverse component of the electric field has already been used in several models of the NPSWs~\cite{Walasik14a,Walasik15a}. It provides similar results than the more accurate approach where all the electric field components are considered in the optical Kerr effect~\cite{Walasik14a,Walasik15a,Walasik14}, even if at high power its results differ from the ones obtained with the full model taking into account all the electric field components in the nonlinear term~\cite{Walasik15a,Walasik15b}. This approximation allows us to use the fixed power algorithm in the finite element method (FEM) to compute the nonlinear stationary solutions and their nonlinear dispersion curves~\cite{Rahman90,ferrando03spatial-soliton-pcf,Drouart08,Walasik14}. In the core, the approximated relation between the nonlinear parameter $\alpha$ and the coefficient $n_2$, that appears in the definition of an intensity dependent refractive index $\Re e(n_{core}) = \Re e(\varepsilon_{l,core})^{1/2} + n_2 I_{core}$, is $\alpha  = \varepsilon_0 c \Re e(\varepsilon_{l,core}) n_2$,
 where the intensity inside the core is defined as $I_{core} = \varepsilon_0 c |\varepsilon_{core}| |E_x|^2 /(2 n_{eff})$, and the vacuum permittivity is denoted by $\varepsilon_0$. For the TE case, the only nonzero component of the electric field is $E_{y}$ and the permittivity in the core is defined as $\varepsilon_{core} = \varepsilon_{l,core} + \alpha |E_y|^2$, with $\alpha > 0$  (focusing nonlinearity). In general, the TE surface plasmon waves do not exit due to continuity of the component $E_{y}$. But, the fundamental TE mode exists only if the the core thickness beyond a cut-off~\cite{nakano94,Sun09}. 
 
 To compute the linear and nonlinear modes we use our custom-built FEM we developed to study both scalar and vector models for several types of waveguides including nonlinear ones with Kerr regions like microstructured optical fibers and NPSWs~\cite{Drouart08,Walasik14,livre12-FPCF,Gilles09-MOF}. In this study, $\varepsilon_{l,core} = 3.46^2 + i \, 10^{-4}$ and $n_2= 10^{-17} $m$^2$/W, corresponding to amorphous hydrogenated silicon~\cite{Lacava-Fedeli13-hydrogenated-amorphous-silicon,Matres-Fedeli13-fom-amorphous}, and $\varepsilon_{met}=-90+ i \, 10$ corresponding to gold at $\lambda=1.55$ $\mu$m~\cite{Palik98,Rakic98-Opical-metallic}, wavelength for which all the simulations are done.
 The propagation losses will be estimated using the method based on the field profiles and imaginary parts of the permittivities described in~\cite{Ariyasu85,Snyder83}, that is similar to the ones from~\cite{Davoyan09,Walasik15a}. 
We stress that the considered parameter values can affect the critical values  given  in this study but would not affect its main conclusions.
\section{Results for TM stationary waves}
\label{sec:results-TM}
\subsection{Linear case}
\label{sec:results-linear-TM}
\begin{figure}[htbp]
\centering
    \includegraphics[width=0.50\columnwidth,height=0.325\columnwidth,angle=-0,clip=true,trim= 0 0 0 0]{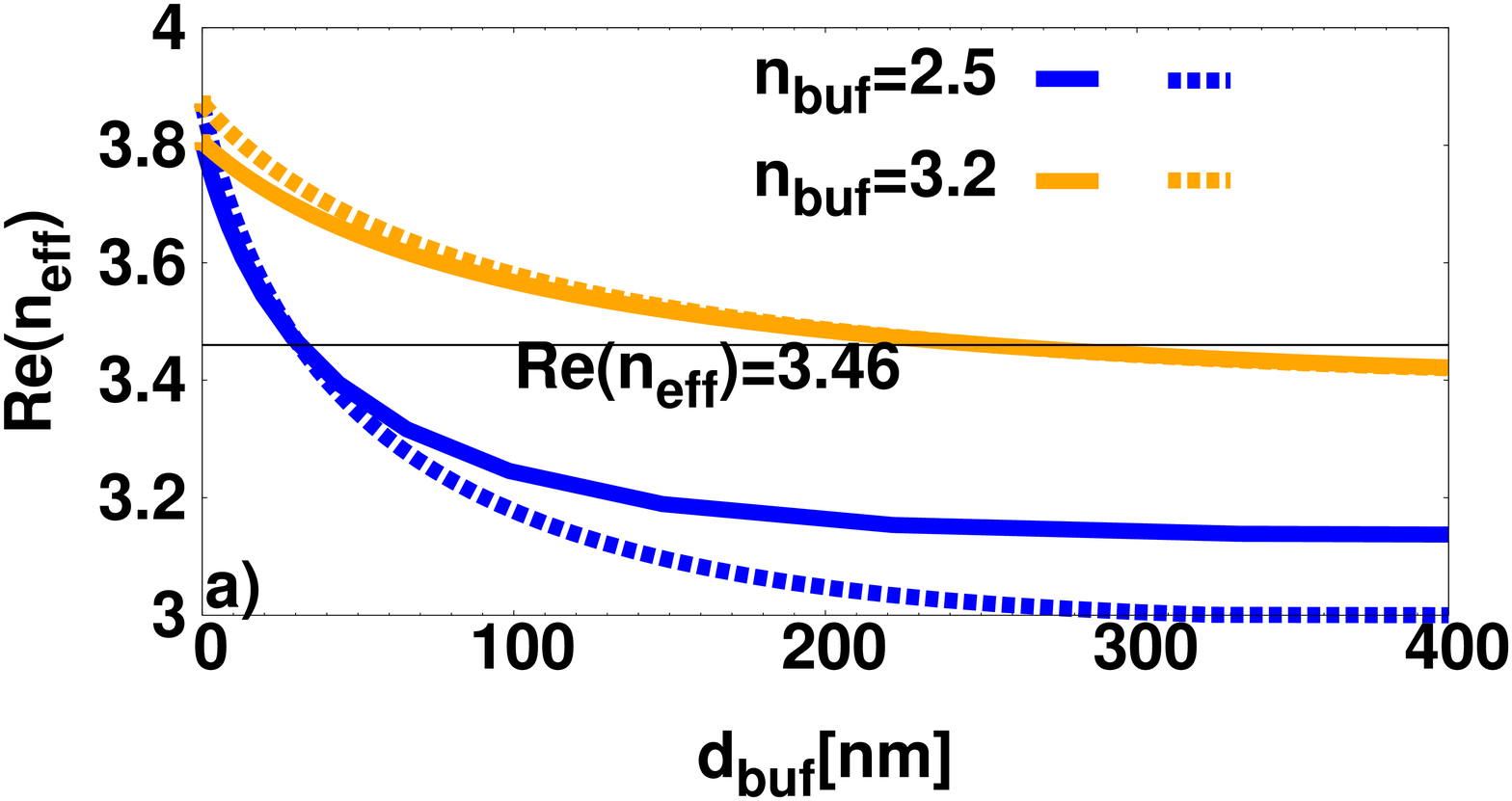}
    \includegraphics[width=0.485	\columnwidth,angle=-0,clip=true,trim= 0 0 0 0]{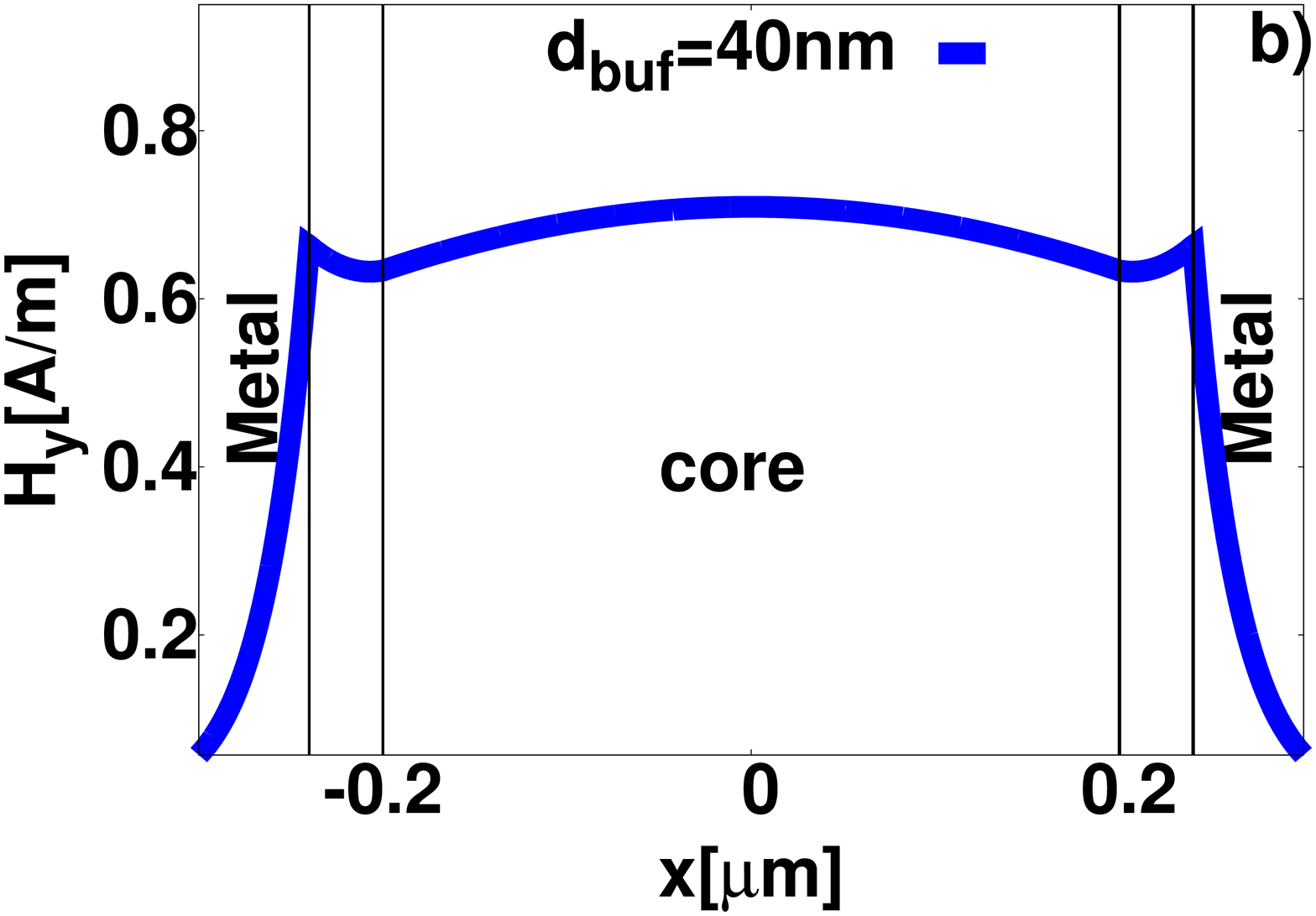}       
    \includegraphics[width=0.485\columnwidth,angle=-0,clip=true,trim= 0 0 0 0]{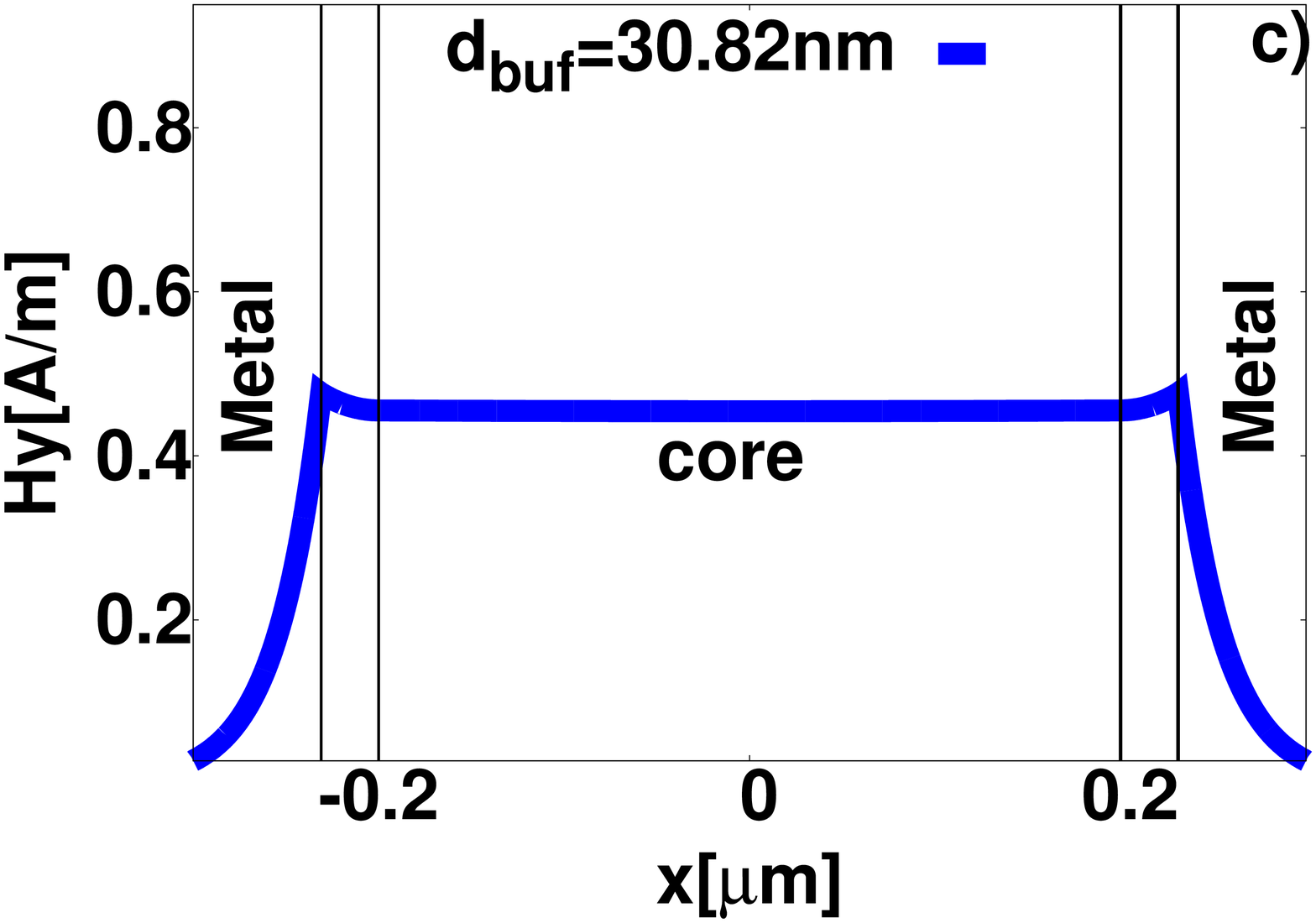}
    \includegraphics[width=0.485\columnwidth,angle=-0,clip=true,trim= 0 0 0 0]{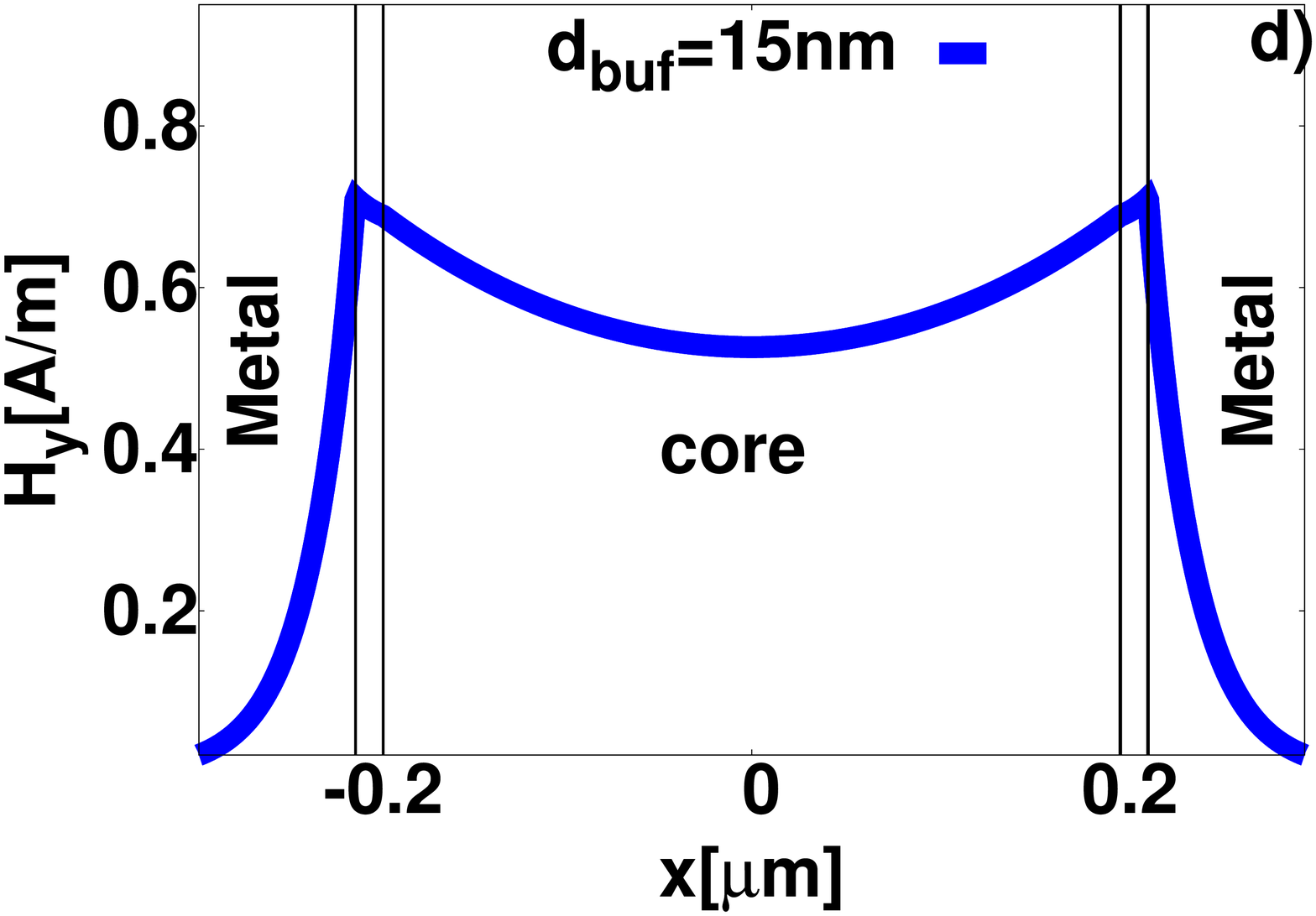}    
  \caption{(a) Linear dispersion curves of the fundamental symmetric TM mode for the  improved NPSW with $d_{core}=400$ (solid lines) and with $d_{core}=300$ nm (dashed lines) as a function  of the thickness of the supplementary buffer dielectric layers $d_{buf}$  for two different layer refractives indices. Typical magnetic field symmetric components $H_{y}(x)$ for $\varepsilon_{buf}=2.5^2$: (b) A cosine-type profile for $d_{buf}=40$ nm such that $\Re e(n_{eff}) < \Re e(\varepsilon_{l,core})^{1/2}$. (c) Flat type profile in the core for $d_{buf}=d^{up}_{buf}=30.82$ nm, with $\Re e(n_{eff})  \approx \Re e(\varepsilon_{l,core})^{1/2}$. (d) Plasmonic type profile for $d_{buf}=15$ nm such that  $\Re e(n_{eff})  > \Re e(\varepsilon_{l,core})^{1/2}$.} 
  \label{fig:linear-profiles}
\end{figure}

For the TM case, we start by the study of the linear case as it is well known that this is a crucial step in the investigation of nonlinear waveguides~\cite{snyder:91}.
Fig.~\ref{fig:linear-profiles}(a) provides the linear dispersion curves fo
\begin{equation}
\label{eqn:lin_disp_half_structure}
\frac{\left( \displaystyle\frac{q}{\epsilon_{buf}}-\displaystyle\frac{r}{\epsilon_{met}} \right)} {\left(\displaystyle \frac{q}{\epsilon_{buf}}+\displaystyle\frac{r}{\epsilon_{met}} \right)}=     \displaystyle\frac{\left( \displaystyle\frac{q}{\epsilon_{buf}}+\displaystyle\frac{p}{\epsilon_{l,core}} \right)} {\left( \displaystyle\frac{q}{\epsilon_{buf}}-\displaystyle\frac{p}{\epsilon_{l,core}} \right)}      ~e^{2k_{0}q \left( d_{buf}\right) }, 
\end{equation}
\begin{subequations}
\label{eqn: p_q_r}
\begin{align}
 p^2=(\Re e(n_{eff}))^{2}-\epsilon_{l,core},\\
 q^2=(\Re e(n_{eff}))^{2}-\epsilon_{buf},\\
 r^2=(\Re e(n_{eff}))^{2}-\epsilon_{met}.
\end{align}
\end{subequations}
Equations \eqref{eqn:lin_disp_half_structure} and \eqref{eqn: p_q_r} give the linear dispersion relation for TM waves for a linear dielectric buffer layer sandwiched between a semi-infinite metal region and a semi-infinite high-index dielectric region. By letting $(\Re e(n_{eff}))^{2} \rightarrow \epsilon_{l,core}$ in Eqns.~\eqref{eqn:lin_disp_half_structure} and \eqref{eqn: p_q_r}, one gets:
\begin{equation}
\begin{aligned}
\label{eqn:full_critical_dbuf_final}
 & d^{up}_{buf}   \approx  \frac{\lambda}{4 \pi \sqrt{Re(\epsilon_{l,core})- \epsilon_{buf} } } \times \\
 & \resizebox{\linewidth}{!}{$\ln \left[ \frac{ Re(\epsilon_{met}) \sqrt{Re(\epsilon_{l,core})- \epsilon_{buf} }- \epsilon_{buf} \sqrt{Re(\epsilon_{l,core})- Re(\epsilon_{met}) } } { Re(\epsilon_{met}) \sqrt{Re(\epsilon_{l,core})- \epsilon_{buf} } +\epsilon_{buf} \sqrt{Re(\epsilon_{l,core})- Re(\epsilon_{met}) }} \right]$.}
 \end{aligned}
\end{equation}
One clearly see that, the critical buffer thickness given by Eq.~\eqref{eqn:full_critical_dbuf_final}, does not depend on the core thickness $d_{core}$ which can be inferred from our FEM simulations depicted in Fig.~\ref{fig:linear-profiles}(a). Using the parameters in Sec.~\ref{sec:models}, $\varepsilon_{buf}=2.5^2$ and $d_{buf}=15$ nm, the full formula given by Eq.~\eqref{eqn:full_critical_dbuf_final} gives $d^{up}_{buf}=31.81$ nm, this result agrees quite well with our FEM simulations (30.8 nm) as it is shown for the blue curves in Fig.~\ref{fig:linear-profiles}(a). An approximated value for $d^{up}_{buf}$ (28.4 nm) can be estimated for $Re(\epsilon_{l,core})<<|Re(\epsilon_{met})|$ as:
\begin{align}
\label{eqn:estimation_critical_dbuf_epscore_LT_epsmet}
d^{up}_{buf}\approx \frac{\lambda}{2 \pi} ~\frac{\epsilon_{buf}}{\sqrt{|Re(\epsilon_{met})|}}~ \frac{1}{Re(\epsilon_{l,core}-\epsilon_{buf})}.
\end{align}
\subsection{Nonlinear case}
\label{sec:results-nonlinear-TM}
\begin{figure}[htbp]
\centering
  \includegraphics[width = 0.49\columnwidth,angle=-0,clip=true,trim= 30 35 50 10]{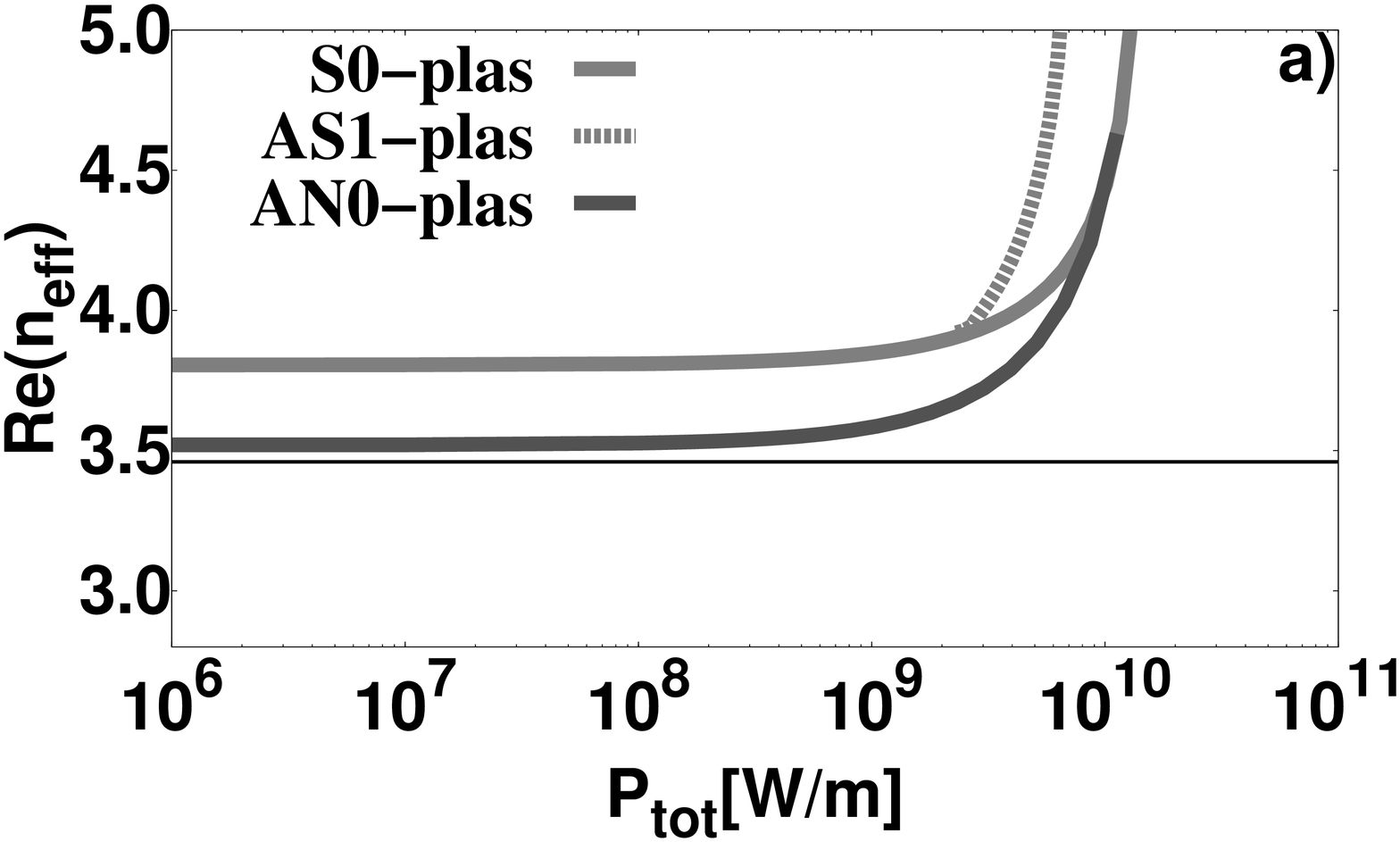}
    \includegraphics[width = 0.49\columnwidth,angle=-0,clip=true,trim= 30 35 50 10]{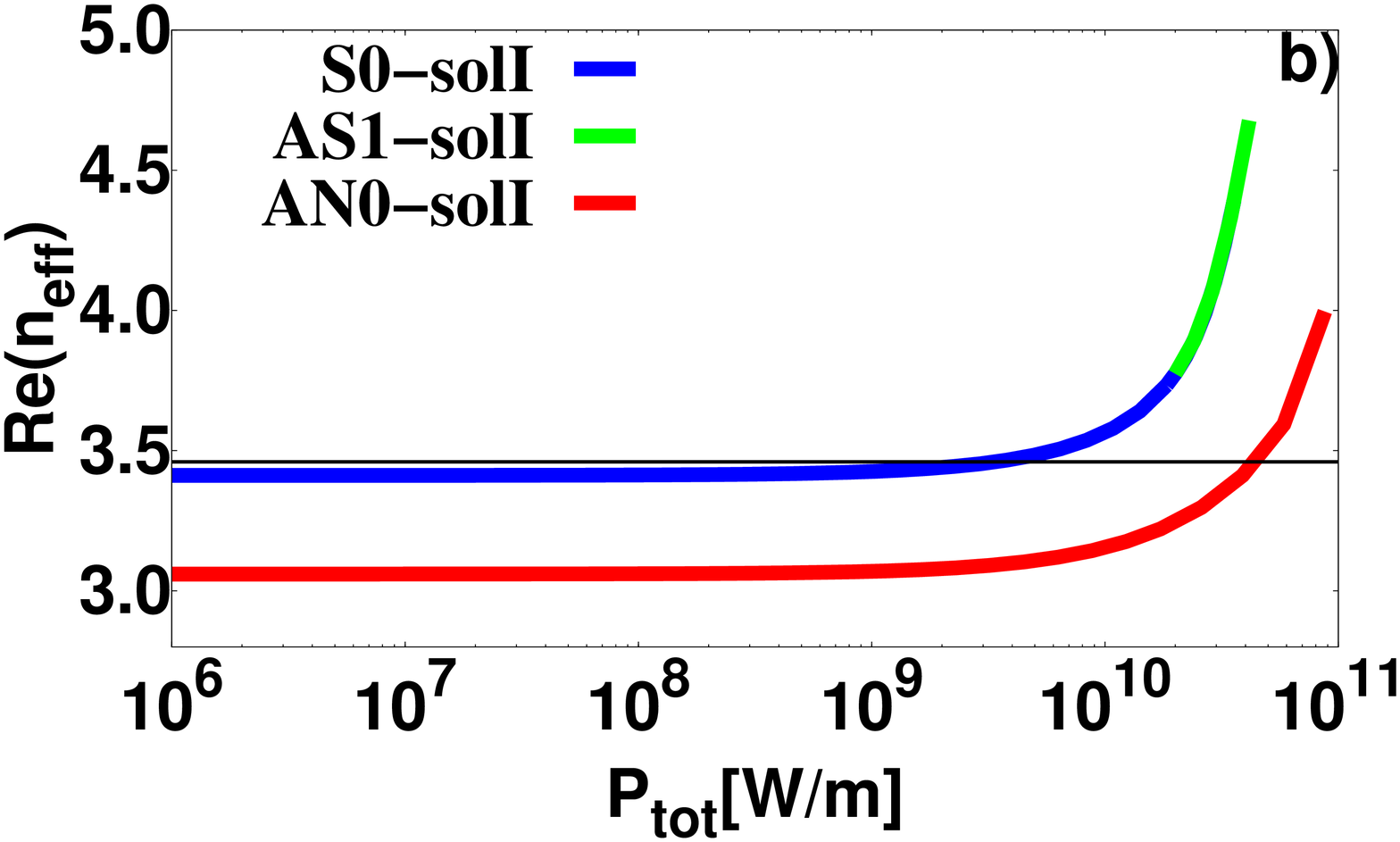}\\
    \includegraphics[width = 0.49\columnwidth,angle=-0,clip=true,trim= 30 35 50 10]{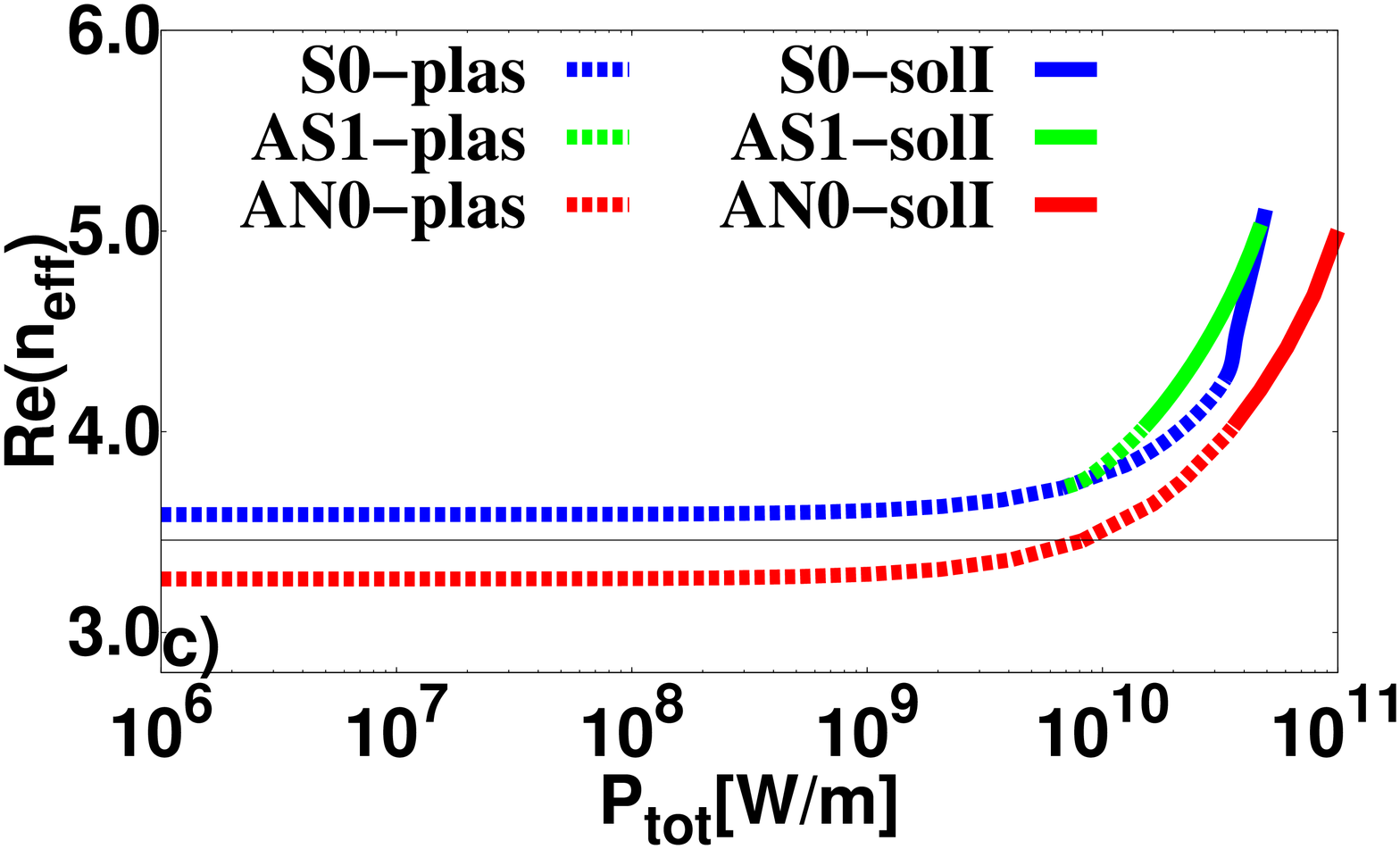}
\includegraphics[width=0.49\columnwidth,angle=-0,clip=true,trim= 30 35 50 10]{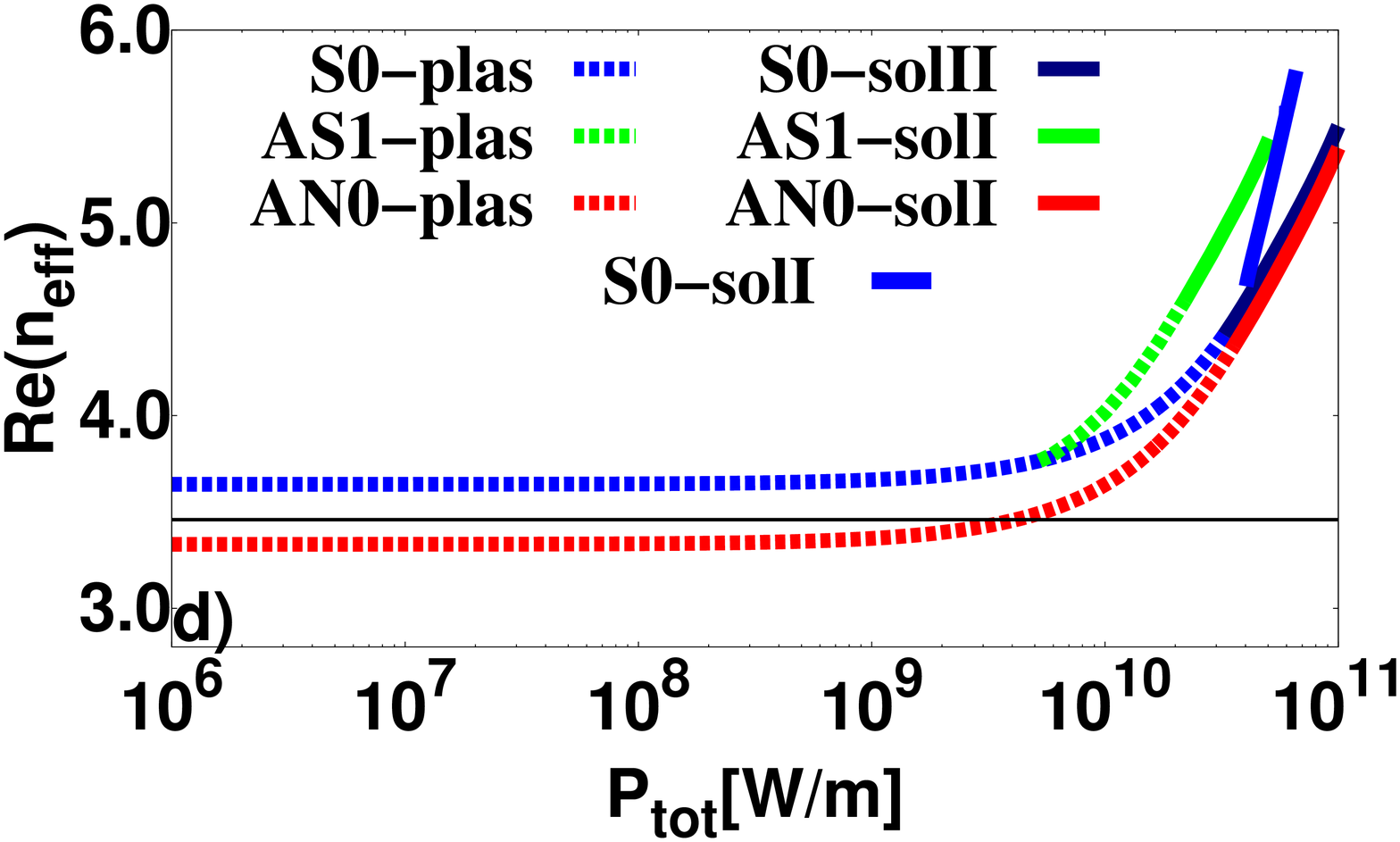}
  \caption{Nonlinear dispersion curves for the simple NPSW (grey curves) and for the improved one (color curves). The blue, green, and red colors indicate the symmetric, asymmetric and antisymmetric nonlinear TM modes, respectively. $d_{core}=400$ nm, $\varepsilon_{buf}=2.5^2$.  (b) for $d_{buf}=40$ nm, (c) for $d_{buf}=15$ nm, and (d) for $d_{buf}=10$ nm.}
  \label{fig:nl-dispersion-curve}
\end{figure}

\begin{figure}[htbp]
\centerline{
\includegraphics[width=0.33\columnwidth,angle=-0,clip=true,trim= 20 0 30 0]{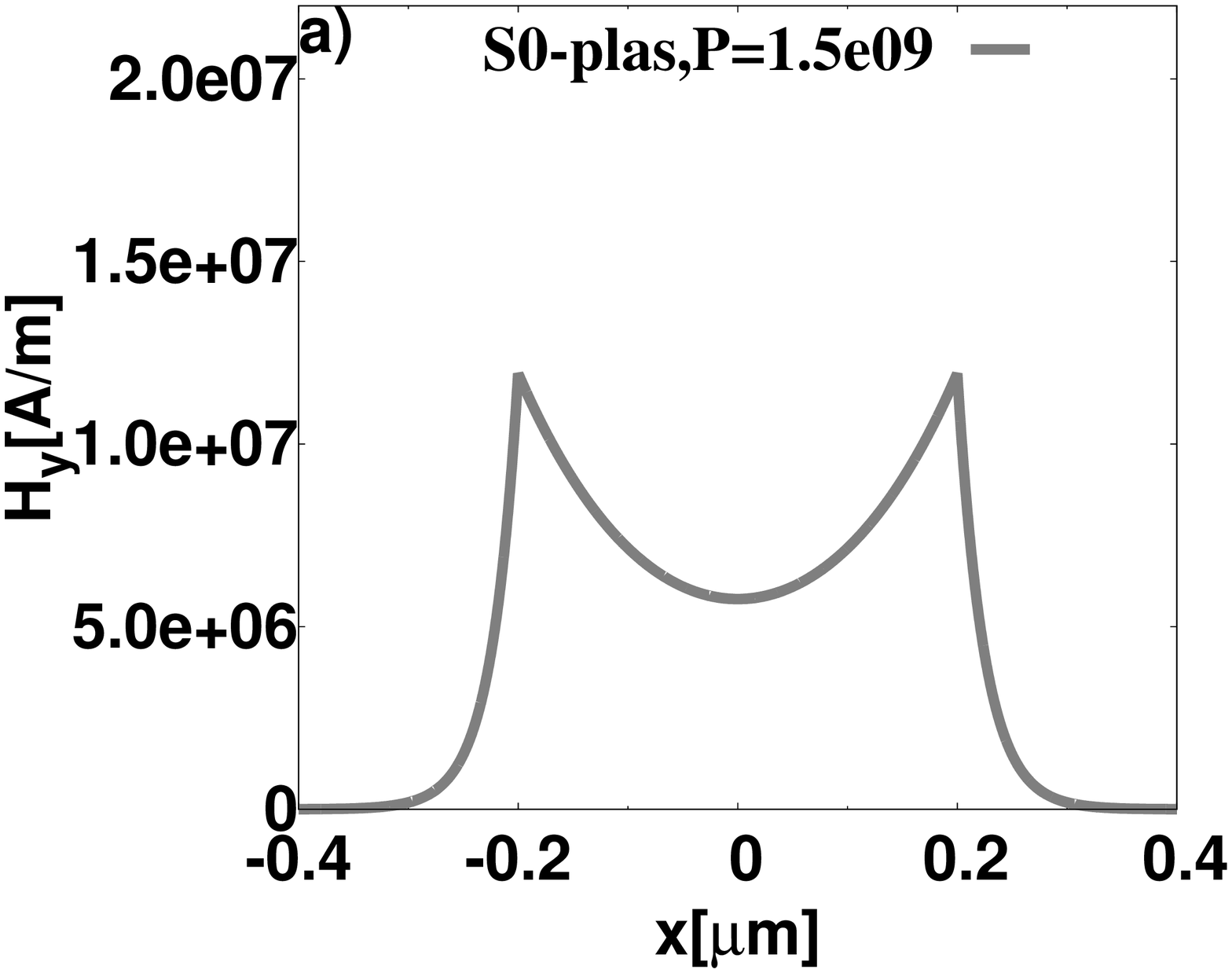}
\includegraphics[width=0.33\columnwidth,angle=-0,clip=true,trim= 20 0 30 0]{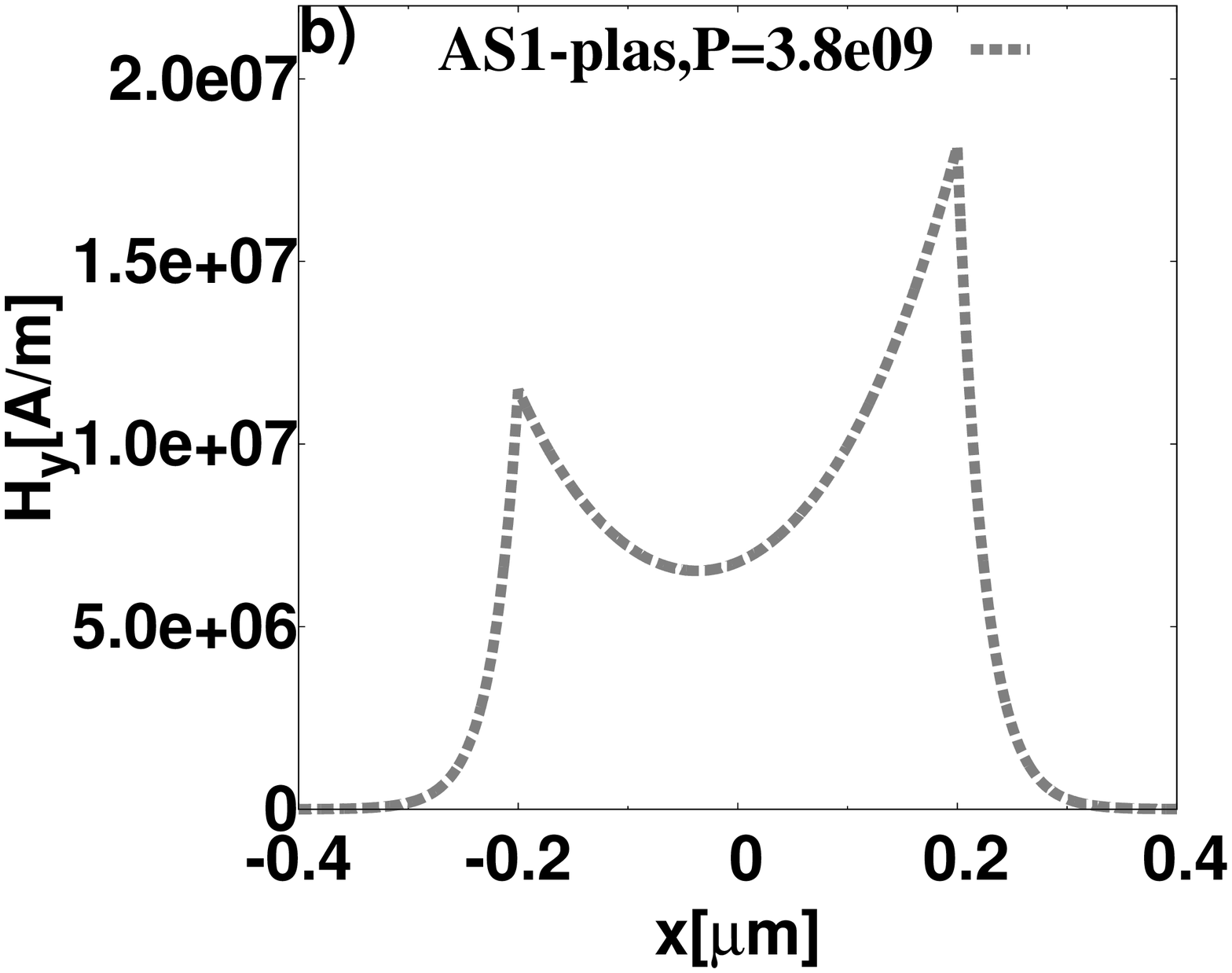}
\includegraphics[width=0.33\columnwidth,angle=-0,clip=true,trim= 20 0 30 0]{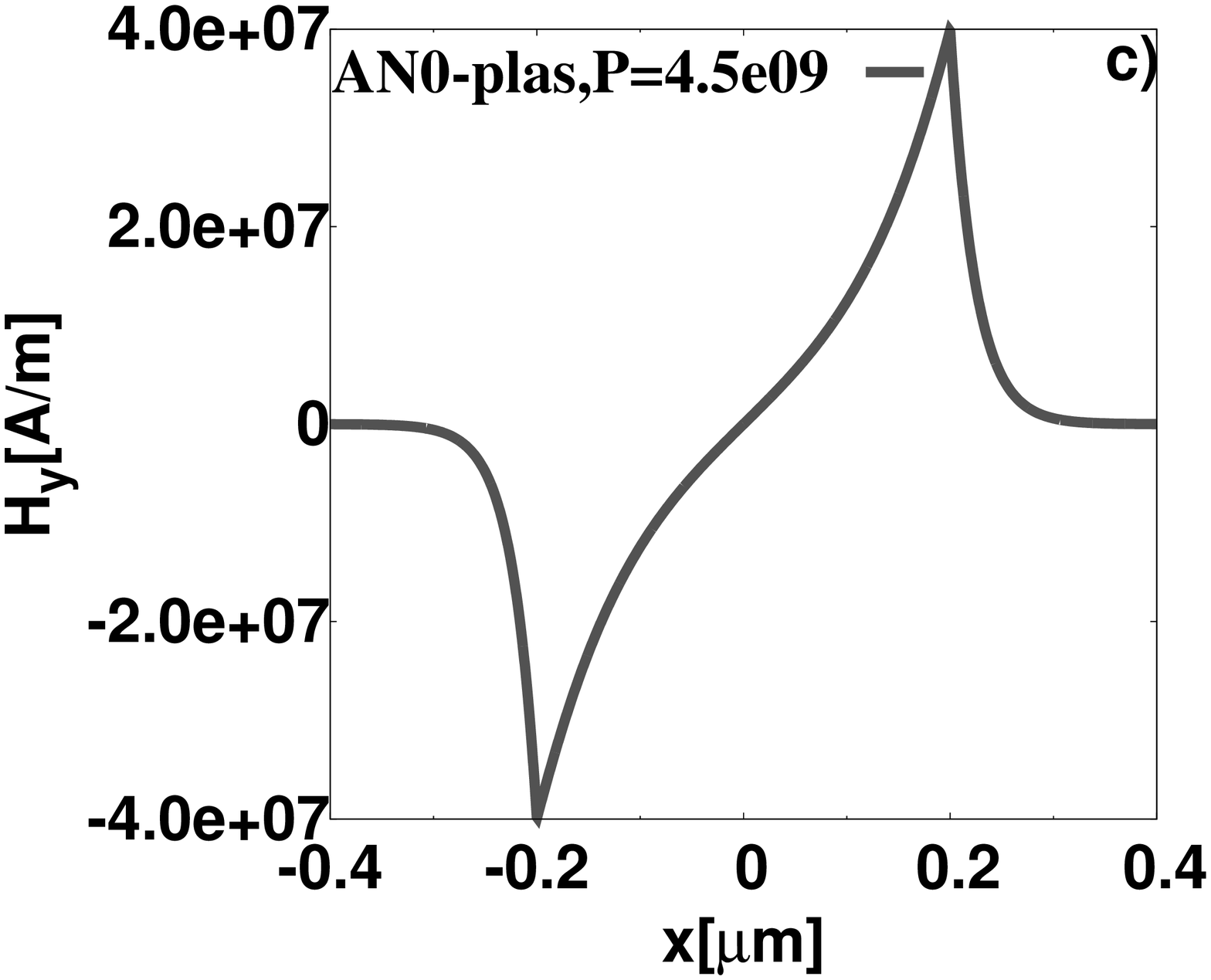}}  
 \centerline{\includegraphics[width=0.33\columnwidth,angle=-0,clip=true,trim= 20 0 30 0]{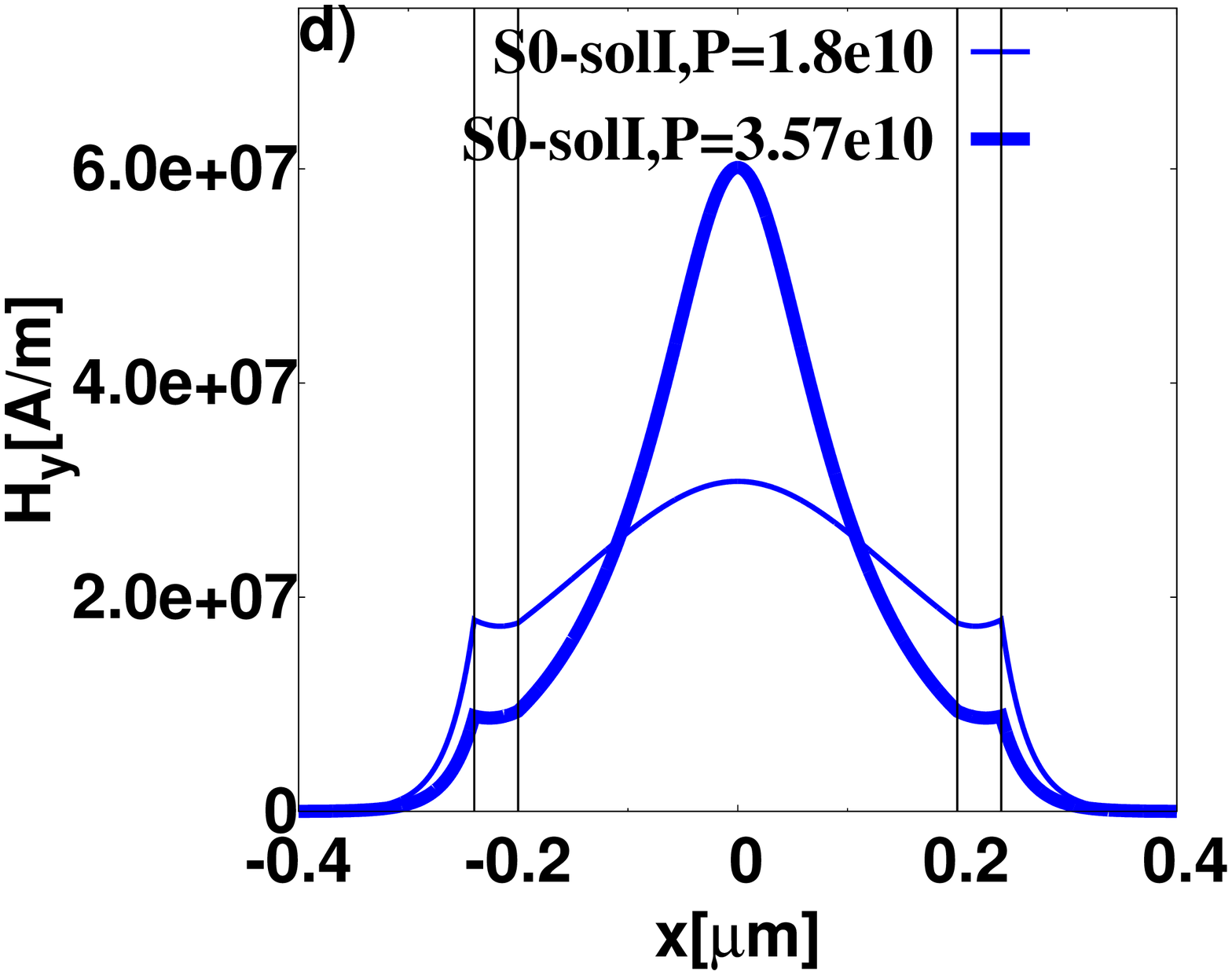}
\includegraphics[width=0.33\columnwidth,angle=-0,clip=true,trim= 20 0 30 0]{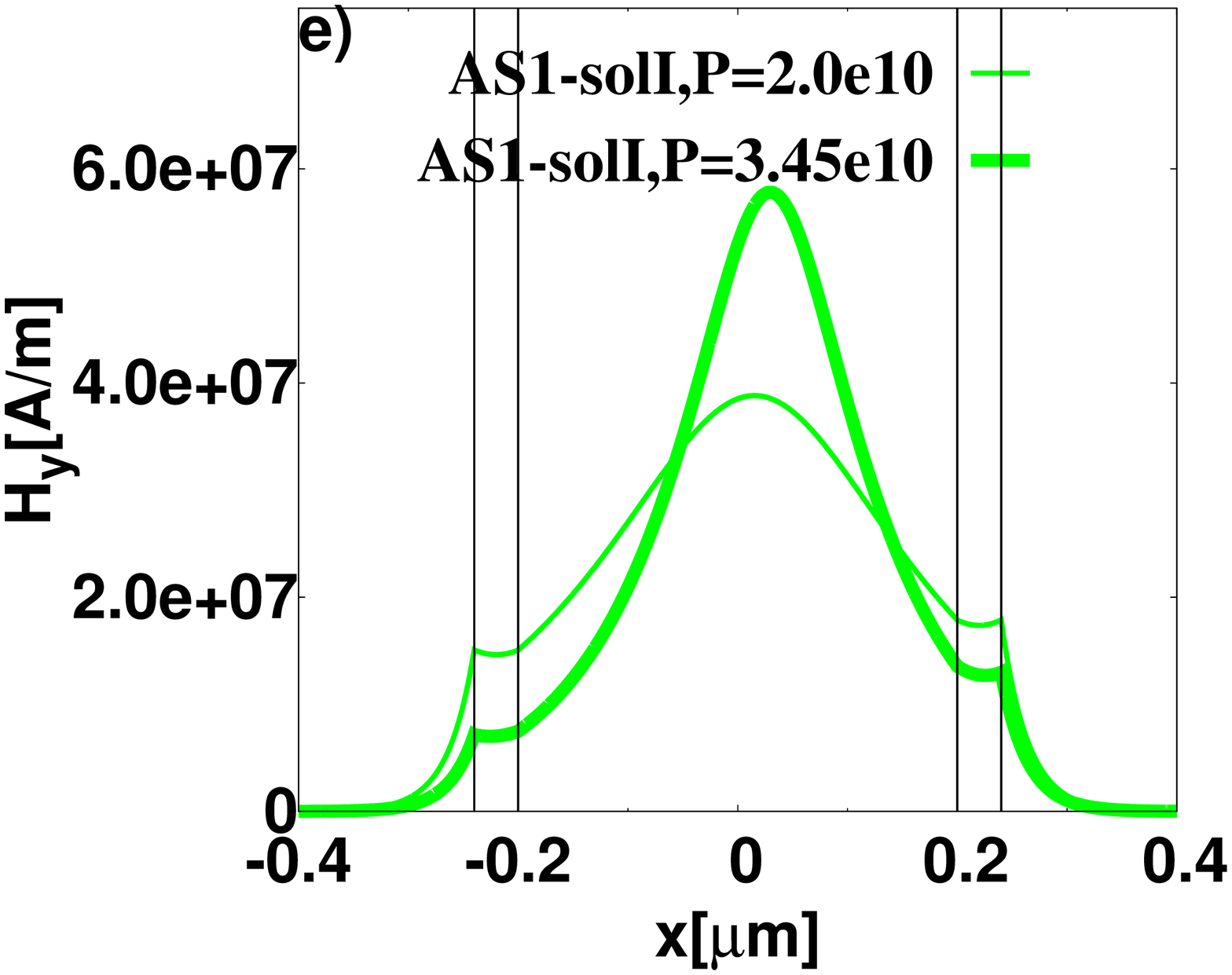}
\includegraphics[width=0.33\columnwidth,angle=-0,clip=true,trim= 20 0 30 0]{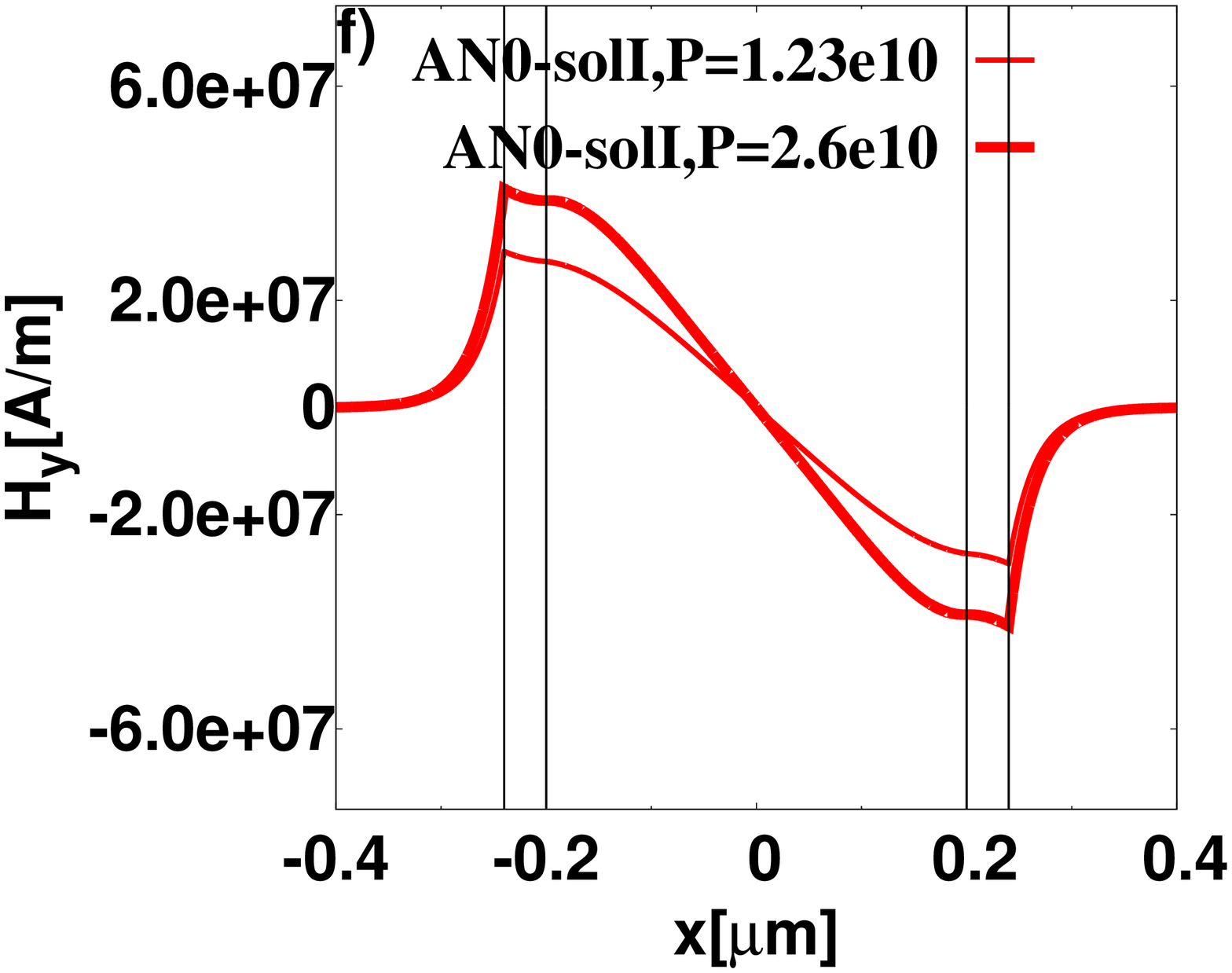}}
\caption{Field profiles $H_y(x)$ for the simple NPSW (top) and for the improved NPSWG (bottom) with $d_{buf}=40$. The symmetric (blue curves), asymmetric (green curves) , and antisymmetric (red curves) TM modes. The power is given in W/m.}
\label{fig:nonlinear-profile-modal-dbuf40nm-transition}
\end{figure}

For the nonlinear case, we study the influence of the supplementary  buffer dielectric layers on the nonlinear dispersion curves. We consider the case in which $\varepsilon_{buf}=2.5^2$ and $d_{core}=400$ nm, which corresponds to the solid blue curve in Fig.~\ref{fig:linear-profiles}(a) with the associated linear field profiles depicted in Fig.~\ref{fig:linear-profiles}(b), (c), and (d). We start with the case such that the effective index of the symmetric linear mode is below the core linear refractive index in which the symmetric linear mode is localized in the high-index core with plasmonic tails in the metal regions (see Fig.~\ref{fig:linear-profiles}(b)). The real part of effective indices of the first three main modes: symmetric, asymmetric and antisymmetric denoted by; S0-solI, AS1-solI, and AN0-solI, respectively are shown in Fig.~\ref{fig:nl-dispersion-curve}(b) with the corresponding field profiles depicted in Fig.~\ref{fig:nonlinear-profile-modal-dbuf40nm-transition}(d), (e), and (f). The results for a simple NPSW made from the improved NPSW by removing the buffer dielectric layers are also shown in Fig.~\ref{fig:nl-dispersion-curve}(a) for comparison (symmetric S0-plas, asymmetric AS1-plas, antisymmetric AN0-plas) and the associated field profiles are given in Fig.~\ref{fig:nonlinear-profile-modal-dbuf40nm-transition}(a), (b), and (c). 
The mode notation used in our study has been chosen to extend coherently the one already introduced for the simple NPSW~\cite{Walasik15a,Walasik15b}. As it can be seen, in the improved NPSW, the bifurcation of the asymmetric mode from the symmetric mode and the relative positions of the three main modes are preserved.  The difference between the nonlinear symmetric mode curve  and the asymmetric nonlinear mode curve is smaller in this new structure (see Fig.~\ref{fig:nl-dispersion-curve}(b)). The reason is that, in this case, the nonlinear symmetric mode profile looks like a symmetric hyperbolic secant spatial soliton and its associated asymmetric mode is simply shifted off the slot center as it is shown in Fig.~\ref{fig:nonlinear-profile-modal-dbuf40nm-transition}(d) and (e).

\begin{figure}[htbp]
\centerline{
\includegraphics[width=0.33\columnwidth,angle=-0,clip=true,trim= 20 0 30 0]{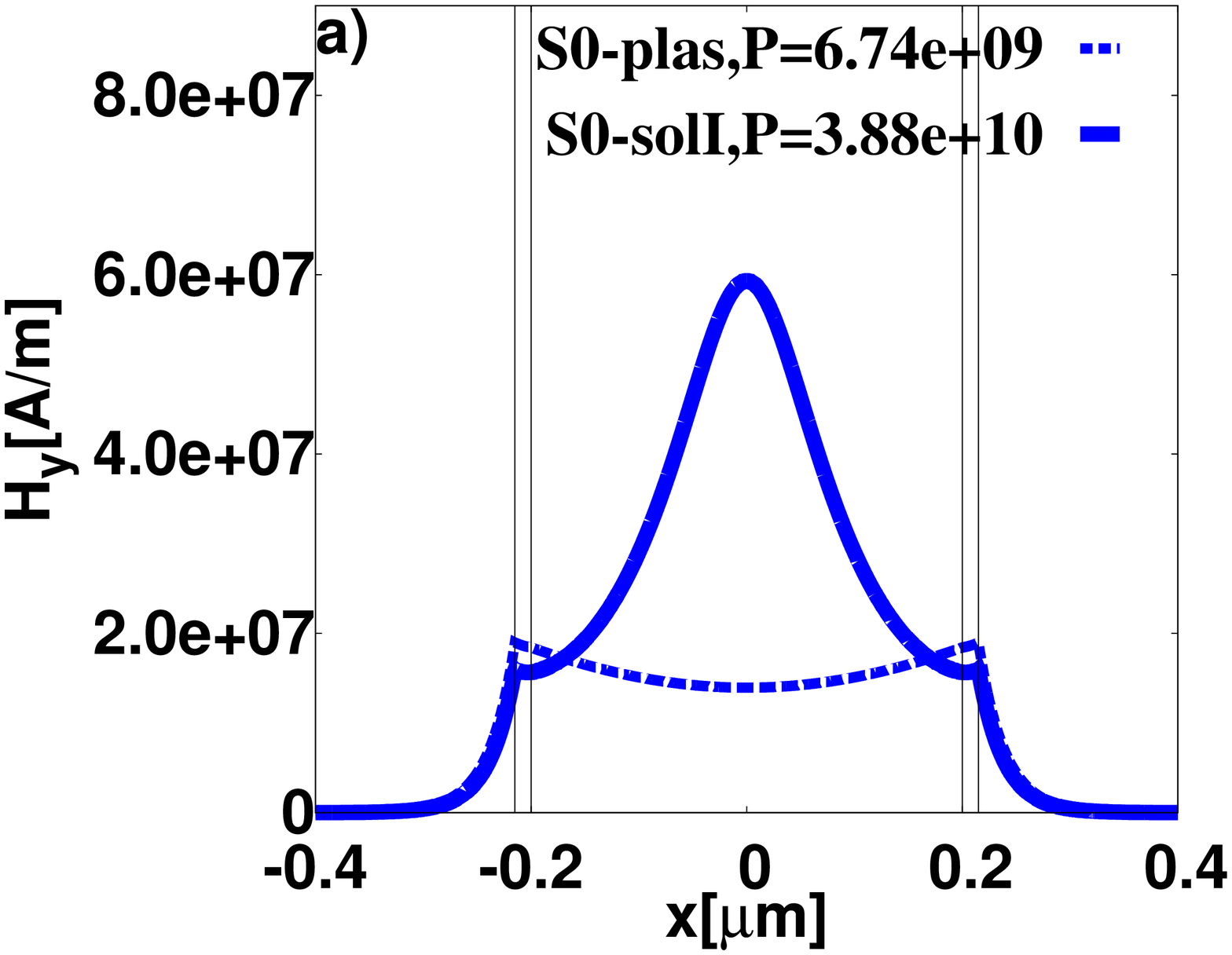}
\includegraphics[width=0.33\columnwidth,angle=-0,clip=true,trim= 20 0 30 0]{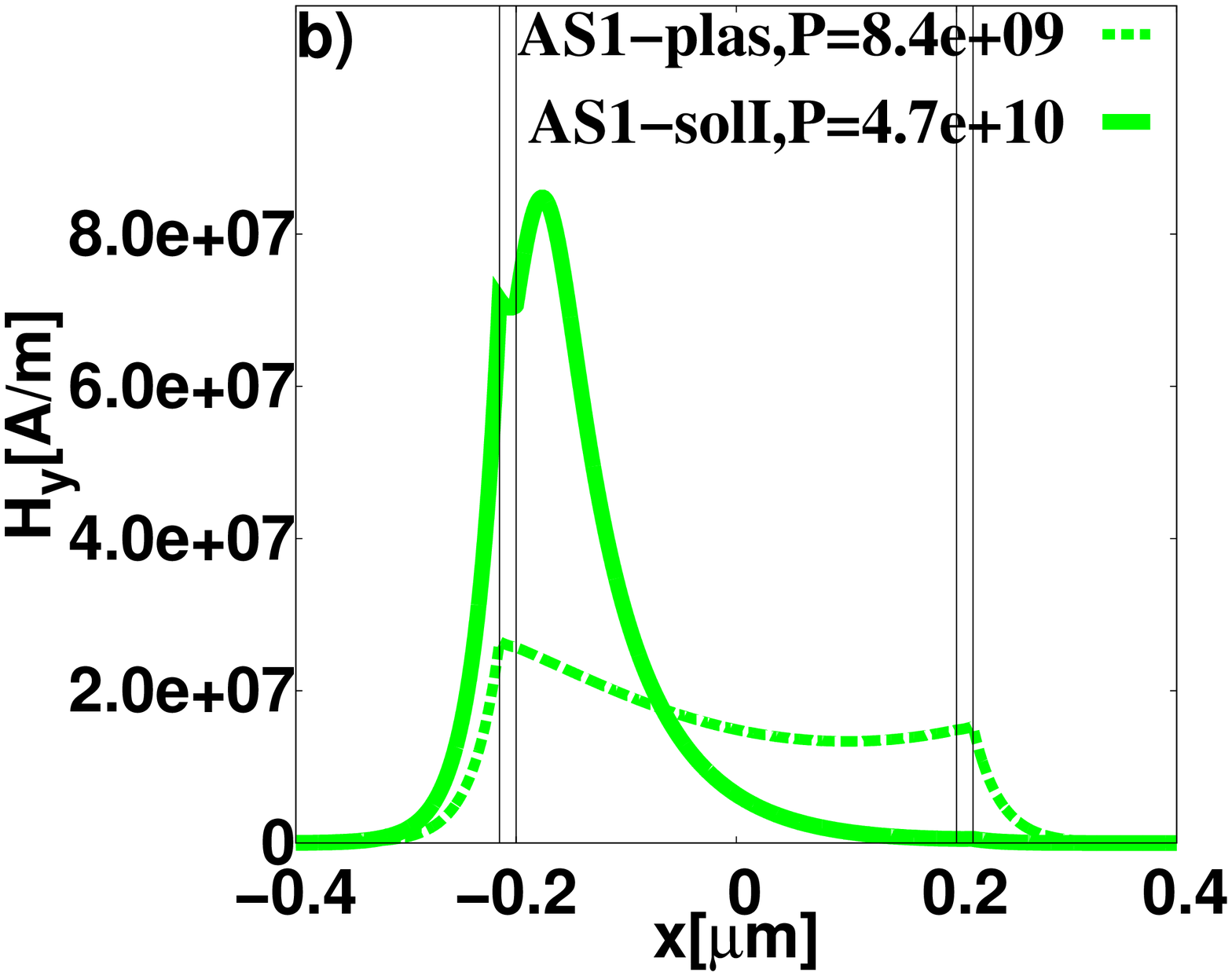}
\includegraphics[width=0.33\columnwidth,angle=-0,clip=true,trim= 20 0 30 0]{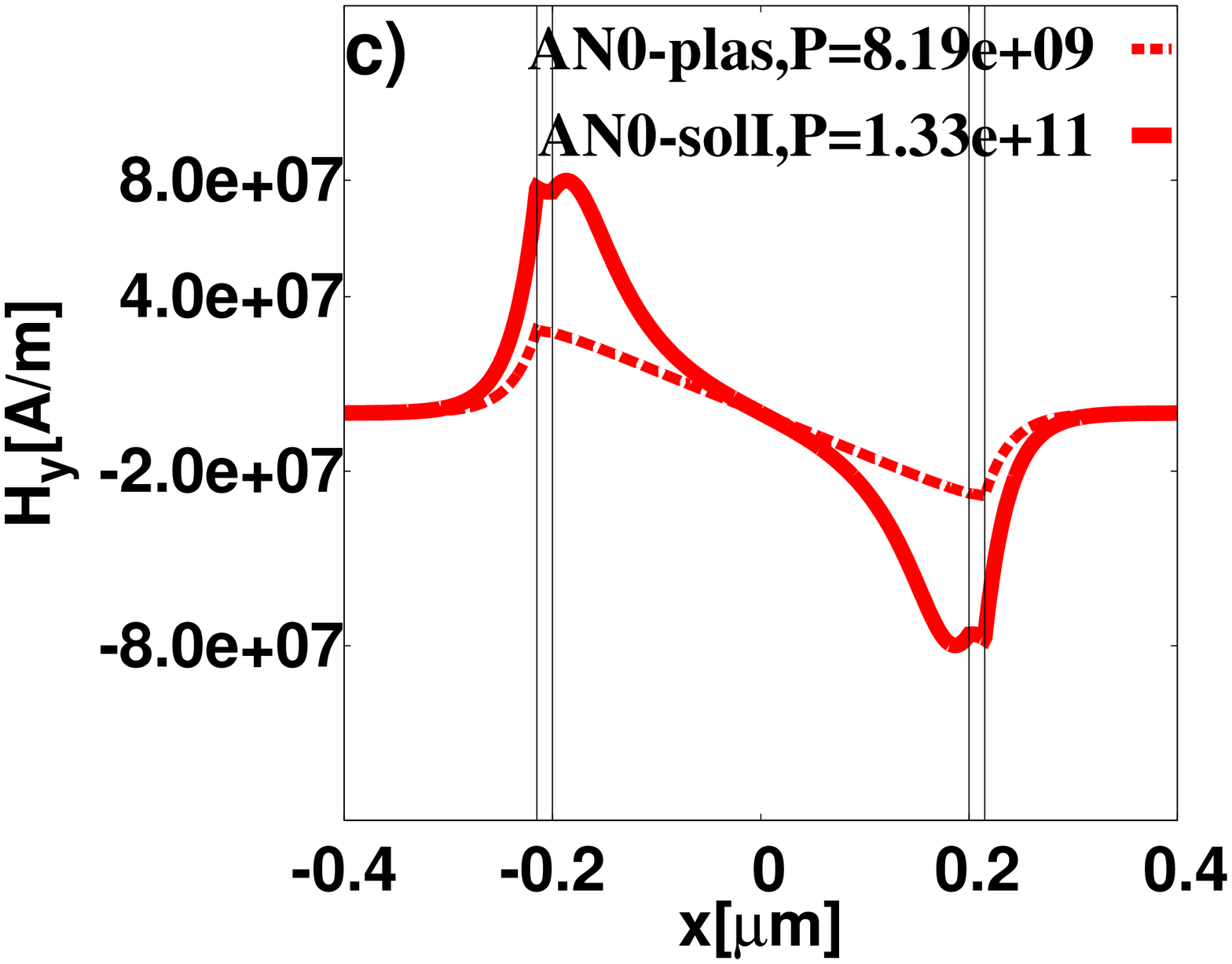}}
\caption{Field profiles $H_y(x)$ for the symmetric (blue curves), asymmetric (green curves), and antisymmetric (red curves) TM modes with $d_{buf}=15$ nm, before (dashed line) and after (solid line) the spatial modal transition induced by the power increase. The power is given in W/m.}
\label{fig:nonlinear-profile-modal-dbuf15nm-transition}
\end{figure}

\begin{figure}[htbp]
  \centerline{
  \includegraphics[width=0.33\columnwidth,angle=-0,clip=true,trim= 20 0 30 0]{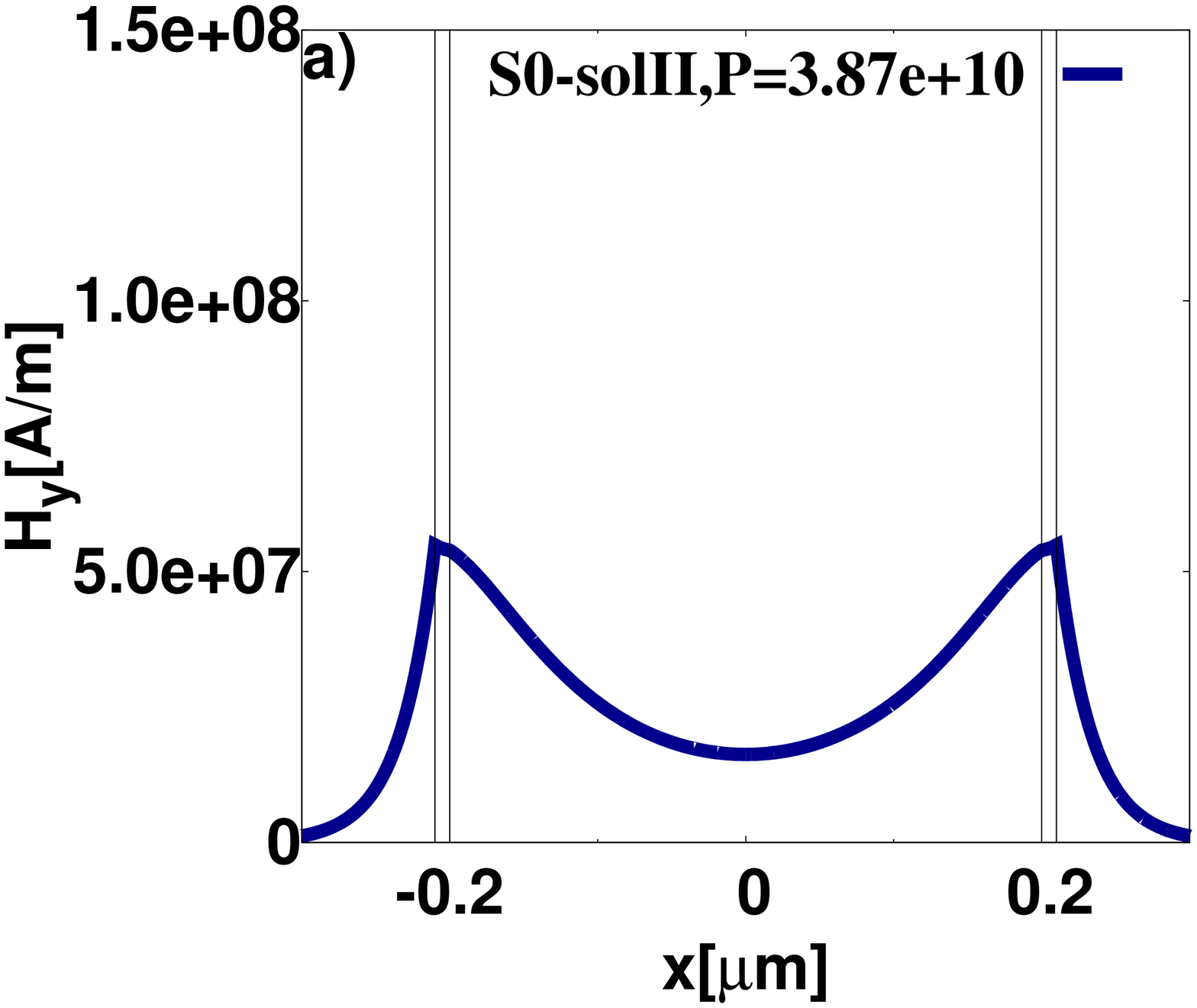}
  \includegraphics[width=0.33\columnwidth,angle=-0,clip=true,trim= 20 0 30 0]{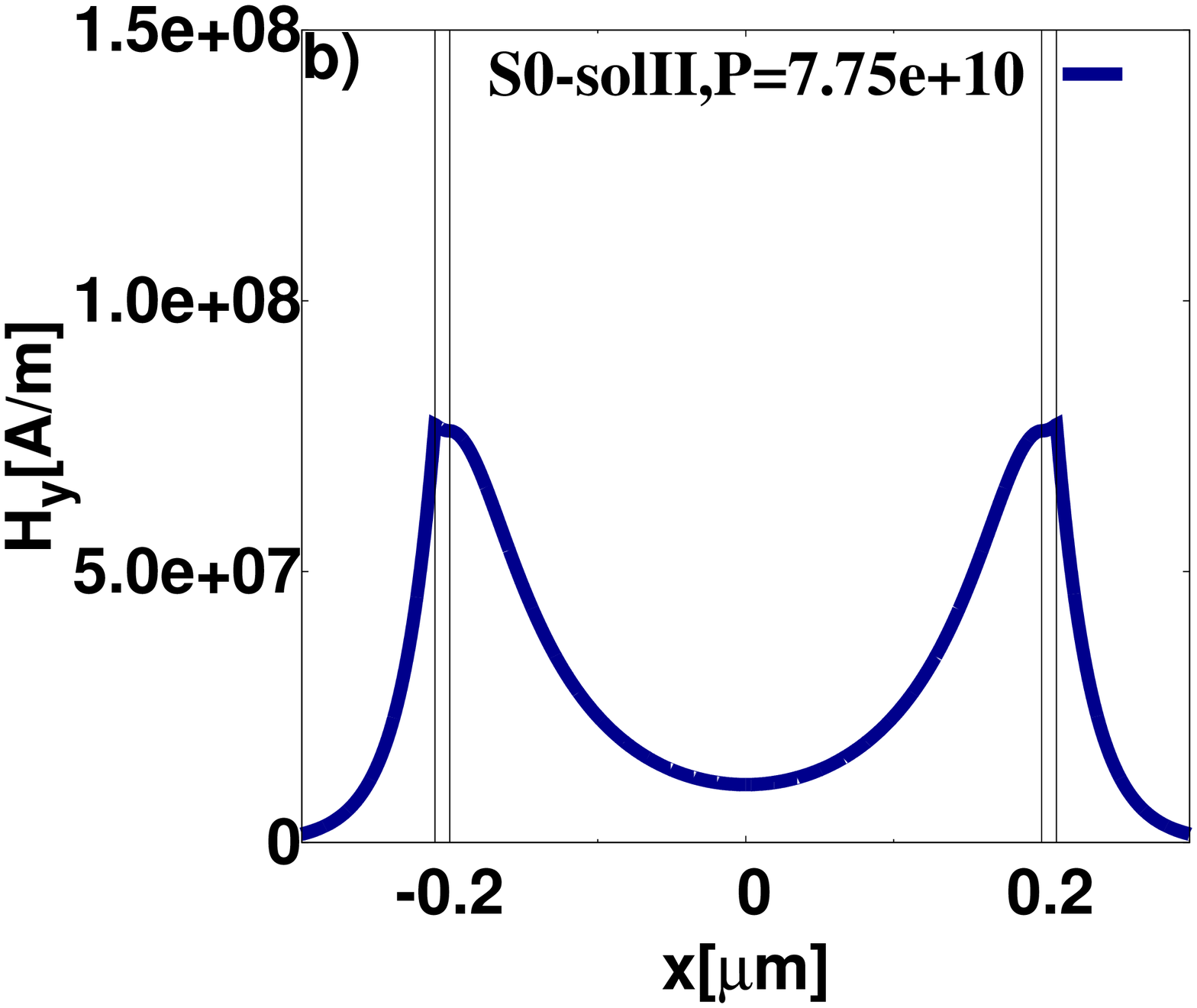}
    \includegraphics[width=0.33\columnwidth,angle=-0,clip=true,trim= 20 0 30 0]{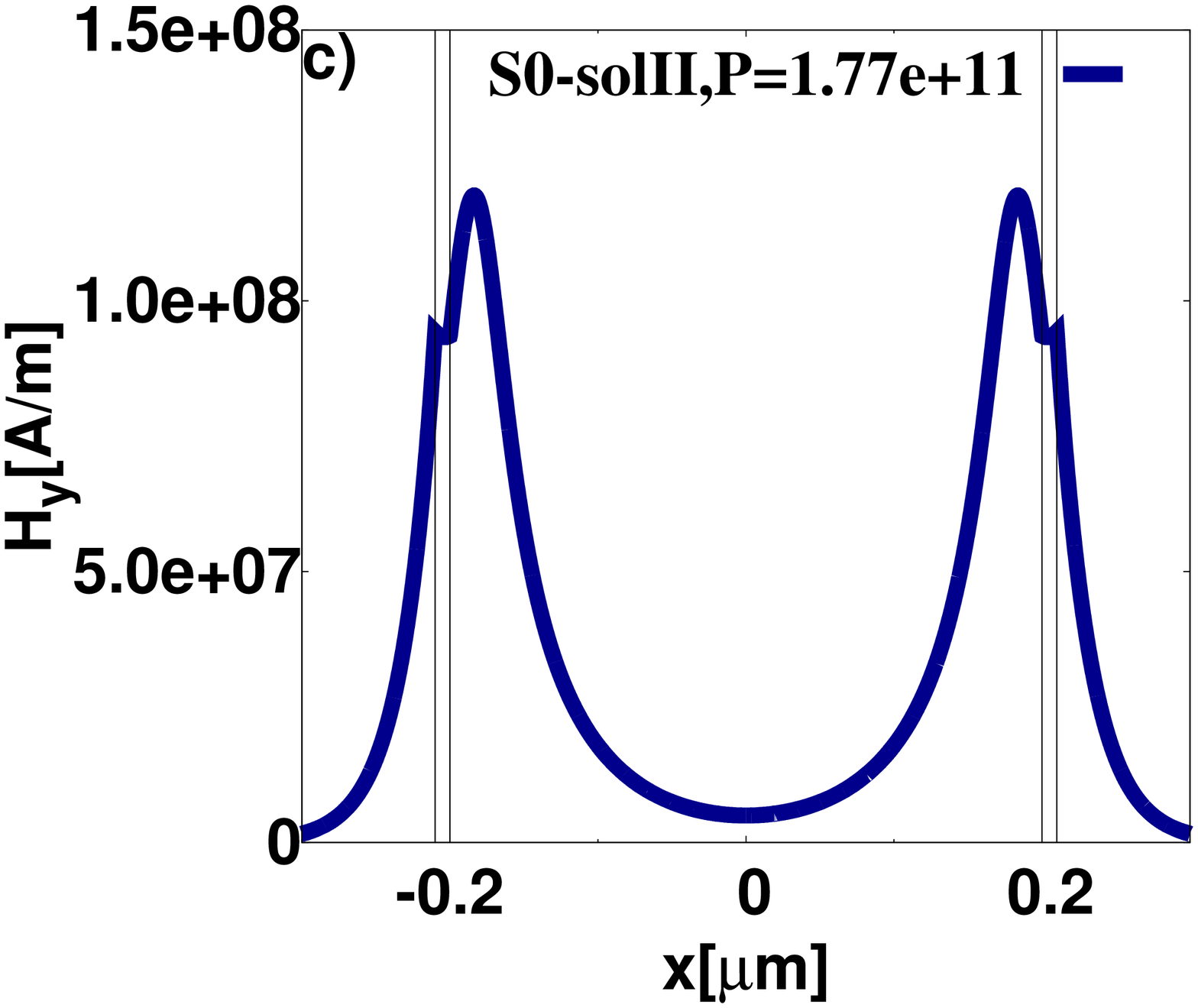}}
   \centerline{ \includegraphics[width=0.33\columnwidth,angle=-0,clip=true,trim= 20 0 30 0]{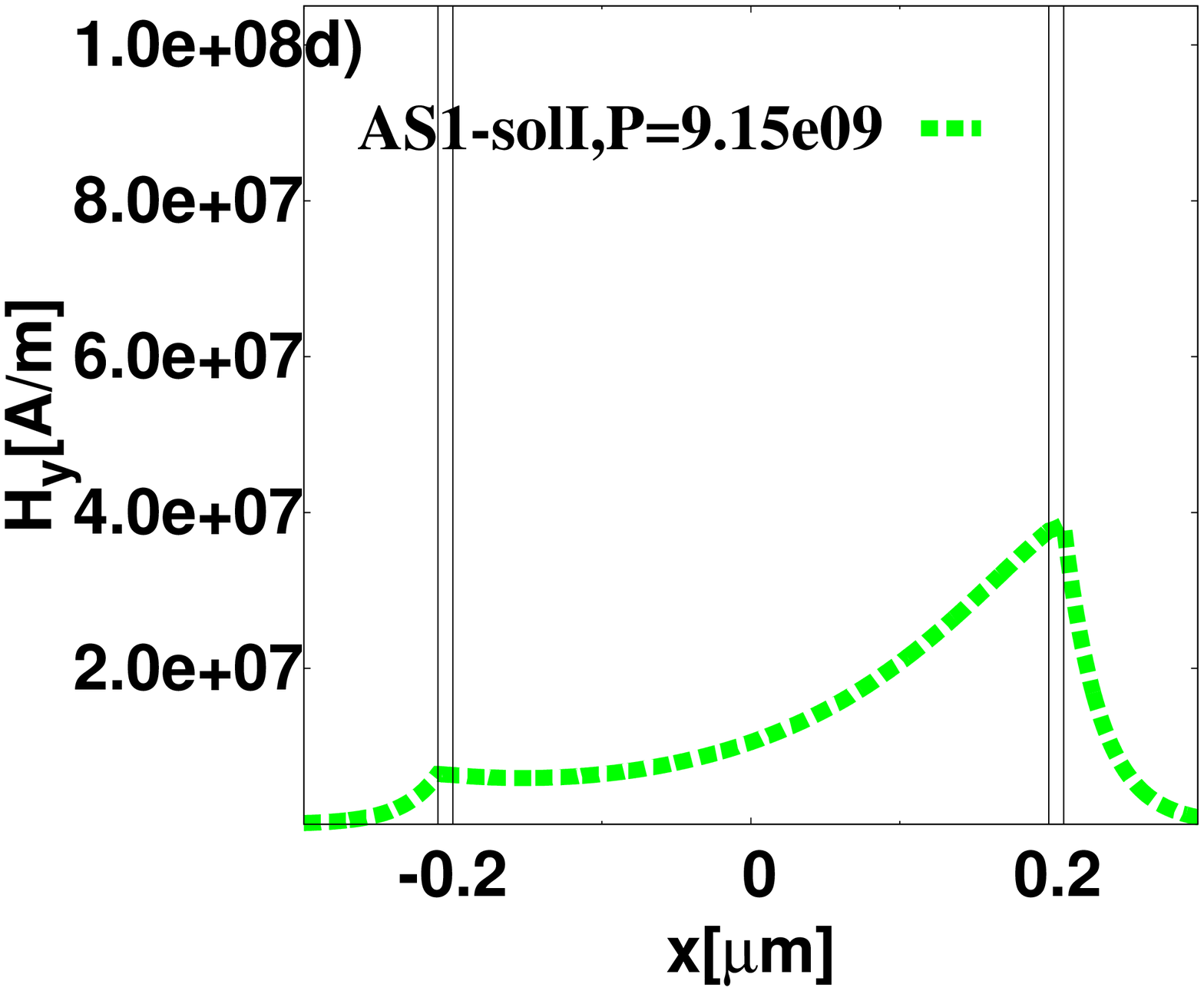}
    \includegraphics[width=0.33\columnwidth,angle=-0,clip=true,trim= 20 0 30 0]{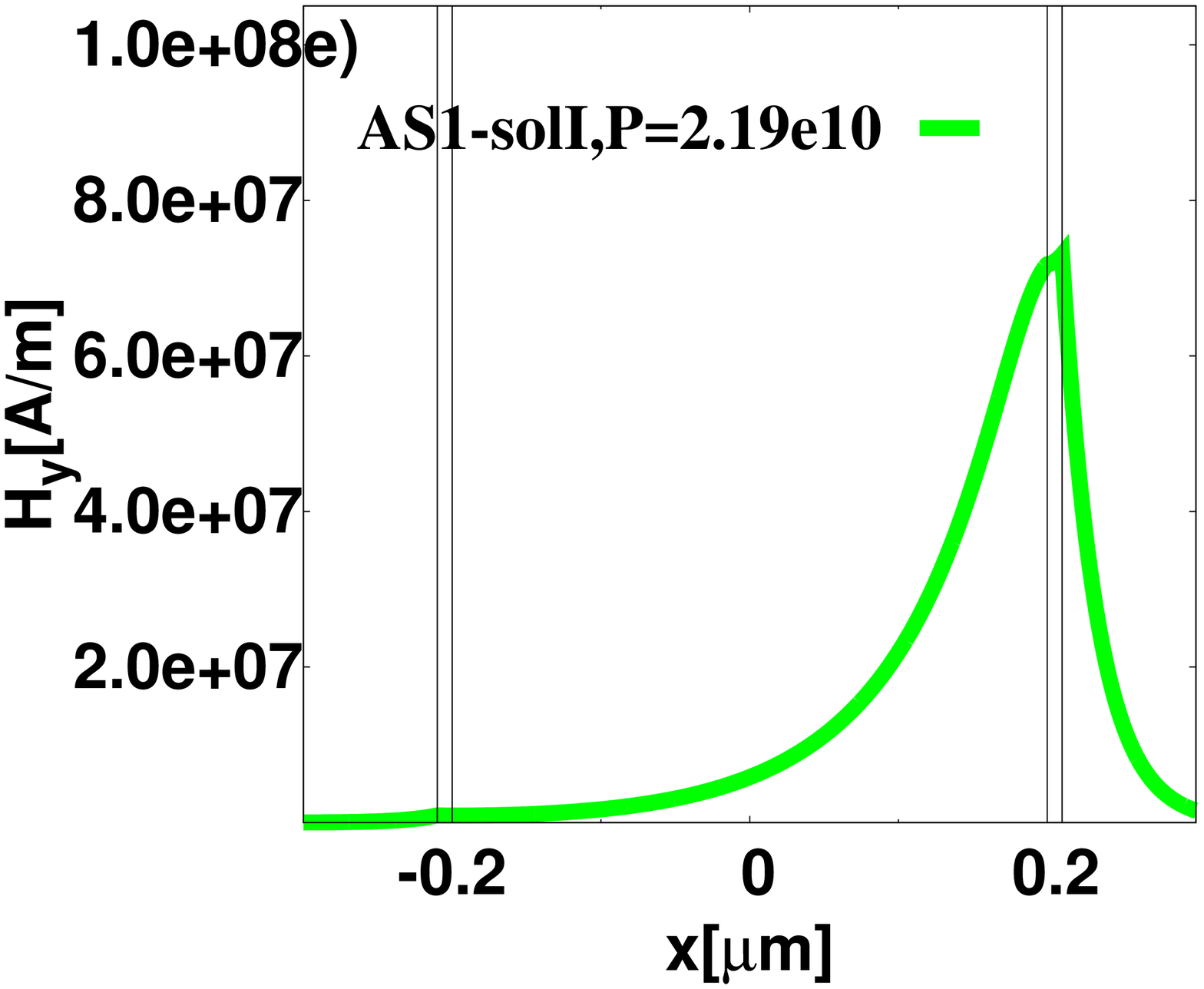}
    \includegraphics[width=0.33\columnwidth,angle=-0,clip=true,trim= 20 0 30 0]{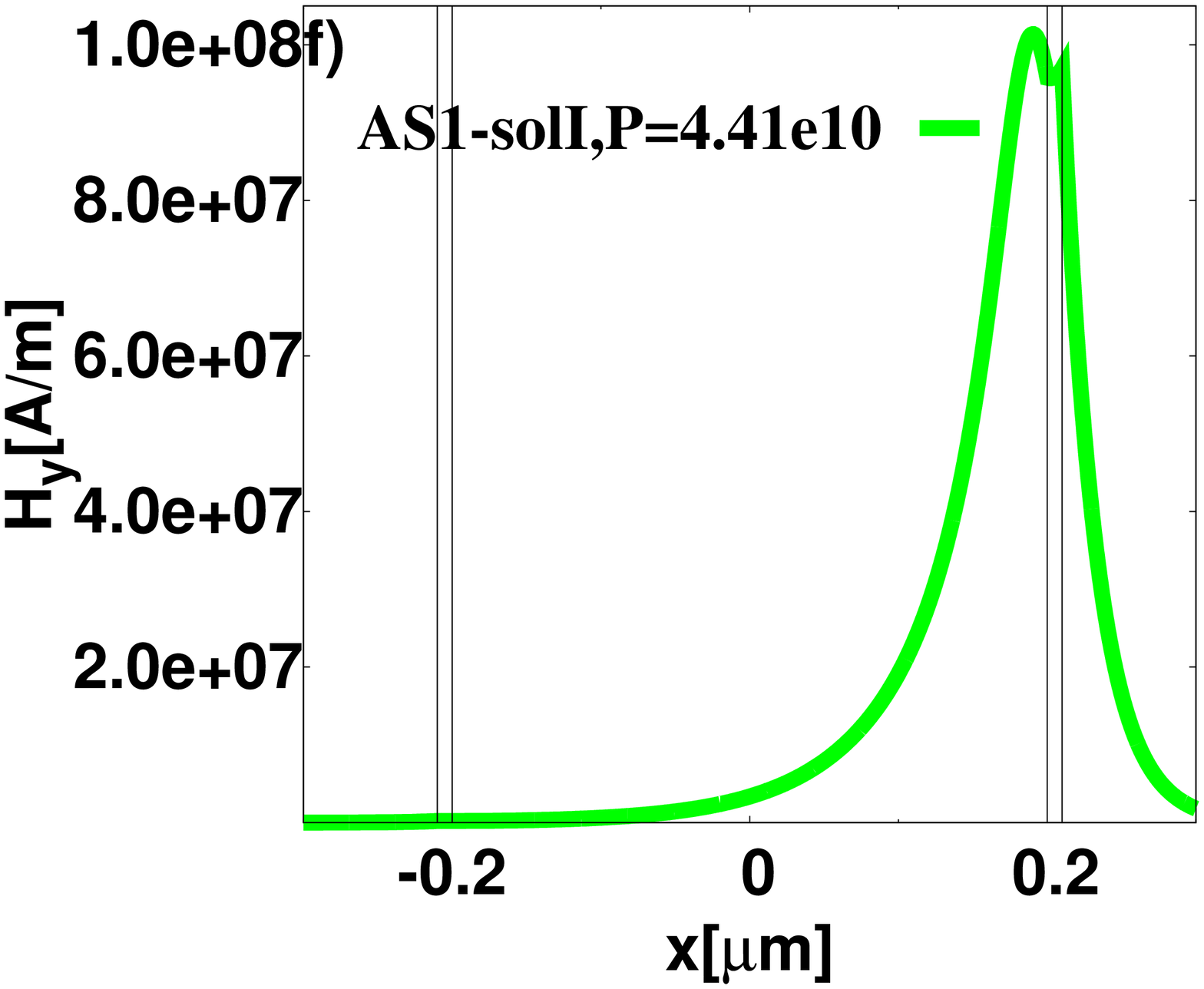}}
    \centerline{\includegraphics[width=0.33\columnwidth,angle=-0,clip=true,trim= 20 0 30 0]{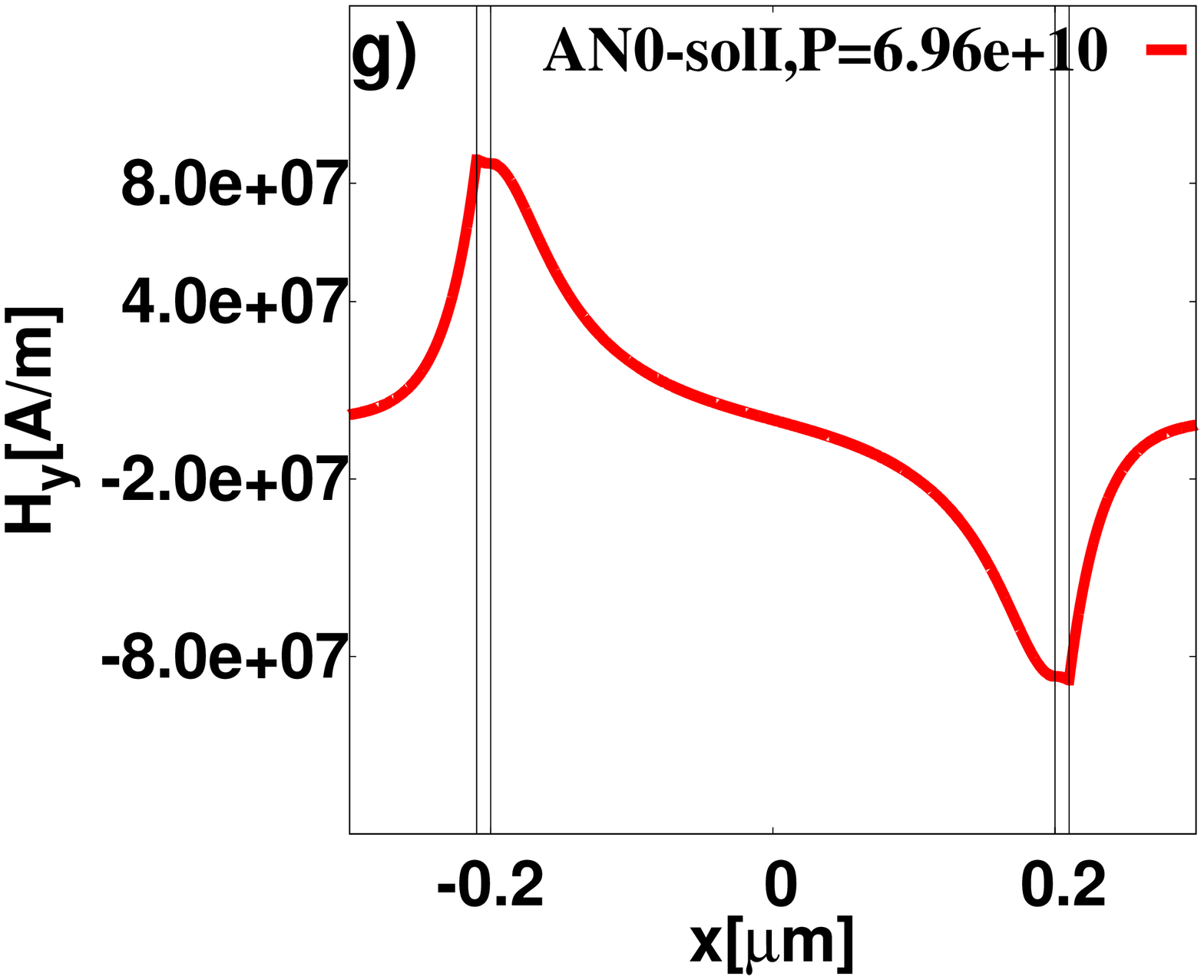}   
    \includegraphics[width=0.33\columnwidth,angle=-0,clip=true,trim= 20 0 30 0]{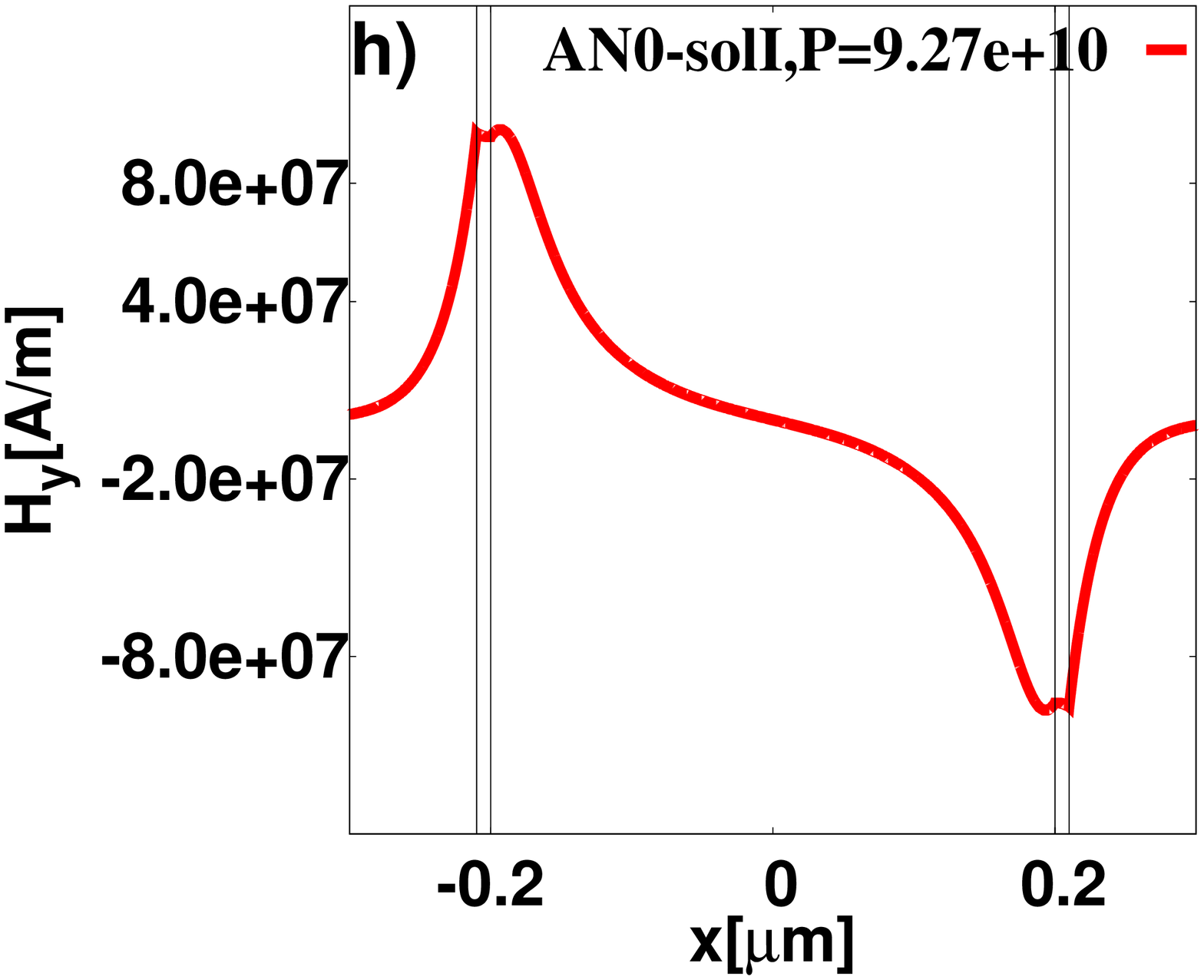}
    \includegraphics[width=0.33\columnwidth,angle=-0,clip=true,trim= 20 0 30 0]{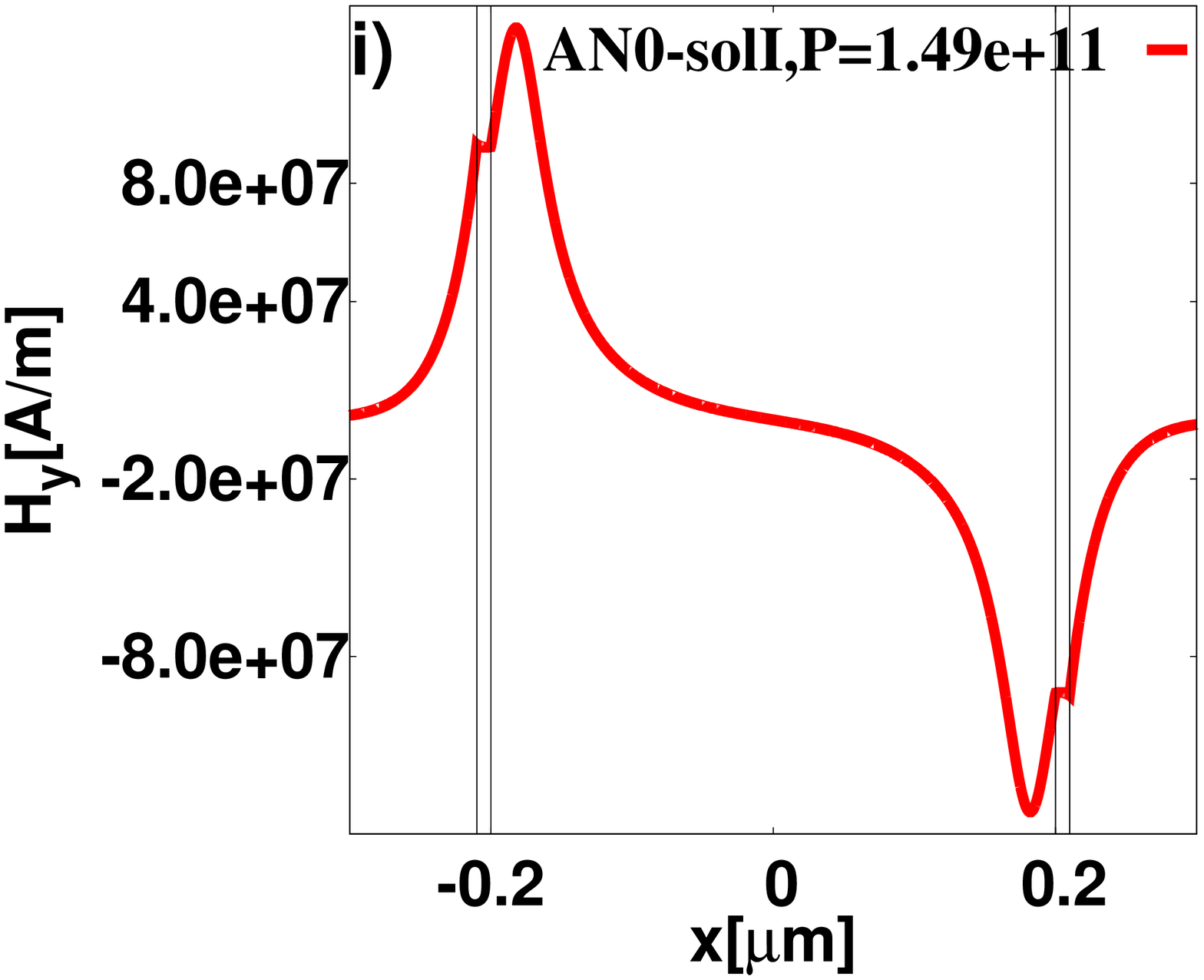}} 
  \caption{Field profiles $H_y(x)$ for the symmetric (blue curves), asymmetric (green curves) , and antisymmetric (red curves) TM modes for $d_{buf}=10$ nm, $d_{core}=400$ nm, and $\varepsilon_{buf}=2.5^2$.}
  \label{fig:nonlinear-profile-modal-dbuf10nm-transition}
\end{figure}

We now study the nonlinear dispersion curves of the improved NPSW such that the effective index of the symmetric linear mode is above the core linear refractive index. The symmetric linear mode is of plasmonic type as it is shown in Fig.~\ref{fig:linear-profiles}(d). The corresponding nonlinear dispersion curves are given in Fig.~\ref{fig:nl-dispersion-curve}(c) and (d) in which the values of the buffer layer thicknesses are $d_{buf}=15$ nm and $d_{buf}=10$ nm, respectively. The global shape  of these dispersion curves is similar to the ones already studied for the three main modes of the simple NPSW (see Fig.~\ref{fig:nl-dispersion-curve}(a)), but the mode field profiles now exhibit a spatial transition from a plasmonic type profile to a solitonic type one. For the asymmetric one: the mode moves gradually from AS1-plas to AS1-solI, and the antisymmetric mode moves from AN0-plas to AN0-solI. The corresponding field profiles for $d_{buf}=15$ nm are shown in Fig.~\ref{fig:nonlinear-profile-modal-dbuf15nm-transition}(b) and (c). For $d_{buf}=10$ nm, the modal transitions for the asymmetric and the antisymmetric modes are shown in Fig.~\ref{fig:nonlinear-profile-modal-dbuf10nm-transition} (green and red curves, respectively). Nevertheless, depending on the thickness of the buffer layers, the nonlinear symmetric mode S0-plas exhibits spatial modal transition toward two different nonlinear symmetric solitonic modes. In Fig.~\ref{fig:nl-dispersion-curve}(c): the symmetric S0-plas mode exhibits spatial transition to symmetric solitonic mode S0-solI; one single peak located in the nonlinear core as it is shown in Fig.~\ref{fig:nonlinear-profile-modal-dbuf15nm-transition}(a). While, in Fig.~\ref{fig:nl-dispersion-curve}(d) in which $d_{buf}=10$ nm, the symmetric mode S0-plas turns gradually to the new nonlinear symmetric mode denoted by S0-solII (dashed blue line to solid dark blue line in Fig.~\ref{fig:nl-dispersion-curve}(d)); it is composed of two solitons located in the core near each interfaces (see dark blue curves in Fig.~\ref{fig:nonlinear-profile-modal-dbuf10nm-transition}). The solid blue curve in Fig.~\ref{fig:nl-dispersion-curve}(d) above the dark blue curve, represents the symmetric nonlinear mode S0-solI which emerges at specific value of the power as a purely nonlinear mode. The field profile is similar to the solid blue profiles in Figs.~\ref{fig:nonlinear-profile-modal-dbuf40nm-transition}(d) and~\ref{fig:nonlinear-profile-modal-dbuf15nm-transition}(a). In general, the symmetric solitonic mode S0-solI looks like the higher order purely nonlinear symmetric mode denoted by SI which has been found in the simple NPSW, but with different boundary conditions at the core interface due to the opposite sign of the real part of the permittivity in the outer regions~\cite{Walasik14a,Walasik15b}. 
\subsubsection{Nonlinear phase diagrams}
\label{subsubsec:Nl-phase-diagrams}
\begin{figure}[htbp]
  \centerline{%
    \includegraphics[width=0.99\columnwidth,angle=-0,clip=true,trim= 0 0 0 0]{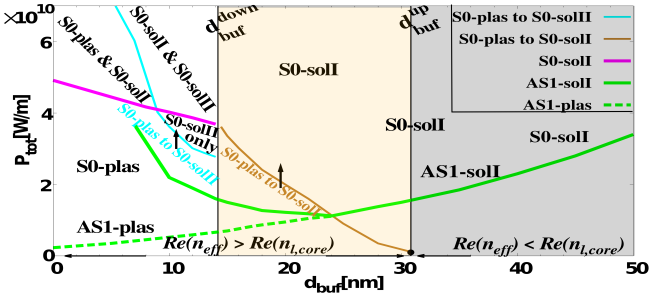}
}
  \caption{Phase diagram of the TM modes of the improved NPSW as a function of dielectric buffer thickness $d_{buf}$ and of the total power $P_{tot}$ for $d_{core}=400$ nm and $\varepsilon_{buf}=2.5^2$. The green curves refer to the first asymmetric modes.  Arrows indicates the corresponding mode transition across the associated curve (thin one).}
  \label{fig:phase-diagram-d_buffer}
\end{figure}

Using numerical simulations providing both the dispersion curves and the associated field profiles like the ones shown above, we obtain the phase diagram of the improved NPSW for the main modes. Such diagram is shown in Fig.~\ref{fig:phase-diagram-d_buffer} where the existence and the type of the first TM symmetric and asymmetric modes are given. The  buffer thickness $d_{buf}$ and the total power $P_{tot}$ are used as parameters. The core thickness  $d_{core}=400$ nm and the permittivity of the buffer layers $\varepsilon_{buf}=2.5^2$, are fixed which provide the solid blue curve of the linear main symmetric mode shown in Fig.~\ref{fig:linear-profiles}, and the nonlinear dispersion curves given in Fig.~\ref{fig:nl-dispersion-curve}. For these parameters, the critical value of the buffer layers $d^{up}_{buf}$  in which the fundamental linear symmetric mode is flat in the core (see Fig.~\ref{fig:linear-profiles}(c)) is $\approx 30.82$ nm. The nonlinear phase diagram depicted in Fig.~\ref{fig:phase-diagram-d_buffer}, provides us with the type and the existence of the main symmetric and asymmetric modes that can be found in the improved NPSW as a function of the buffer layer thickness $d_{buf}$. This phase diagram is divided into three regions. The first region is for $d_{buf} > d^{up}_{buf}$ (right most grey region). In this region, the symmetric linear mode is localized in the high-index core with $\Re e(n_{eff})<\Re e(n_{l,core})$ as it is given in Fig.~\ref{fig:linear-profiles}(b). As the power increases, the main symmetric mode tends to be more localized in the core with plasmonic tails in the metal regions, it is of S0-solI type (symmetric soliton-type with a single peak as it is shown in Fig.~\ref{fig:nonlinear-profile-modal-dbuf40nm-transition}(a)). Due to the nonlinearity, the asymmetric nonlinear mode bifurcates from the nonlinear symmetric mode S0-solI at some critical power value (see solid green curve in Fig.~\ref{fig:phase-diagram-d_buffer}). It is denoted by AS1-solI (soliton peak shifted-off the slot center as it is depicted in Fig.~\ref{fig:nonlinear-profile-modal-dbuf40nm-transition}(b)). Above the bifurcation threshold curve, both the nonlinear symmetric mode S0-solI and the nonlinear asymmetric mode AS1-solI exist as it is  shown in the nonlinear dispersion diagram obtained for $d_{buf}=40$ nm in Fig.~\ref{fig:nl-dispersion-curve}(b). 

The second region is for $d^{down}_{buf}<d_{buf}<d^{up}_{buf}$ (yellow region). In this region, the symmetric linear mode is of plasmonic type (see the field profile in Fig.~\ref{fig:linear-profiles}(d)). Below the brown curve labelled "S0-plas to S0-solI" in Fig.~\ref{fig:phase-diagram-d_buffer}, the nonlinear symmetric mode is of plasmonic type S0-plas. With the increase of power, the nonlinear symmetric mode S0-plas exhibits a spatial modal transition from S0-plas to S0-solI (see Fig.~\ref{fig:nonlinear-profile-modal-dbuf15nm-transition}(a)), in which the mode becomes more localized in the nonlinear core (single spatial soliton peak in the nonlinear core with plasmonic tails in the metal regions as it is given in Fig.~\ref{fig:nonlinear-profile-modal-dbuf15nm-transition}(a)). Along  the brown curve, the nonlinear symmetric mode is flat in the core (the arrow indicates the transition from S0-plas to S0-solI). Above $d_{buf} \simeq 24$ nm (but below $d^{up}_{buf}$), the solid green curve above the brown thin line in Fig.~\ref{fig:phase-diagram-d_buffer}, represents the limit curve for the nonlinear asymmetric mode AS1-solI: below the curve, it does not exist. It emerges directly from the symmetric nonlinear mode S0-solI as expected (see Fig.~\ref{fig:nonlinear-profile-modal-dbuf40nm-transition}(e) for an example of such field profile). For $d^{down}_{buf}<d_{buf}<24$ nm, the dashed green curve is the limit curve of the asymmetric plasmonic  mode AS0-plas: below the curve, it does not exist. It emerges at low powers from the symmetric one, S0-plas (see the dashed green curve in Fig.~\ref{fig:nl-dispersion-curve}(c)). At higher powers (see the solid green line in Fig.~\ref{fig:phase-diagram-d_buffer}), it moves to an asymmetric mode of solitonic type AS1-solI (a peak being located in the core and near one of its interface, see solid line profile in Fig.~\ref{fig:nonlinear-profile-modal-dbuf15nm-transition}(b)). Fig.~\ref{fig:nl-dispersion-curve}(c) represent an example of the dispersion curves, and Fig.~\ref{fig:nonlinear-profile-modal-dbuf15nm-transition} shows the corresponding field profiles in this second region for $d_{buf}=15$ nm. The nonlinear dispersion curves in Fig.~\ref{fig:nl-dispersion-curve}(c) can be seen as a vertical cut at $d_{buf}=15$ nm in the nonlinear phase diagram depicted in Fig.~\ref{fig:phase-diagram-d_buffer}.

The third region: for $0<d_{buf}<d^{down}_{buf}$ (left most white region). The symmetric linear mode is of plasmonic type as it can be inferred from Fig.~\ref{fig:linear-profiles}(a). Below the cyan curve labelled "S0-plas to S0-solII", the nonlinear symmetric mode is of plasmonic type S0-plas. Above the cyan curve, the new symmetric nonlinear solution S0-solII emerges, at high powers, from the S0-plas mode; it is composed of two solitons located in the core near each interfaces (see the dark blue curves in the first row of Fig.~\ref{fig:nonlinear-profile-modal-dbuf10nm-transition}). The transition curve  between the S0-plas region and the S0-solII region is shown in cyan in the phase diagram shown in Fig.~\ref{fig:phase-diagram-d_buffer}. It is worth mentioning that, the symmetric nonlinear mode S0-solII (see first row of Fig.~\ref{fig:nonlinear-profile-modal-dbuf10nm-transition}) is different from the asymmetric mode shown in Fig.~\ref{fig:nonlinear-profile-modal-dbuf15nm-transition}(b). Furthermore, this mode exists for core thicknesses as small as $100$ nm and it can not be considered in this case as equivalent to the mirror-imaged of the mode existing in the single interface configuration~\cite{Ferrando13}. In this third region, the nonlinear symmetric mode S0-solI (single peak in the core) appears as a purely nonlinear mode without any reference in the linear case. The limit curve for the emergence of the S0-solI is shown in thick pink in the phase diagram. It is worth mentioning that, the appearance of the S0-solI is similar to the appearance of the nonlinear symmetric higher order mode denoted by SI for the simple NPSW~\cite{Walasik14a,Walasik15b}, but with different boundary conditions at the core interface due to the opposite sign of the real part of the permittivity in the outer regions. For $d_{buf}=0$ nm (simple NPSW), we recover the same type of modes as expected. It is important to point out that in the region above  the cyan curve and below the pink one, only the S0-solII mode exists. The asymmetric nonlinear plasmonic mode AS1-plas emerges at low power from the nonlinear symmetric mode S0-plas (dashed green curve), and moves gradually to the nonlinear symmetric mode AS1-solI (solid green curve). For an example of the nonlinear dispersion curves with the corresponding field profiles in this third region, see Fig.~\ref{fig:nl-dispersion-curve}(d) and Fig.~\ref{fig:nonlinear-profile-modal-dbuf10nm-transition}, respectively. This example can be seen as a vertical cut in Fig.~\ref{fig:phase-diagram-d_buffer} at $d_{buf}=10$ nm. 

As we mentioned before, the linear symmetric mode for $d_{buf}=d^{up}_{buf} \approx 30.82$ nm is flat in the core (see Fig.~\ref{fig:linear-profiles}(c) for an example of such mode profile). For the nonlinear case, the mode is still nearly flat in the core up to some critical value of the power such that above this value a peak emerges in the core, and the symmetric mode becomes of S0-solI type. The critical power is shown as black circle in Fig.~\ref{fig:phase-diagram-d_buffer} which corresponds to $P_{tot}\approx 10^{9}$ [W/m]. The asymmetric mode AS1-solI bifurcates directly from the symmetric mode S0-solI as expected which is indicated by the crossing between the solid green curve and the the  vertical thin line at $d_{buf}=d^{up}_{buf}$ in Fig.~\ref{fig:phase-diagram-d_buffer}. The results for the simple NPSW correspond to the vertical line at $d_{buf}=0.0$ nm. At low power, the symmetric mode is of plasmonic type S0-plas (see Fig.\ref{fig:nonlinear-profile-modal-dbuf40nm-transition}(a)). The bifurcation of the asymmetric nonlinear mode AS1-plas from the symmetric mode S0-plas is represented by the beginning of the dashed green curve at $d_{buf}=0.0$ nm. Above this point, both the symmetric S0-plas, and the asymmetric mode AS1-plas exist as it is shown in Fig.~\ref{fig:nl-dispersion-curve}(a). The beginning of the pink curve at  $d_{buf}=0.0$ nm corresponds to the appearance of the symmetric higher order mode denoted by SI~\cite{Walasik15a,Walasik15b}. This mode looks like the symmetric mode S0-solI (see solid blue profile Figs.~\ref{fig:nonlinear-profile-modal-dbuf40nm-transition} and~\ref{fig:nonlinear-profile-modal-dbuf15nm-transition}), but with different boundary conditions at the core interface as we mentioned before.
 \begin{figure}[htbp]
  \centerline{%
    \includegraphics[width=0.99\columnwidth,angle=-0,clip=true,trim= 0 0 0 0]{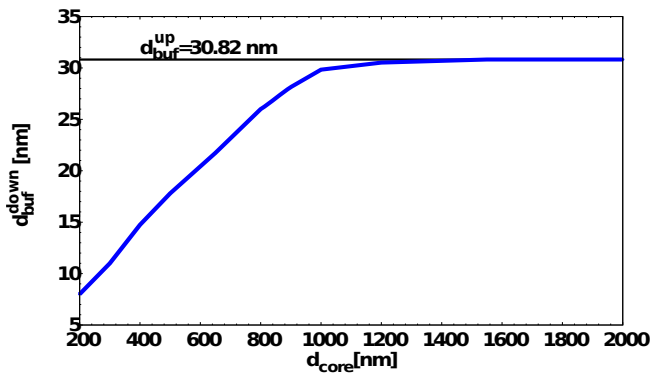}	}
  \caption{Evolution of the critical value $d^{down}_{buf}$ as a function of the core thickness $d_{core}$ for $\varepsilon_{buf}=2.5^2$ and $d_{buf}=15$ nm. The horizontal line represents the critical value $d^{up}_{buf}$.}
  \label{fig:evolution_critical_buffer_values}
 \end{figure}
In Fig.~\ref{fig:evolution_critical_buffer_values}, we study the influence of the core thickness $d_{core}$ on the lower critical value $d_{buf}^{down}$.  The horizontal line represents the upper critical value $d^{up}_{buf}$. The permittivities being fixed, we found that, the upper critical value $d^{up}_{buf}$ does not change with the core thickness as expected (see Eq.~\eqref{eqn:full_critical_dbuf_final}). While, $d_{buf}^{down}$ increases with $d_{core}$, such that they coincide when the core thickness reaches the critical value $d^{\lambda}_{core} \simeq 1550$ nm (one wavelength), above this core thickness, the critical value $d_{buf}^{down}$ will no longer change with $d_{core}$.
We also provide elsewhere the phase diagram of the improved NPSW using $d_{core}$ and $P_{tot}$ as parameters, for a given fixed values of $d_{buf}$ and $\varepsilon_{buf}$~\cite{elsawy16-improvedNPSWG}. We found that the threshold for the Hopf bifurcation, from which the asymmetric mode emerges, strongly decreases  with the core thickness as it was already found in the simple NPSW~\cite{Walasik15a,Walasik15b}. We provided the nonlinear phase diagram describing the existence and the type of the main TM modes in our improved structure. We have seen the importance of the inclusion of the supplementary buffer layers on the types of solutions that propagate in the NPSWs. Nevertheless, these supplementary layers slightly shift the bifurcation threshold toward higher powers as it is depicted in Fig.~\ref{fig:phase-diagram-d_buffer}. While, the increase of the core thickness, reduces the bifurcation threshold~\cite{elsawy16-improvedNPSWG} as expected from the simple NPSWs.~\cite{Walasik15a,Walasik15b}.
\section{Study of the losses for TM waves}
\label{sec:loss_study}
\begin{figure}[htbp]
\centering
  \includegraphics[width = 0.49\columnwidth,angle=-0,clip=true,trim= 30 35 80 10]{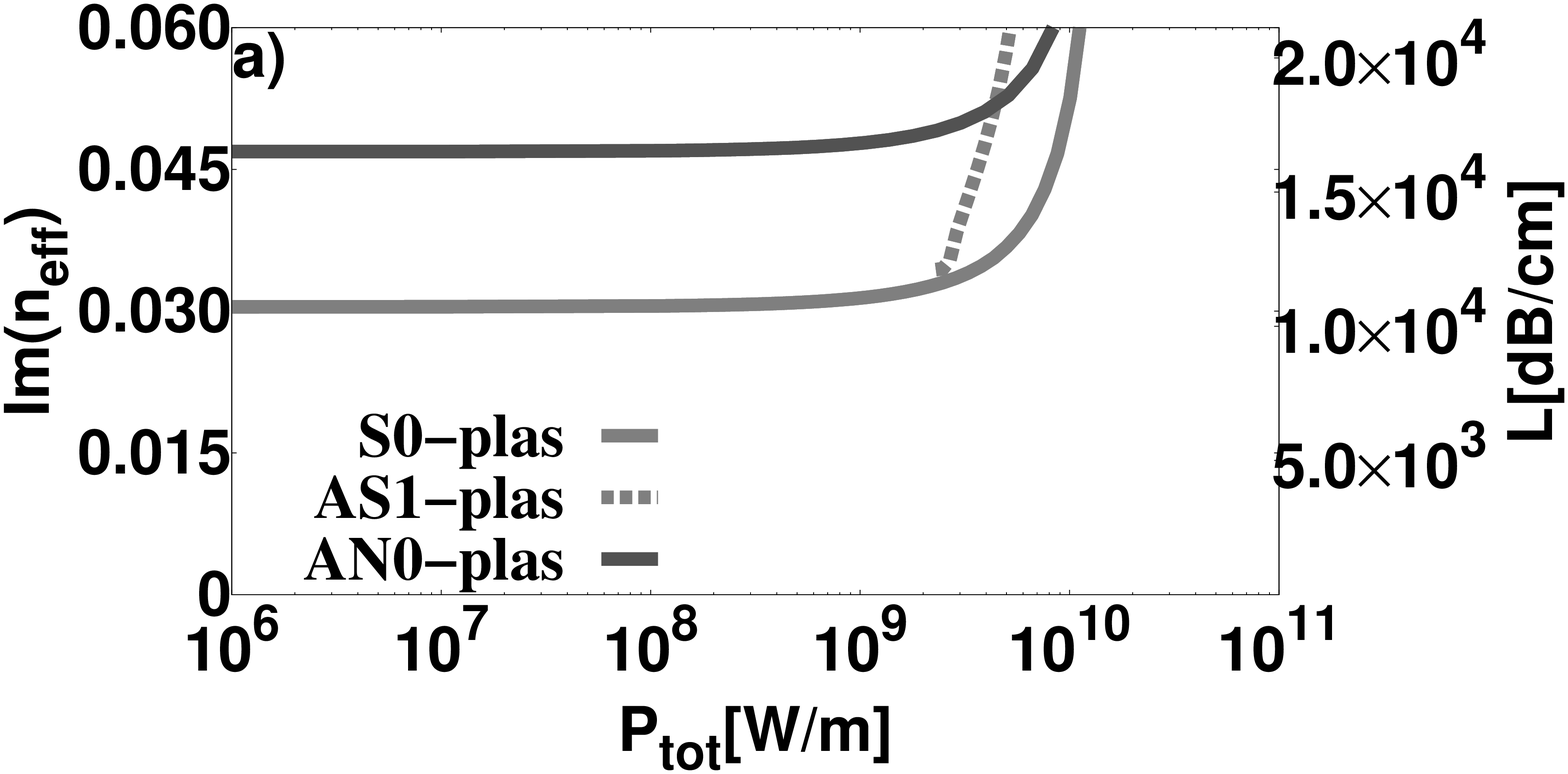}
    \includegraphics[width = 0.49\columnwidth,angle=-0,clip=true,trim= 30 35 80 10]{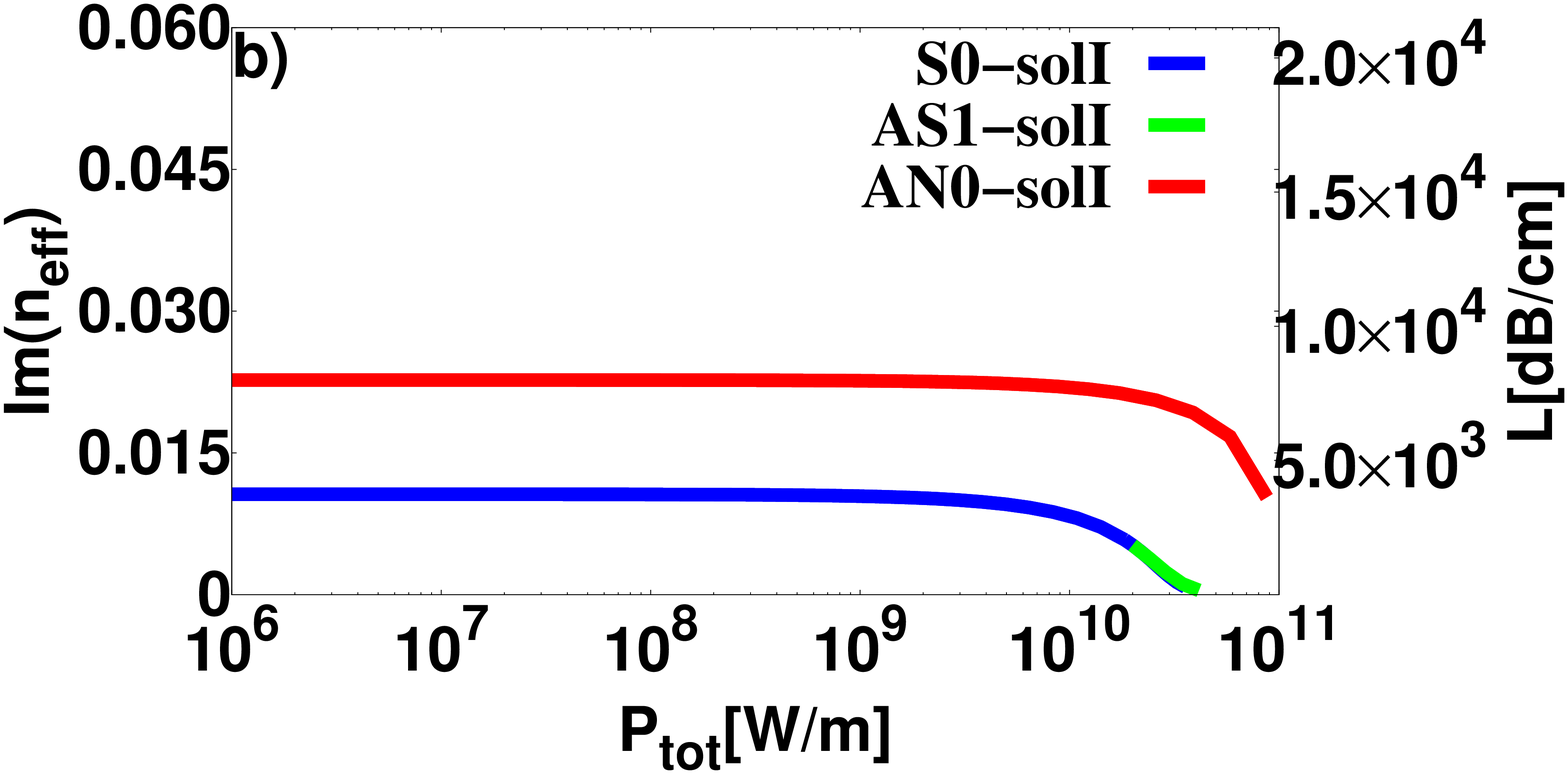}
    \includegraphics[width = 0.49\columnwidth,angle=-0,clip=true,trim= 30 35 80 10]{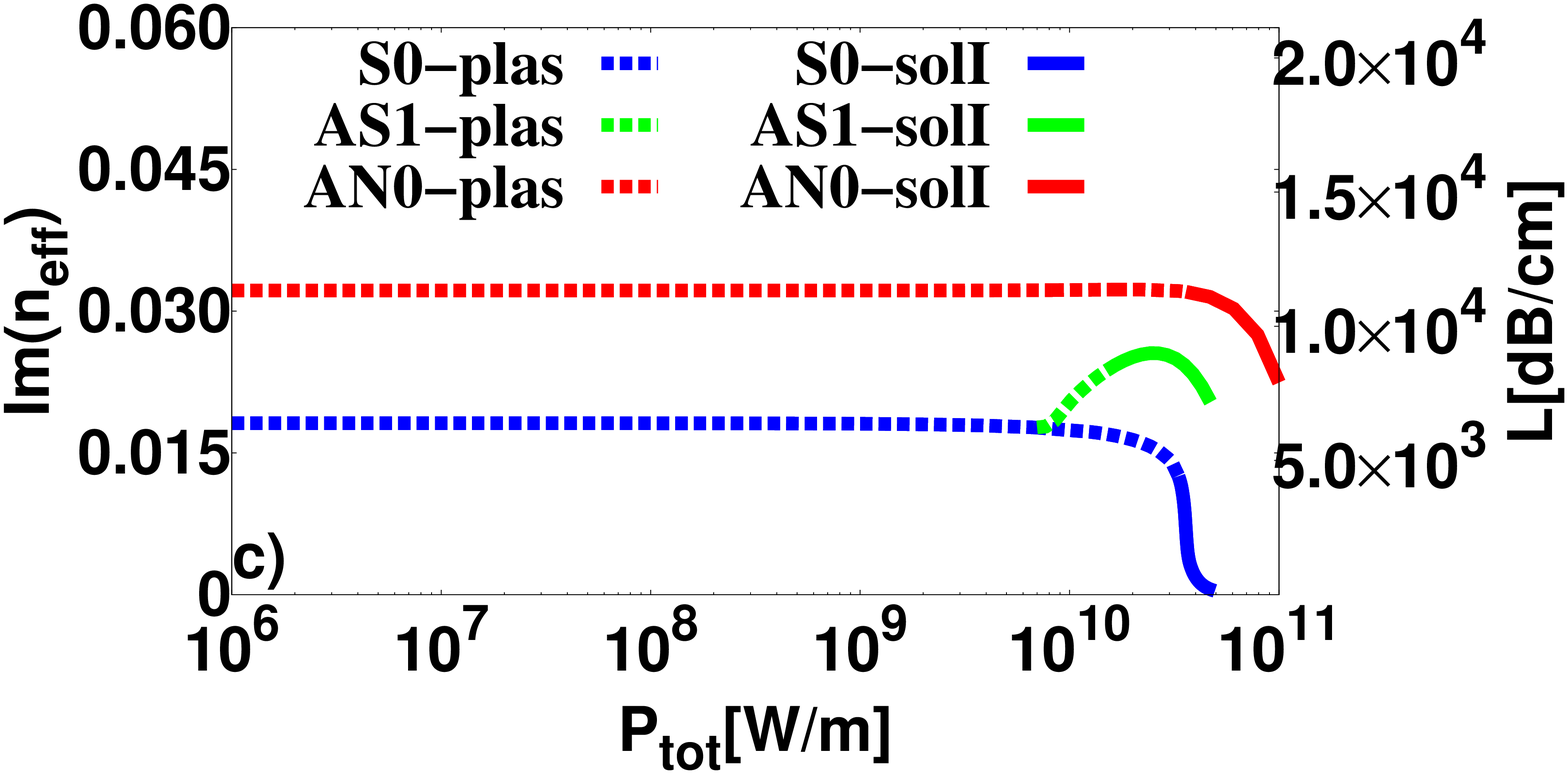}
\includegraphics[width=0.49\columnwidth,angle=-0,clip=true,trim= 30 35 80 10]{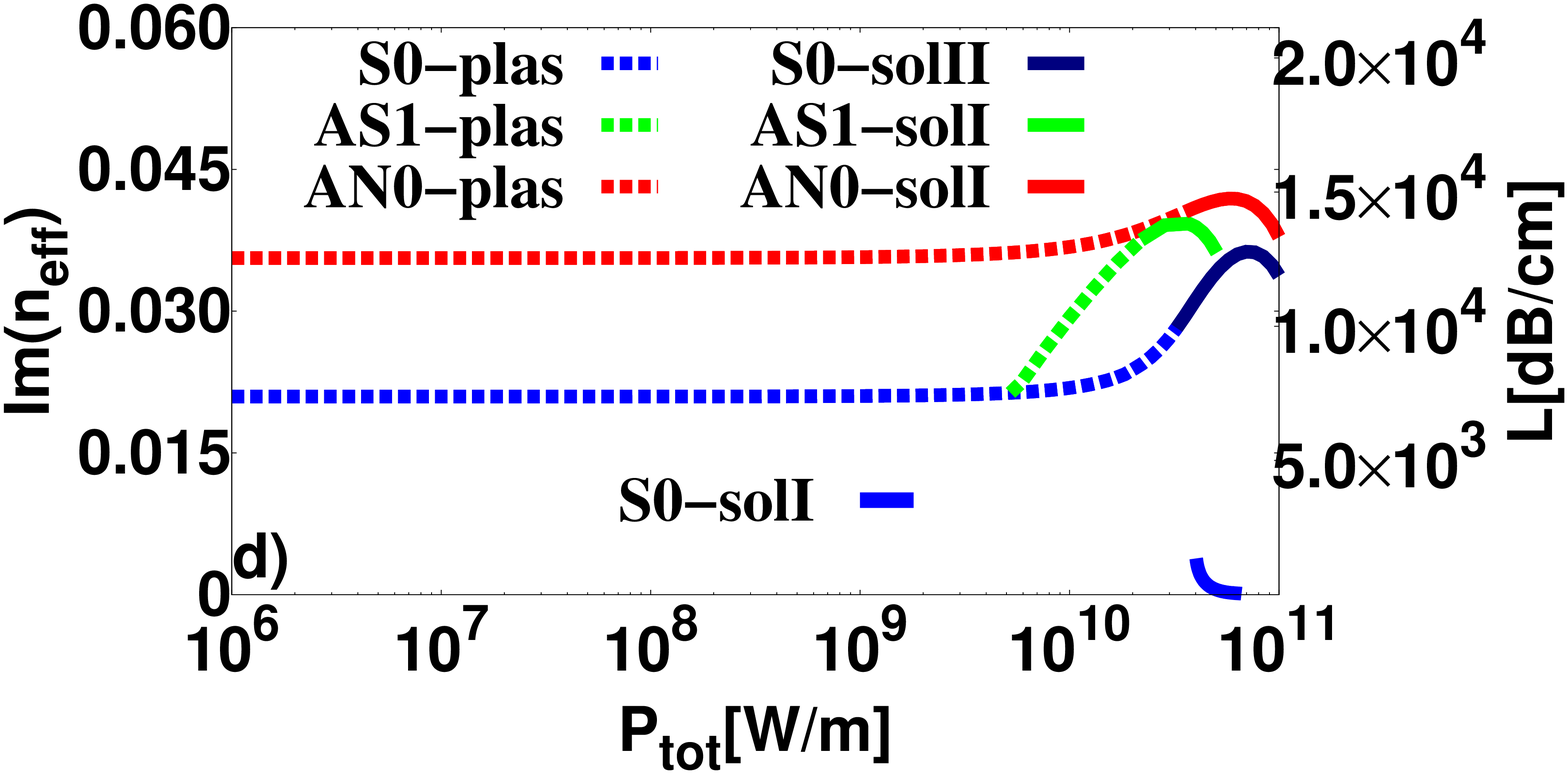}
  \caption{Losses for the simple NPSW (grey curves) and for the improved one (color curves). The blue, green, and red colors indicate the symmetric, asymmetric and antisymmetric nonlinear TM modes, respectively. $d_{core}=400$ nm and $\varepsilon_{buf}=2.5^2$.  (b) for $d_{buf}=40$ nm, (c) for $d_{buf}=15$ nm, and (d) for $d_{buf}=10$ nm. For the real part of the effective indices see Fig.~\ref{fig:nl-dispersion-curve}. For the field profiles see Figs.~\ref{fig:nonlinear-profile-modal-dbuf40nm-transition},~\ref{fig:nonlinear-profile-modal-dbuf15nm-transition}, and~\ref{fig:nonlinear-profile-modal-dbuf10nm-transition}.}
  \label{fig:loss-nl-dispersion-curve}
\end{figure}

In the previous section~\ref{sec:results-nonlinear-TM}, we have shown the influence of the supplementary buffer layers on the type of modes (for the TM case) that propagate in the simple NPSW. We demonstrated that, even for a very thin buffer dielectric layers, the type of the nonlinear modes propagating in the simple NPSW is modified (see phase diagram in Fig.~\ref{fig:phase-diagram-d_buffer} and the nonlinear dispersion curves depicted in Fig.~\ref{fig:nl-dispersion-curve}). In this section, we study the influence of the buffer layers on the overall losses. The imaginary part of the effective indices of the main modes with the corresponding values of the losses are shown in Fig.~\ref{fig:loss-nl-dispersion-curve} which corresponds to Fig.~\ref{fig:nl-dispersion-curve} for the real part of the effective indices. The losses are estimated using the approach based on the imaginary part of the permittivity and the field profiles. This method is  described in the case of linear waveguides \cite{Snyder83} and has also been used for the nonlinear studies \cite{Walasik12,Walasik14}. 

For the simple slot configuration, the losses of the three main modes increase with the power as it is shown in Fig.~\ref{fig:loss-nl-dispersion-curve}(a). For $d_{buf}=40$ nm, in which the symmetric linear mode is core localized and the nonlinear modes are of solitonic type (see Fig.~\ref{fig:nonlinear-profile-modal-dbuf40nm-transition}), we found that the losses are at least three times smaller than the ones of the simple NPSW, and that they decrease with the power unlike the simple NPSW (see Fig.~\ref{fig:loss-nl-dispersion-curve}(b)). We now study the losses for the improved NPSW in which the effective index of the symmetric linear mode is above the core linear refractive index. The corresponding curves are given in Fig.~\ref{fig:loss-nl-dispersion-curve}(c) and (d). As we have seen before in Fig.~\ref{fig:nl-dispersion-curve}(c) and (d), the mode field profiles now exhibit a spatial transition from a plasmonic type profile to a solitonic type one. Furthermore, at low powers, the losses are reduced, at least, by a factor two compared to the associated simple NPSW. For $d_{buf}=15$ nm, the losses decrease with the power for the main symmetric and antisymmetric modes leading to low loss solutions at high powers (see blue and red curves in Fig.~\ref{fig:loss-nl-dispersion-curve}(c) and the associated field profiles in Fig.~\ref{fig:nonlinear-profile-modal-dbuf15nm-transition}(a) and (c)). For the asymmetric mode,  loss first increases with power and then decreases. This loss increase is linked to the fact that, in this case, the field first tends to increase near one of the metal interface, increasing the fraction of field in a highly lossy region. Then, the mode moves gradually from a plasmonic type to a solitonic type asymmetric profile, and when the peak of the soliton is far enough from the metal interface the losses start to decrease as it is shown in Fig.~\ref{fig:nonlinear-profile-modal-dbuf15nm-transition}(b). For $d_{buf}=10$ nm, losses of the three main modes first, increase with the power and then decrease. The increase of the losses can be understood by looking at the field profiles in Fig.~\ref{fig:nonlinear-profile-modal-dbuf10nm-transition}, in which the fields tend to increase near the metal regions. With the increase of the power, the modes move gradually from plasmonic type to solitonic type: S0-plas to S0-solII (for the fundamental symmetric mode), AS1-plas to AS1-solI (for the asymmetric mode), and S0-plas to AN0-solI (for the antisymmetric mode), in which the soliton peaks emerge near the interfaces, and when the soliton peaks are far enough from the metal regions, the losses start to decrease. The symmetric higher order mode S0-solI (see solid  blue line in Fig.~\ref{fig:loss-nl-dispersion-curve}(d)), has lower loss and the loss decrease with the increase of the power as expected due to the localization of the  soliton peak in the nonlinear core (see solid blue curve in Fig.~\ref{fig:nonlinear-profile-modal-dbuf15nm-transition}(a) for an example of such field profile). It is worth mentioning, that the increase of the losses in Fig.~\ref{fig:loss-nl-dispersion-curve}(d), is similar to the simple NPSW in Fig.~\ref{fig:loss-nl-dispersion-curve}(a). While, due to the inclusion of very thin buffer layers $d_{buf}=10$ nm, the behaviour of the losses change at high power (where the field profiles exhibit spatial transitions from plasmonic to solitonic type solutions), which proves the importance of the inclusion of the dielectric buffer layers between the nonlinear core and the metal regions in the simple NPSW.
\section{Stability study for TM waves}
\label{sec:stability}
In this section we study the stability properties for the TM waves in the most significant cases. In a previous work, using two different numerical methods including the FDTD,  we have already shown that the results about the stability of the main symmetric and asymmetric modes  derived in reference~\cite{Mitchell93}  from  theoretical arguments in the framework of the weak guidance approximation in nonlinear waveguides are still valid, at least numerically, in simple NPSWs except just around the bifurcation point where subtle behaviours occur~\cite{Walasik15b}.
Here, we have again used the nonlinear capabilities of the FDTD method implemented in the \textsc{meep} software~\cite{Oskooi10,Rodriguez07}. The metal permittivity is described by a Drude model to obtain the fixed negative value used at the studied wavelength in the stationnary study.
\begin{figure}[htbp]
  \centerline{\includegraphics[width=1.0\columnwidth,clip=true,trim= 60 5 150 15]{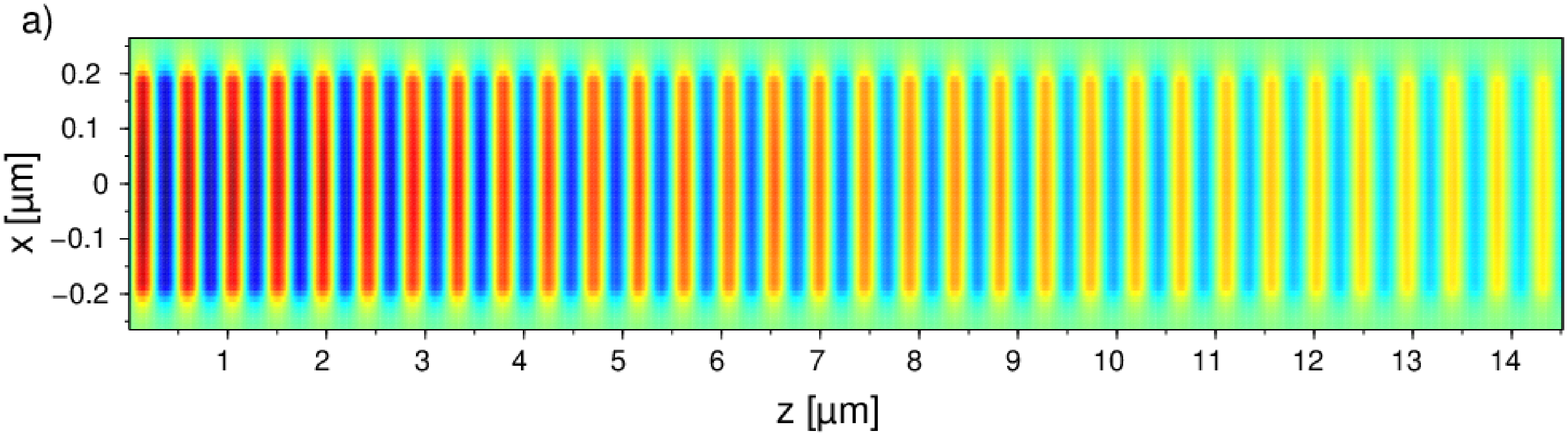}}
  \centerline{\includegraphics[width=1.0\columnwidth,clip=true,trim= 60 5 150 15]{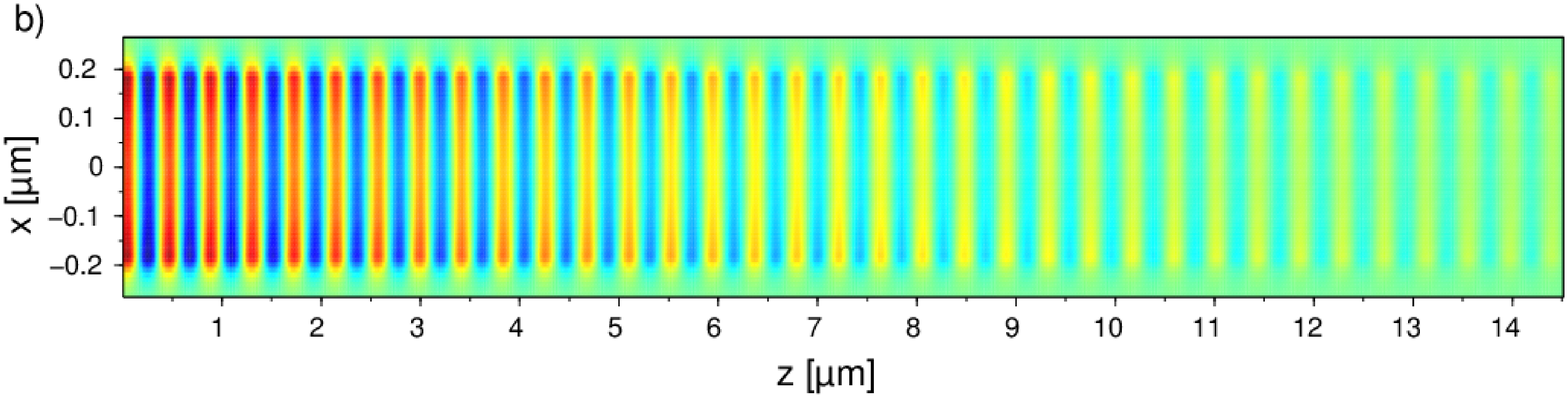}}
  \caption{Evolution during the propagation of the $H_y$ field profile for $d_{core}=300$nm,  $d_{buf}=40$~nm. (a) S0-solI symmetric solitonic type mode  at a power such that  $\langle \Delta n \rangle = 0.043$ with  $\varepsilon_{buf} = 2.5^2$. (b) S0-plas symmetric plasmonic type mode at a power such that  $\langle \Delta n \rangle = 0.015$ with  $\varepsilon_{buf} = 3.2^2$. $\langle \Delta n \rangle$ is the spatial average along the core of the nonlinear refractive index change~\cite{Walasik14a,Walasik15b}, other parameters as in the text.}  \label{fig:improved-NPWS-stab-mode-fdtd}
\end{figure}
As it can be seen in Fig.~\ref{fig:improved-NPWS-stab-mode-fdtd}(a), for the improved NPSW, the symmetric mode S0-solI is stable at low power. The S0-plas mode (plasmonic type) is also stable as  it is depicted in Fig.~\ref{fig:improved-NPWS-stab-mode-fdtd}(b). The asymmetric mode  AS1-solI is also stable for the tested powers above the bifurcation threshold (data not shown).  Due to the losses of the metal, and as pointed out in~\cite{salguiero14complex-modes-plasmonic-nonlinear-slot-waveguides,Walasik15b}, the temporal nonlinear solutions are actually not self-coherent~\cite{snyder:91} as it can be seen  in Fig.~\ref{fig:improved-NPWS-stab-mode-fdtd} with  the decrease of  the field intensity along the propagation.
\section{TE stationary waves}
\label{sec:TE_modes}
In this section we study the transverse electric TE polarized waves in the improved structure depicted in Fig.~\ref{fig:geom-5layers-NPSW}. In this case the electromagnetic fields have only three nonzero components $E_{y}$, $H_{x}$, and $H_{z}$. The only nonzero component of the electric field is the longitudinal component $E_{y}$. Consequently, the TE surface plasmon waves do not exist in isotropic planar linear metal/dielectric structures. However, the fundamental transverse electric mode for linear (metal/dielectric/metal) structures exists if the core width is beyond a cut-off thickness~\cite{nakano94,Sun09}. For this mode, the maximum intensity is located in the core of the waveguide, thus losses in the TE case are much smaller than their counterparts in the TM case in which the field penetrates into the metal.
\subsection{Linear TE case}
\label{subsec:linear_TE_case}
\begin{figure}[htbp]
\centerline{
  \includegraphics[width=0.495\columnwidth,clip=true,trim= 0 0 0 0]{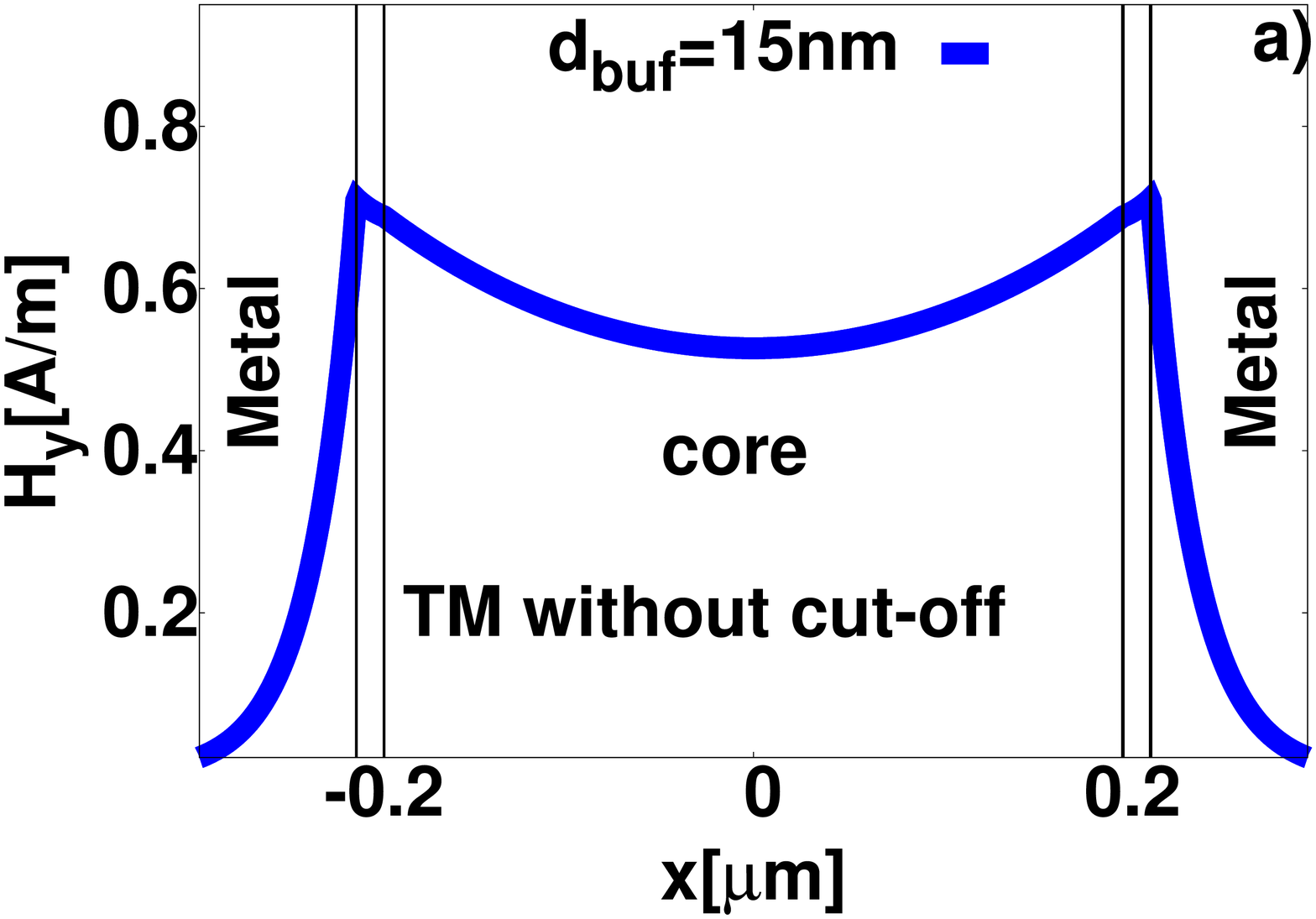}
  \includegraphics[width=0.495\columnwidth,clip=true,trim= 0 0 0 0]{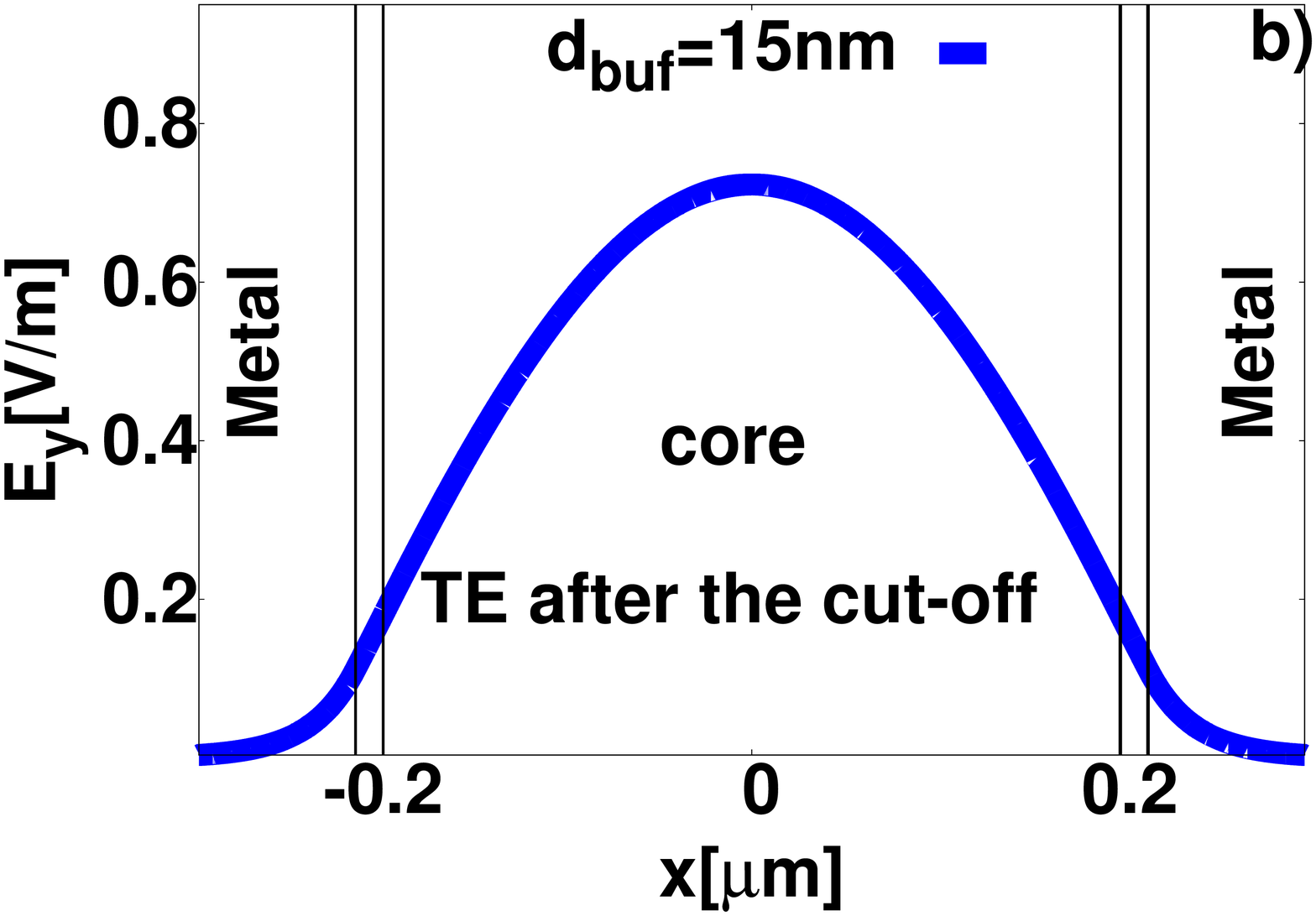}}
  \centerline{\includegraphics[width=0.495\columnwidth,clip=true,trim= 0 0 0 0]{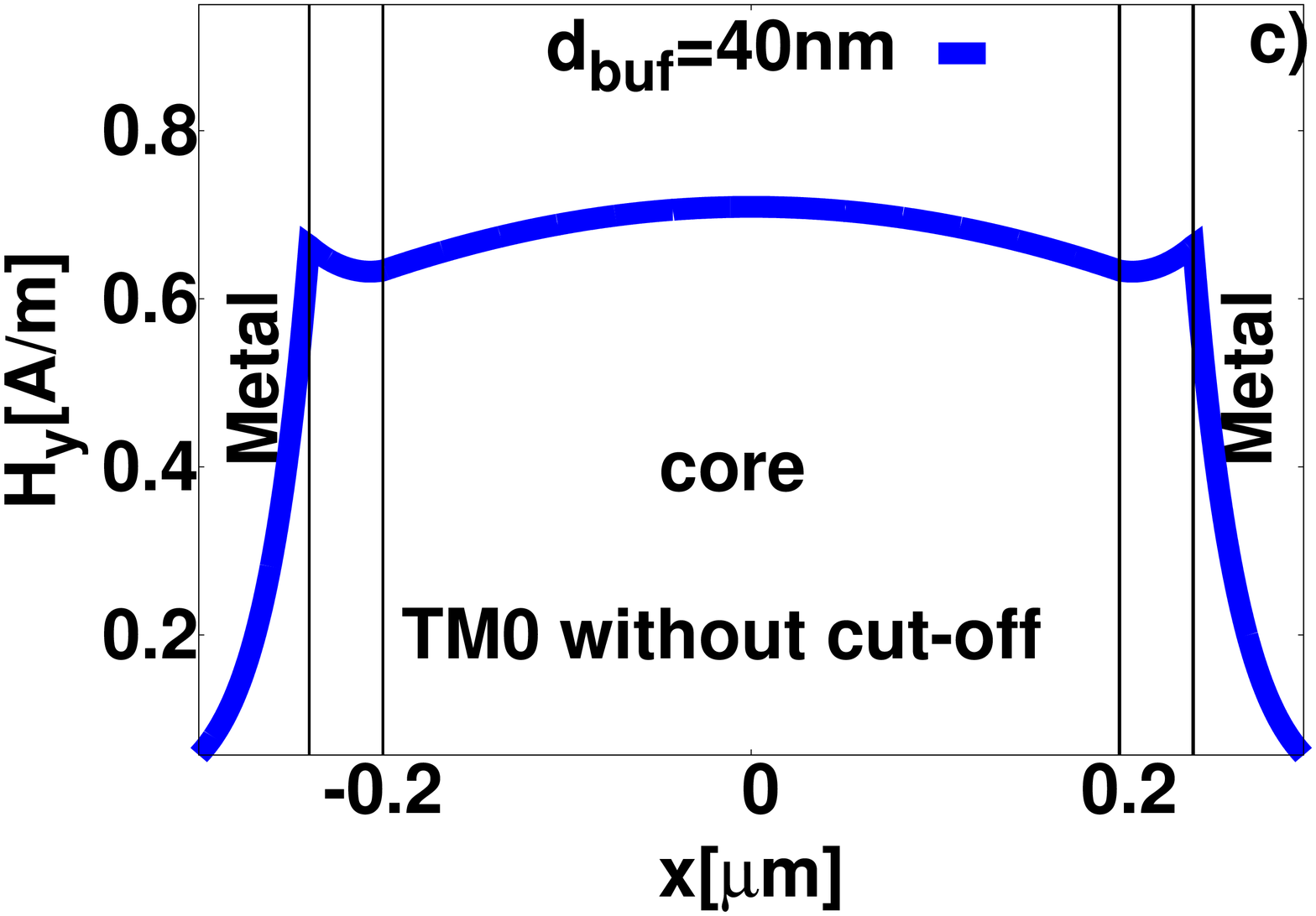}
  \includegraphics[width=0.495\columnwidth,clip=true,trim= 0 0 0 0]{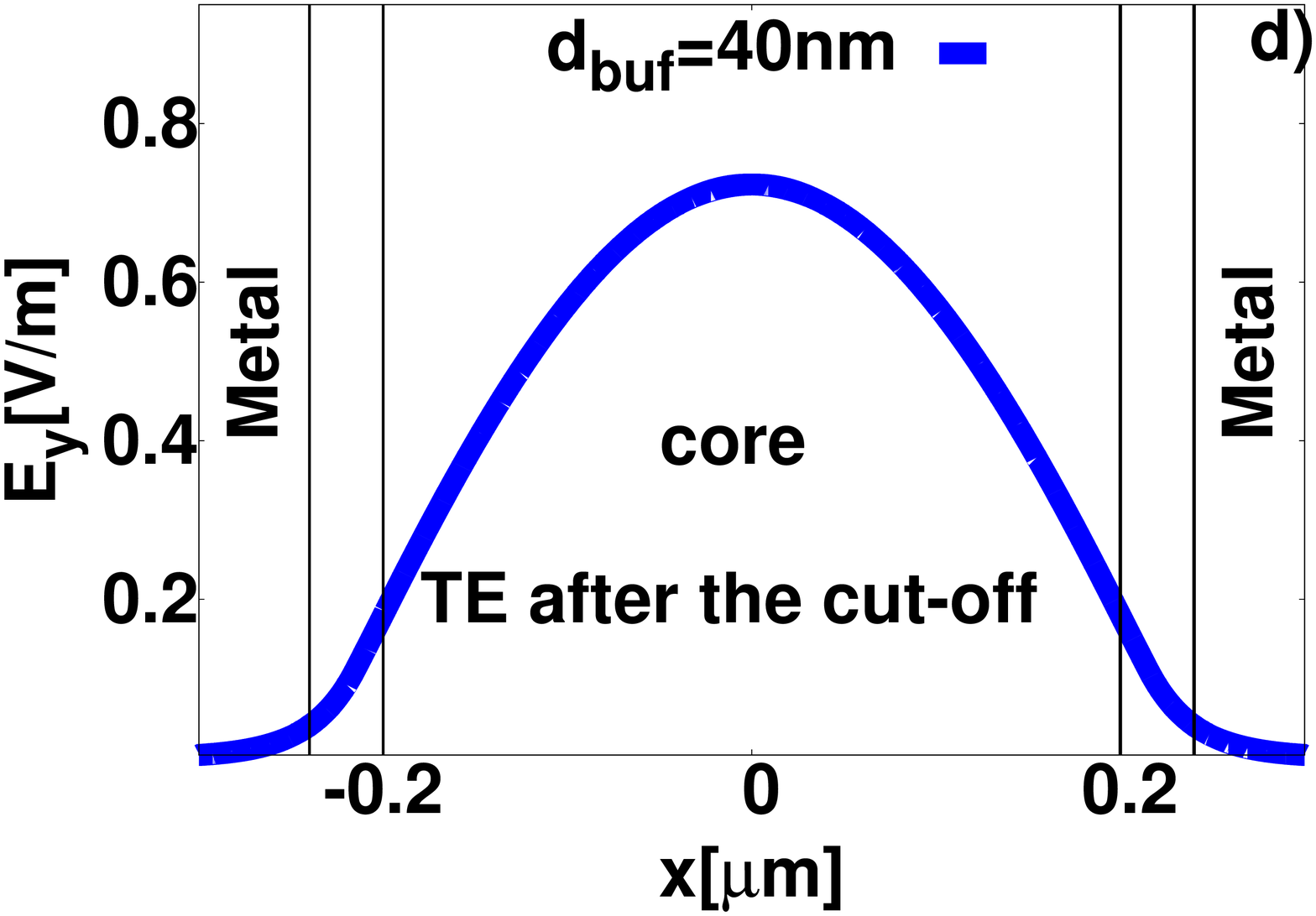}}
  \caption{Distribution of the symmetric linear profiles for fundamental symmetric linear modes with $d_{core}=400$ nm and $\varepsilon_{buf} = 2.5^2$ for two different buffer layers thicknesses $d_{buf}=15$ nm and $d_{buf}=40$ nm. (a) and (c) for fundamental linear TM modes without cut-off thickness. (b) and (d) for linear fundamental TE mode above the cut-off thickness of $d_{core}$.}  \label{fig:symmetric_linear_profiles_TE_TM}
\end{figure}

As we saw in Sec.~\ref{sec:results-linear-TM} for the TM case, the linear case is a fundamental step in understanding the nonlinear behaviour. For this reason, we start by the linear case in our study of the TE modes. In Fig.~\ref{fig:symmetric_linear_profiles_TE_TM}, we compare the fundamental symmetric linear profiles for both TM and TE waves for two different values of $d_{buf}$. We choose the core thickness $d_{core}=400$ nm, which is above the cut-off core thickness of the symmetric linear TE mode (there is no cut-off thickness for the linear TM mode). The linear TE mode being core localized for small values of $d_{buf}$, even for large $d_{buf}$ it stays core localized (see Fig.~\ref{fig:symmetric_linear_profiles_TE_TM}(b) and (d)). Consequently, unlike the TM mode, it does not exhibit any spatial transition as a function of $d_{buf}$. While, for the TM case (see Fig.~\ref{fig:symmetric_linear_profiles_TE_TM}(a) and (c)) the buffer layer thicknesses $d_{buf}$, play an important role in distinguishing between two different types of modes (plasmonic and core localized modes) as we mentioned before in Sec.~\ref{sec:results-linear-TM}. 

\begin{figure}[htbp]
  \centerline{%
    \includegraphics[width=1.0\columnwidth,angle=-0,clip=true,trim= 0 0 0 0]{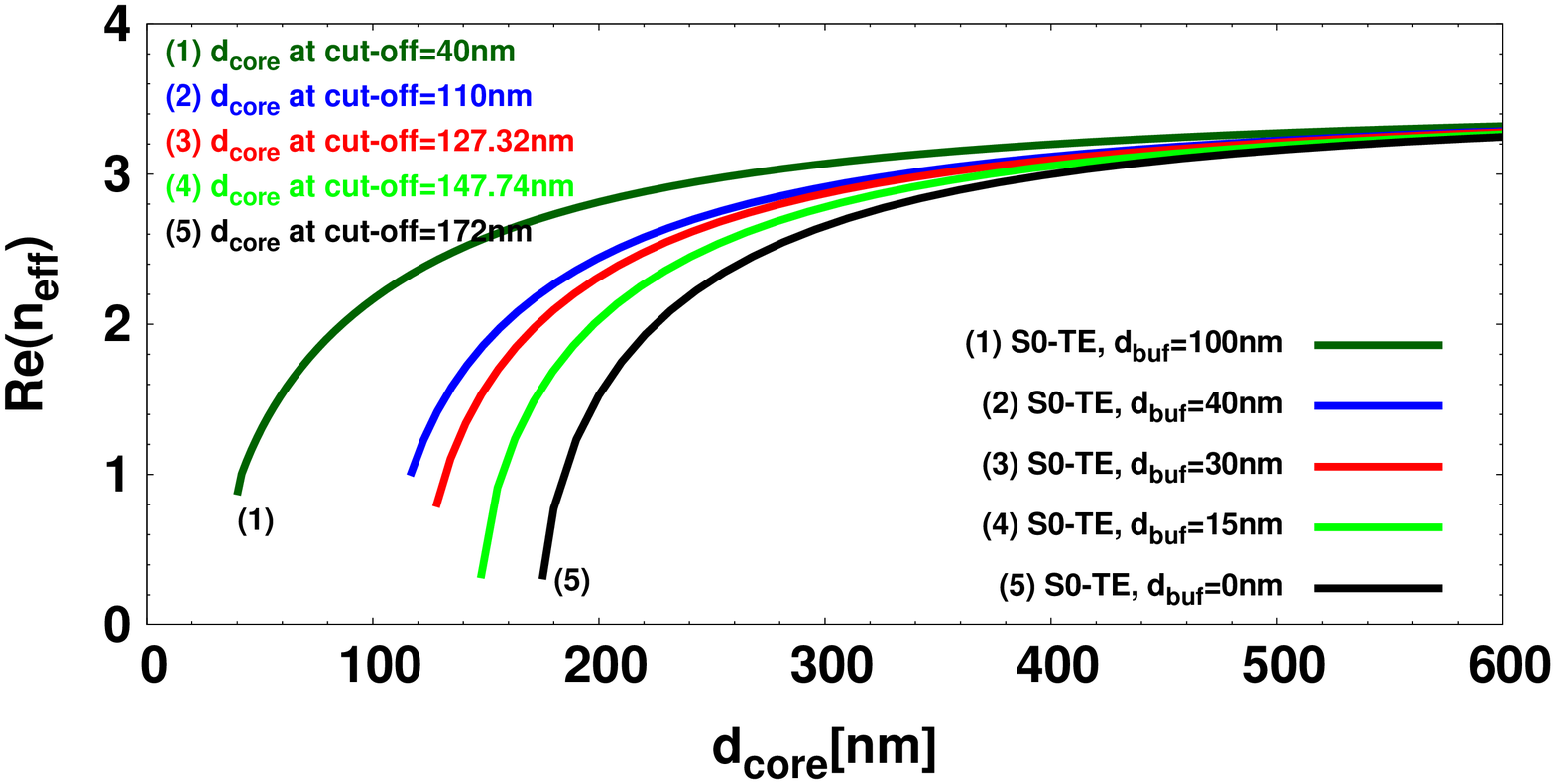}	}
  \caption{Linear dispersion relation for the symmetric fundamental transverse electric mode S0-TE as a function of the core thickness $d_{core}$ for $\varepsilon_{buf} = 2.5^2$. The color curves represent the linear dispersion relation for different thicknesses of the buffer layer $d_{buf}$. The cut-off thickness for $d_{core}$ is given in the top region of the figure for each plotted curve.}
  \label{fig:linear_dispersion_curve_TE0}
 \end{figure}

\begin{figure}[htbp]
    \includegraphics[width=1.0\columnwidth,height=0.55\columnwidth,angle=-0,clip=true,trim= 0 0 0 0]{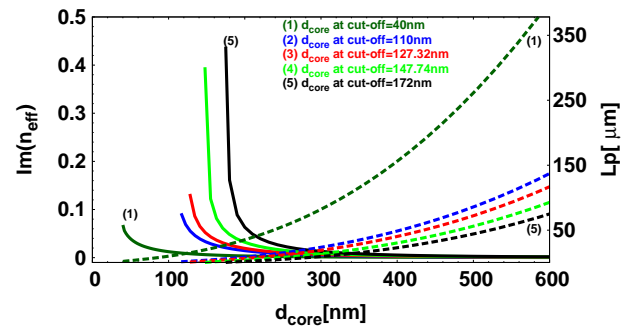}	
  \caption{$Im(n_{eff})$ and the propagation distances for the linear symmetric fundamental transverse electric mode S0-TE as a function of the core thickness $d_{core}$ for $\varepsilon_{buf} = 2.5^2$. The color curves represent the results for different values of $d_{buf}$. The solid lines are for $Im(n_{eff})$, and the dashed lines are for the propagation length $L_{p}$ (see Fig.~\ref{fig:linear_dispersion_curve_TE0} last sentence).}
  \label{fig:loss_linear_dispersion_curve_TE0}
 \end{figure}

The real and the imaginary parts of the effective index for the fundamental symmetric linear TE mode S0-TE (see Fig.~\ref{fig:symmetric_linear_profiles_TE_TM}(b) and (d) for an example of such field profile) are given in Fig.~\ref{fig:linear_dispersion_curve_TE0} and~\ref{fig:loss_linear_dispersion_curve_TE0}, respectively, as a function of the core thickness $d_{core}$ for five different values of $d_{buf}$. For each value of $d_{buf}$, there exist a cut-off thickness for $d_{core}$ such that below this value the fundamental symmetric linear TE mode does not exit. The cut-off core thickness decreases with the increase of the buffer layer thicknesses $d_{buf}$ such that the black  curve (which represents the results obtained of the improved structure by removing the supplementary buffer layers i.e the simple NPSW) has the highest cut-off core thickness $\approx 172$ nm, while the dark green curve has the lowest cut-off value $\approx 40$ nm for $d_{buf}=100$ nm. It is worth mentioning that, for $d_{core}>600$ nm all the curves merge together regardless the value of the buffer thicknesses, because the core becomes very thick and the S0-TE mode moves away from the metal cladding.
In Fig.~\ref{fig:loss_linear_dispersion_curve_TE0}, we show the imaginary parts of the fundamental symmetric linear TE mode S0-TE with the corresponding propagation length $L_{p}$. The meaning of this figure is similar to Fig.~\ref{fig:linear_dispersion_curve_TE0}. The solid curves for $Im(n_{eff})$ and the dashed ones for the propagation length $L_{p}= \lambda /\left(4\pi Im(n_{eff}) \right)$. For a fixed value of $d_{core}$, the improved structure supports a fundamental symmetric linear TE mode with longer  propagation length than the one obtained from the simple structure. The dark green curve (for $d_{buf}=100 $ nm), has the longest propagation length  and the black curve has the shortest propagation length (for $d_{buf}=0.0$ nm). 
\subsection{Nonlinear TE case}
\label{sebsec:nl_TE_case}
In this part, we study the nonlinear dispersion curves for the improved NPSW depicted in Fig.~\ref{fig:geom-5layers-NPSW}, for TE polarized waves. In Subsec.~\ref{subsec:linear_TE_case}, we have shown that, as soon as the cut-off core thickness is exceeded, the supplementary buffer layers for the TE case have no significant effects on the linear field profiles, unlike the TM case, as it is shown in Fig.~\ref{fig:symmetric_linear_profiles_TE_TM}(b) and (d). Nevertheless, the inclusion of the linear buffer layers between the nonlinear core and the metal regions allows us to reduce the cut-off thickness of $d_{core}$ for the fundamental symmetric linear TE mode (see Fig.~\ref{fig:linear_dispersion_curve_TE0}), and to increase its propagation length (see Fig.~\ref{fig:loss_linear_dispersion_curve_TE0}). For the nonlinear TE case, we consider $d_{core}=400$ nm, $\varepsilon_{buf} = 2.5^2$, and $d_{buf}=15$ nm (for the other parameters see Sec.~\ref{sec:models}), these parameters are used for linear field profile shown in Fig.~\ref{fig:symmetric_linear_profiles_TE_TM}(b). We chose the core thickness $d_{core}=400$ nm to be above the cut-off thickness which is $\approx147$ nm for these parameters (see green curve in Figs.~\ref{fig:linear_dispersion_curve_TE0} and~\ref{fig:loss_linear_dispersion_curve_TE0}). The only non-zero component of the electric field is the continuous component $E_{y}$ therefore, the nonlinear permittivity of Kerr-type in the core is written as $\varepsilon_{core} = \varepsilon_{l,core} + \alpha |E_y|^2$, with $\alpha > 0$  (focusing nonlinearity). We use the fixed power algorithm in the FEM~\cite{Drouart08,Walasik14,livre12-FPCF} to compute the nonlinear dispersion curves and the associated field profiles, in which the input is the total power and the outputs are the effective index and the corresponding field component ($E_{y}(x)$ in this case). 
\begin{figure}[htbp]
  \centerline{%
    \includegraphics[width=0.8\columnwidth,angle=-0,clip=true,trim= 30 30 30 30]{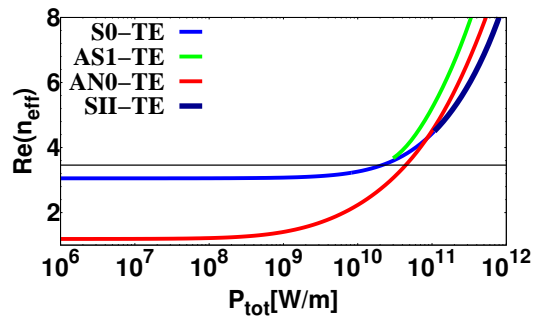}}
  \caption{Nonlinear dispersion curves of the improved NPSW for the TE polarized waves. The blue, green, and red colors indicate the symmetric, asymmetric and antisymmetric nonlinear TE modes, respectively. $d_{core}=400$ nm, $\varepsilon_{buf}=2.5^2$, and $d_{buf}=15$ nm.}
  \label{fig:nl_dispersion_curve_TE0}
 \end{figure}

\begin{figure}[htbp]
\centerline{
\includegraphics[width=0.33\columnwidth,angle=-0,clip=true,trim= 30 0 50 10]{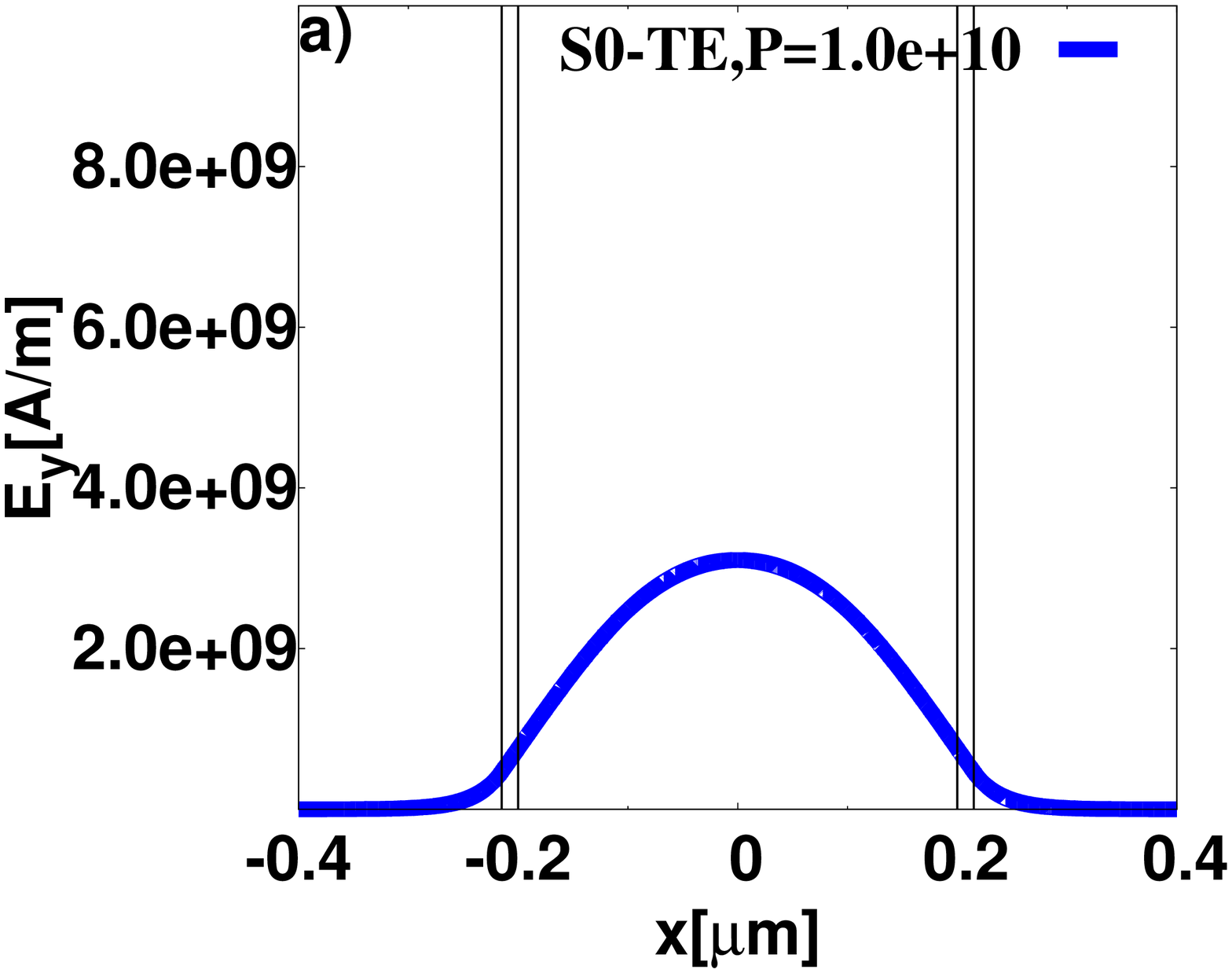}
\includegraphics[width=0.33\columnwidth,angle=-0,clip=true,trim= 30 0 50 10]{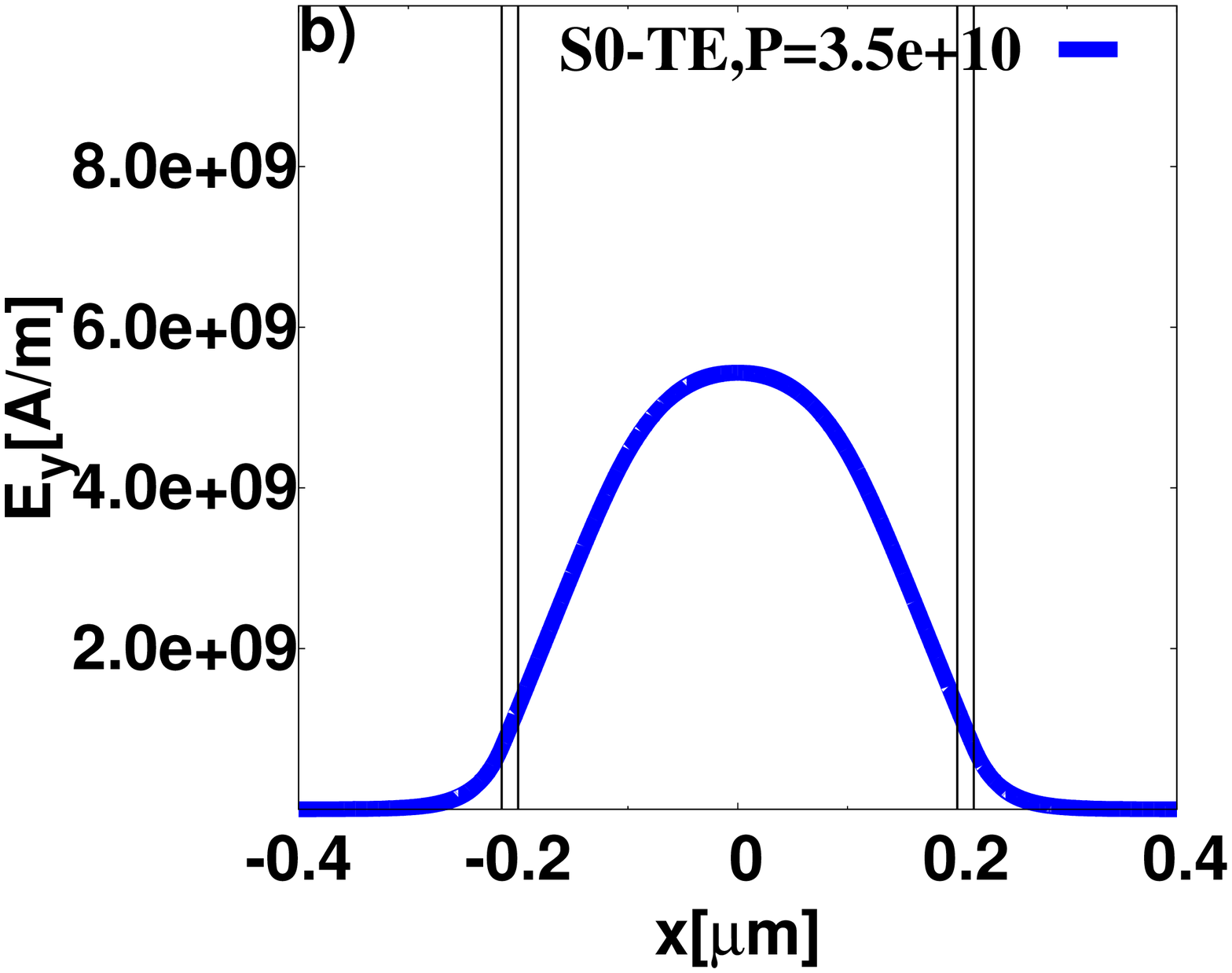}
\includegraphics[width=0.33\columnwidth,angle=-0,clip=true,trim= 30 0 50 10]{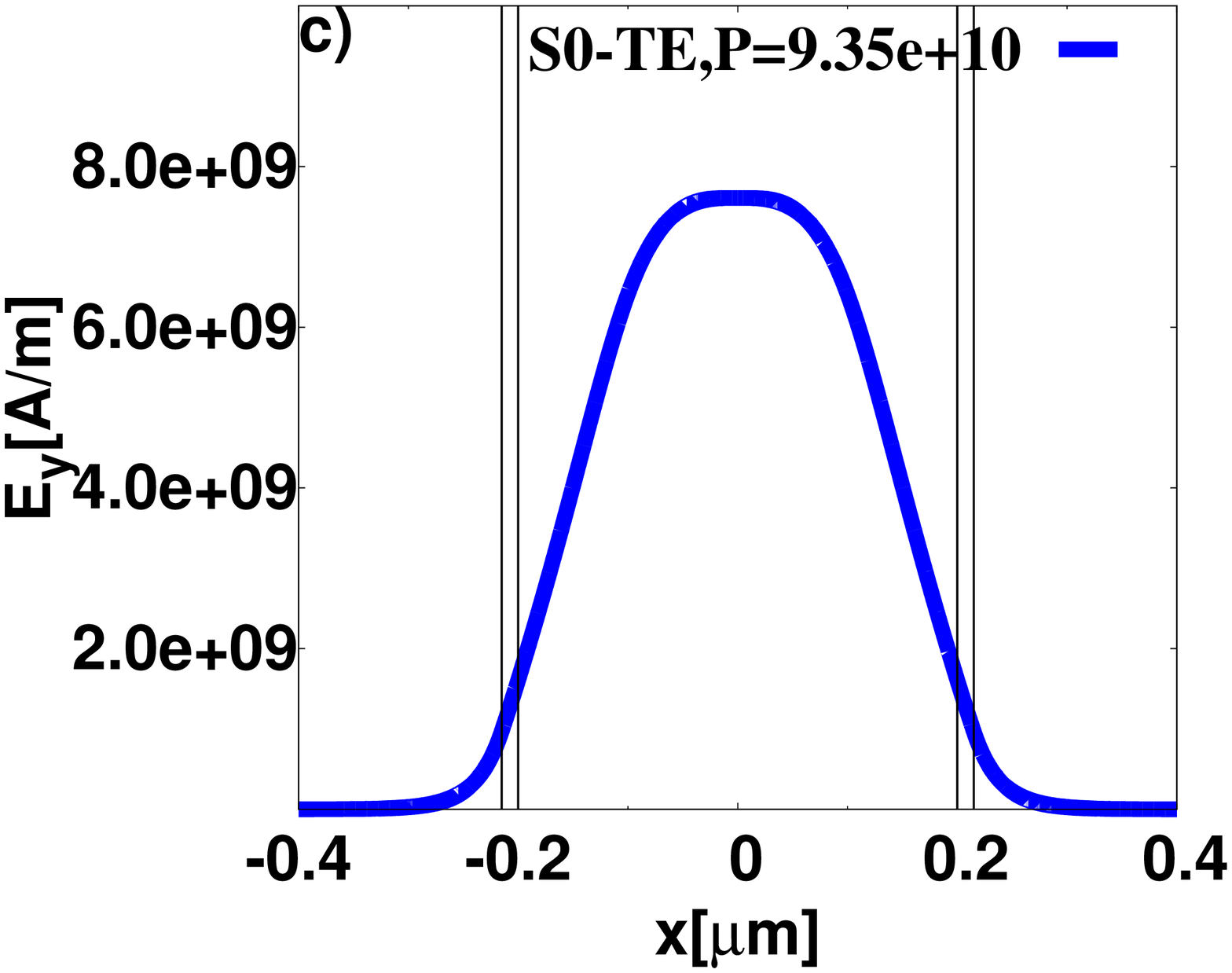}}  
 \centerline{\includegraphics[width=0.33\columnwidth,angle=-0,clip=true,trim= 30 0 50 10]{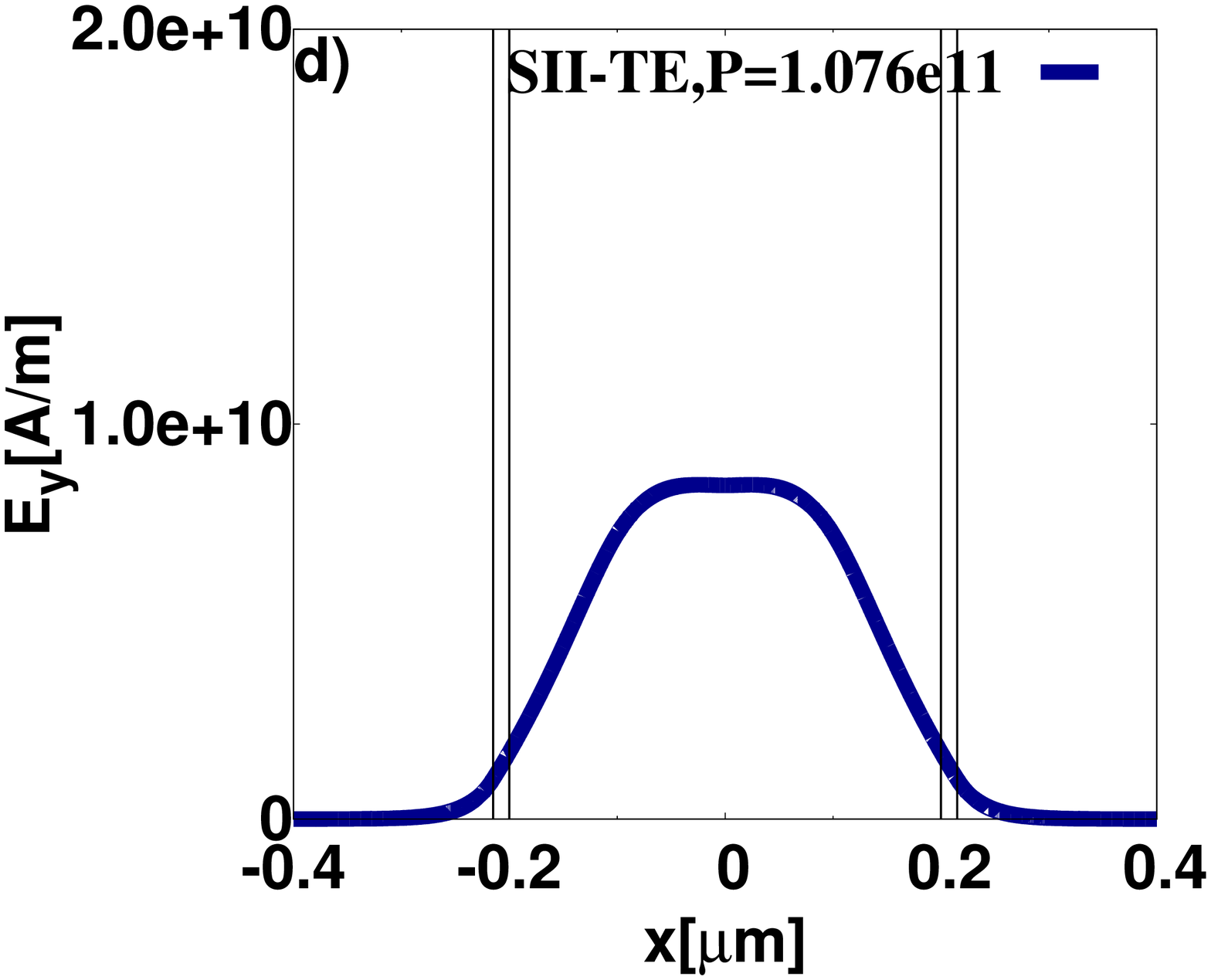}
\includegraphics[width=0.33\columnwidth,angle=-0,clip=true,trim= 30 0 50 10]{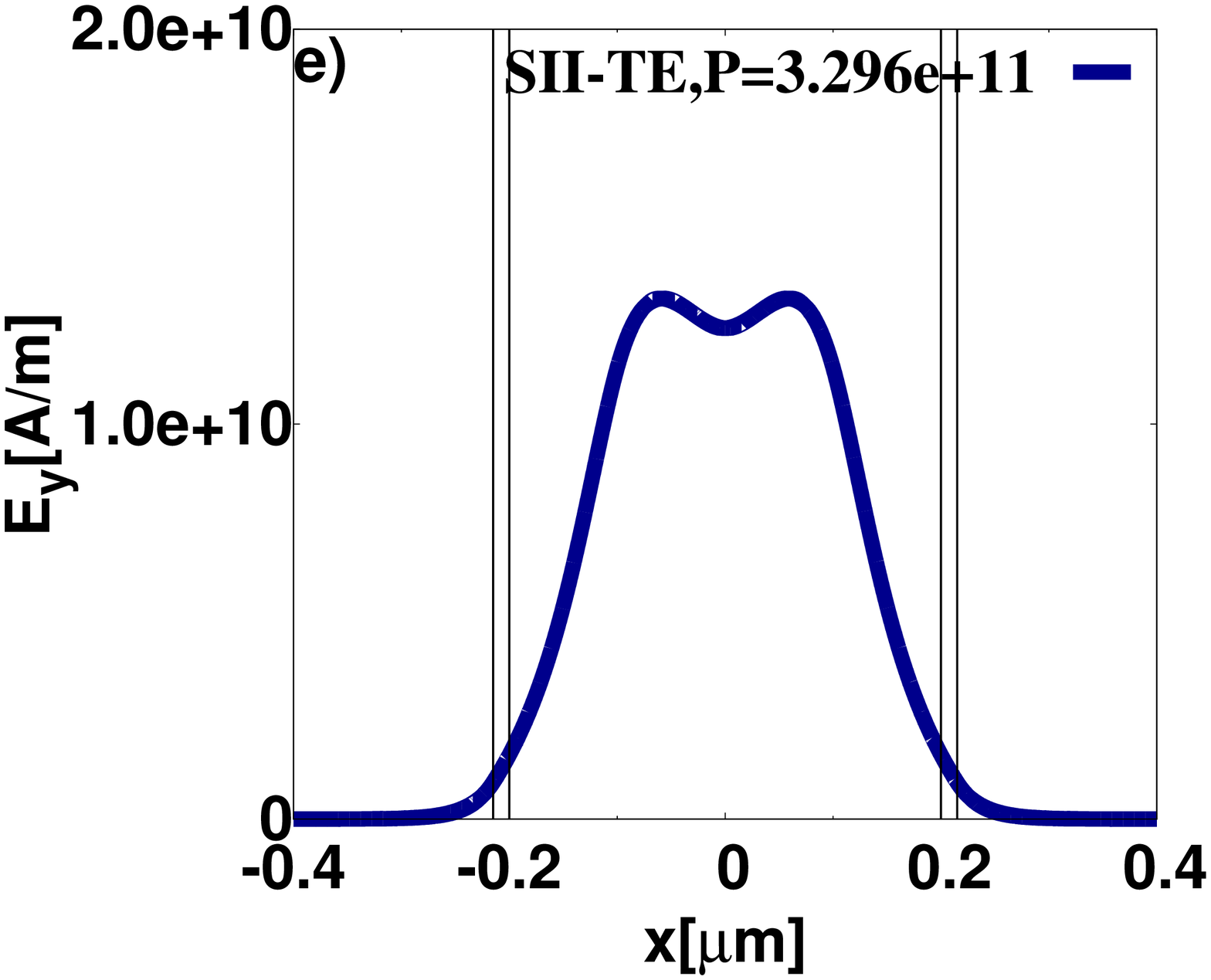}
\includegraphics[width=0.33\columnwidth,angle=-0,clip=true,trim= 30 0 50 10]{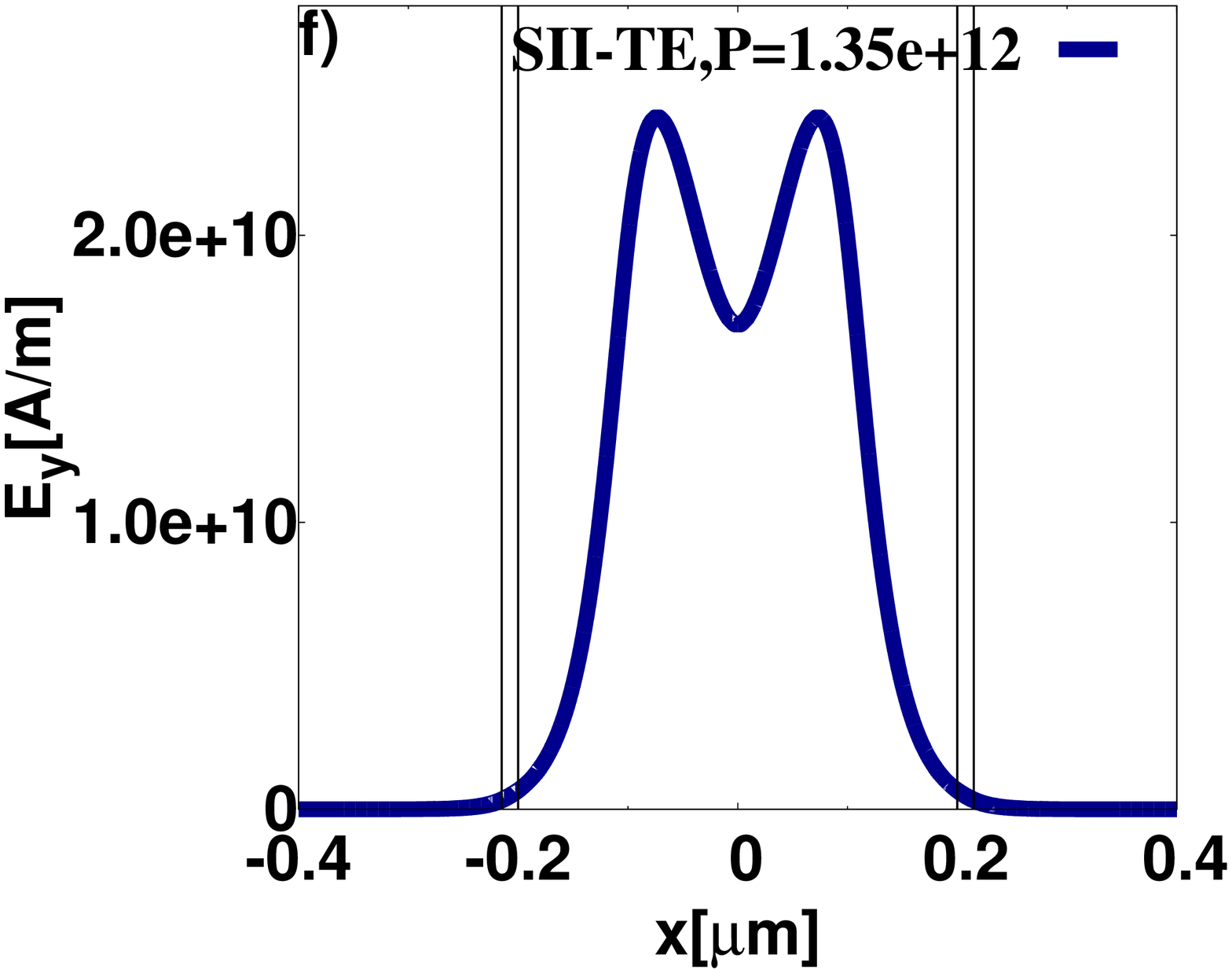}}
\caption{Spatial transition of the symmetric nonlinear field profiles $E_y(x)$ for the improved NPSW in the TE case. (top) for S0-TE and (bottom) for SII-TE with $d_{core}=400$ nm, $\varepsilon_{buf}=2.5^2$, and $d_{buf}=15$ nm. The power is given in W/m.}
\label{fig:TEsymmetric-nonlinear-profile-modal-dbuf15nm-transition}
\end{figure}

\begin{figure}[htbp]
\centerline{
\includegraphics[width=0.33\columnwidth,angle=-0,clip=true,trim= 30 0 50 10]{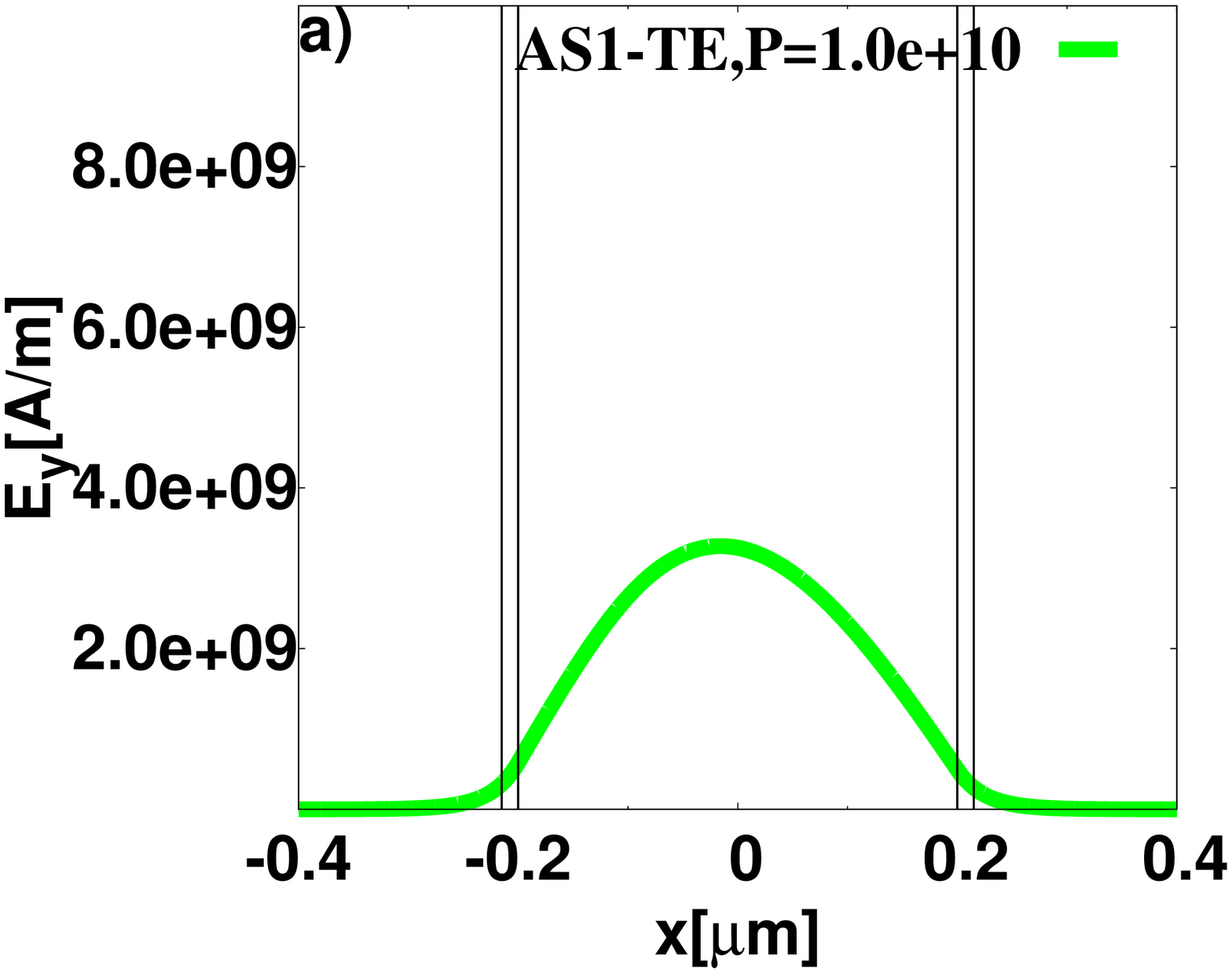}
\includegraphics[width=0.33\columnwidth,angle=-0,clip=true,trim= 30 0 50 10]{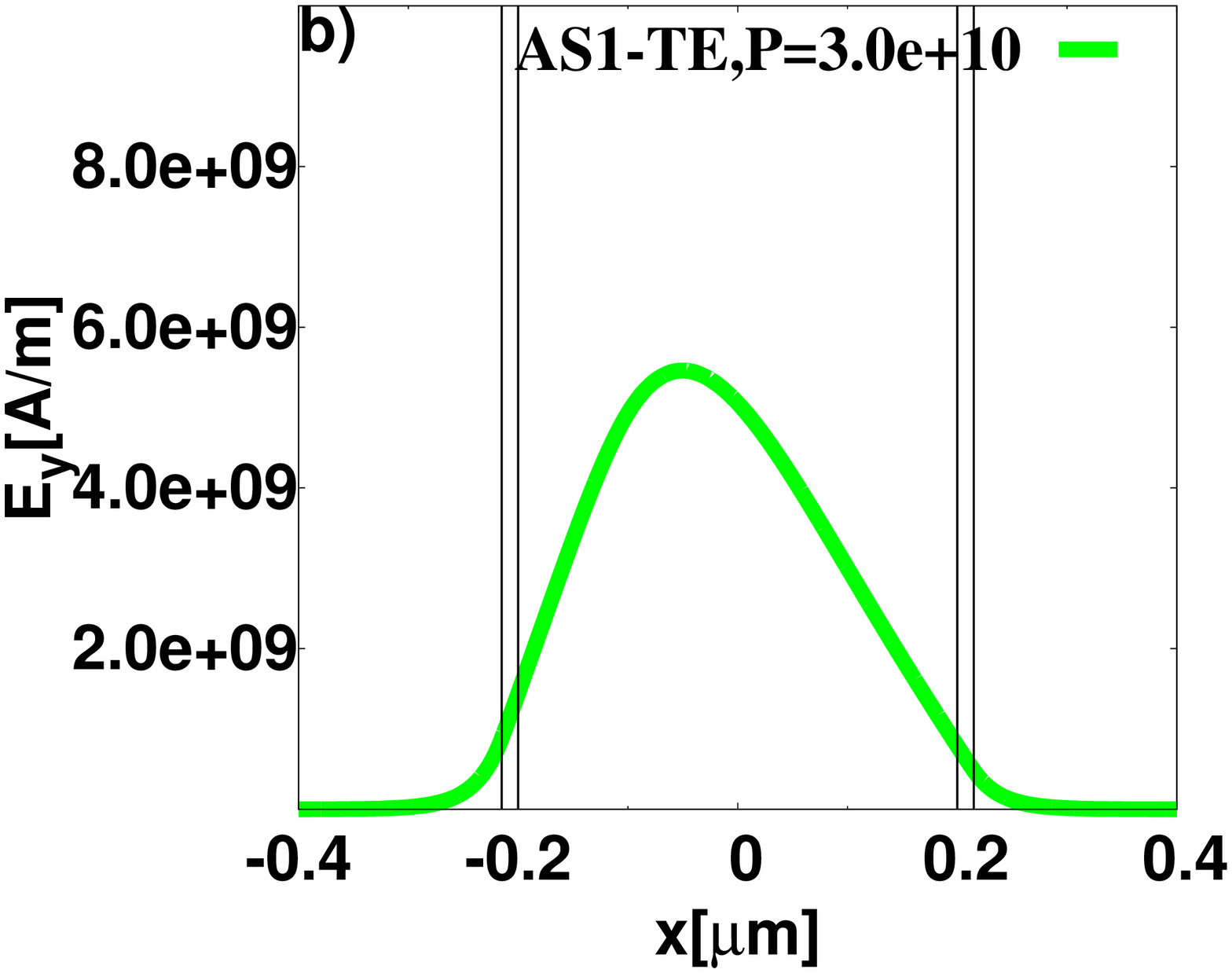}
\includegraphics[width=0.33\columnwidth,angle=-0,clip=true,trim= 30 0 50 10]{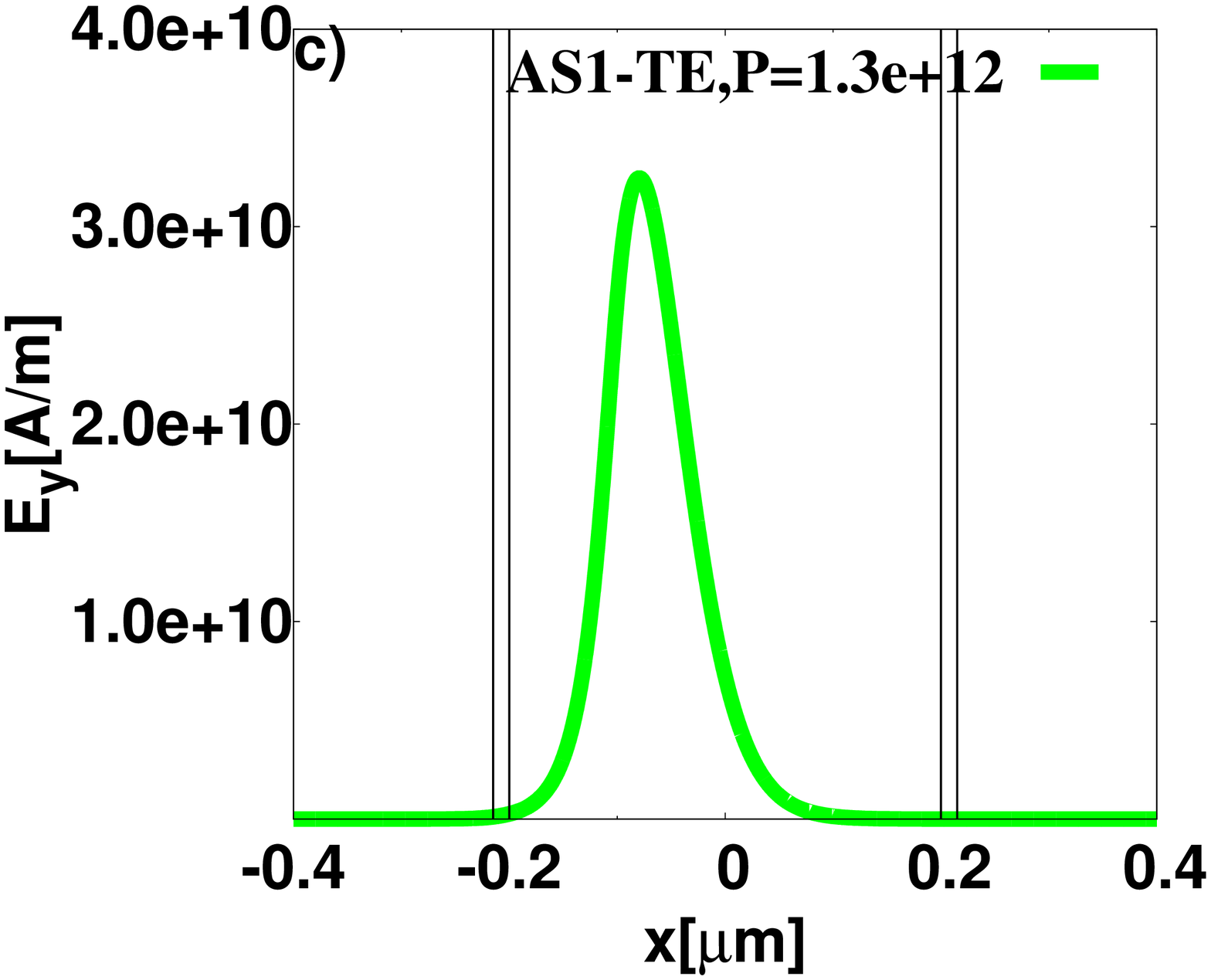}}  
 \centerline{\includegraphics[width=0.33\columnwidth,angle=-0,clip=true,trim= 30 0 50 10]{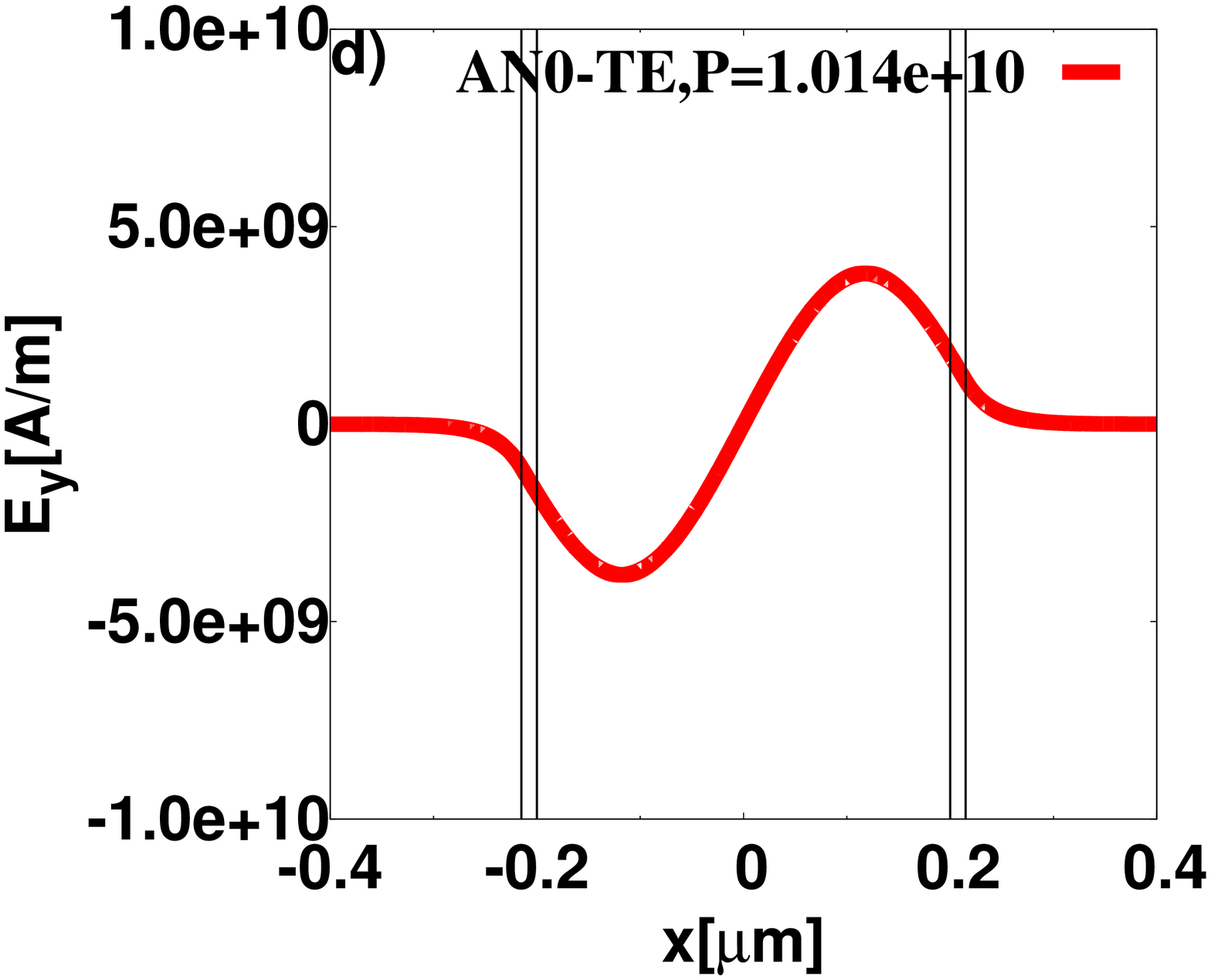}
\includegraphics[width=0.33\columnwidth,angle=-0,clip=true,trim= 30 0 50 10]{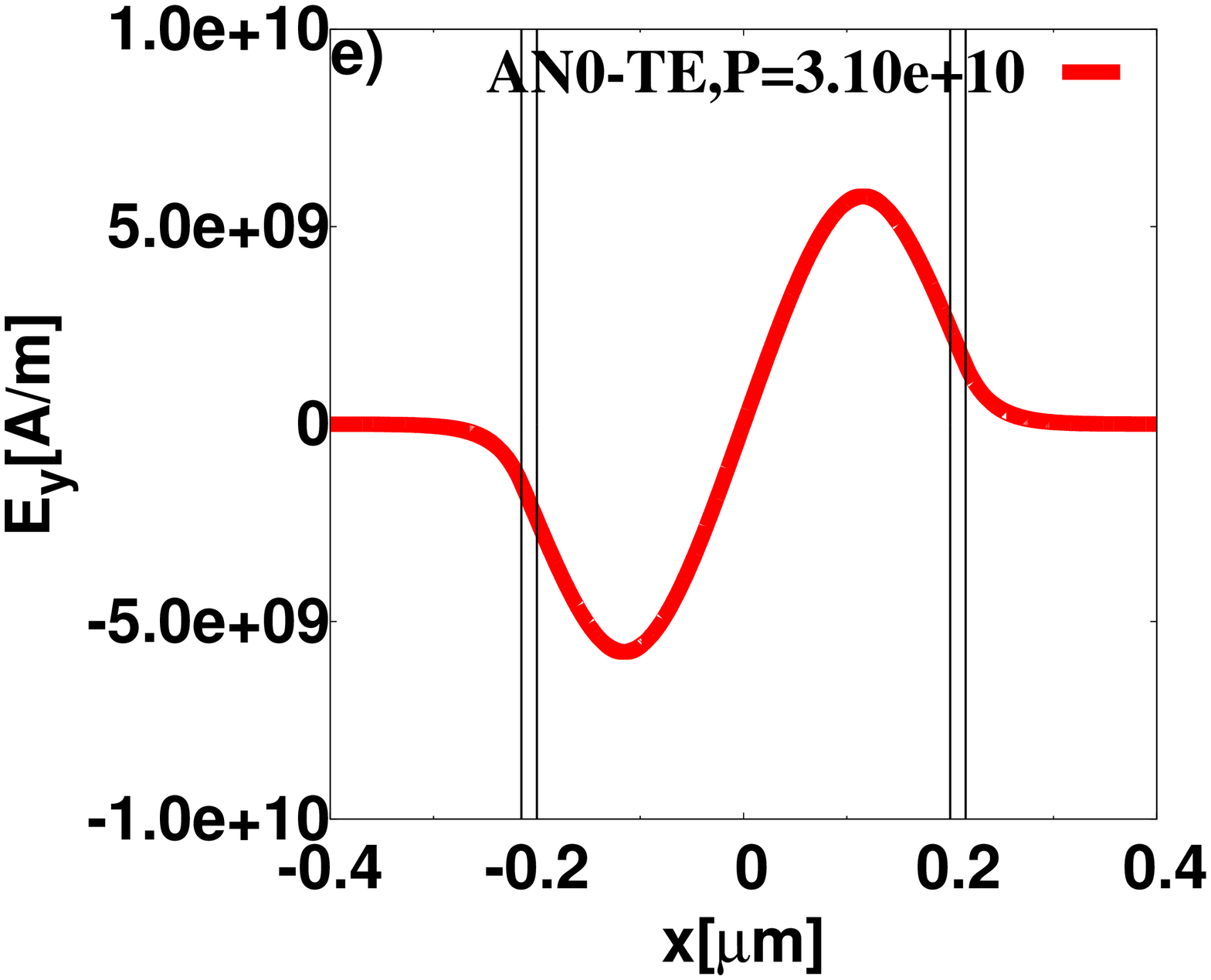}
\includegraphics[width=0.33\columnwidth,angle=-0,clip=true,trim= 30 0 50 10]{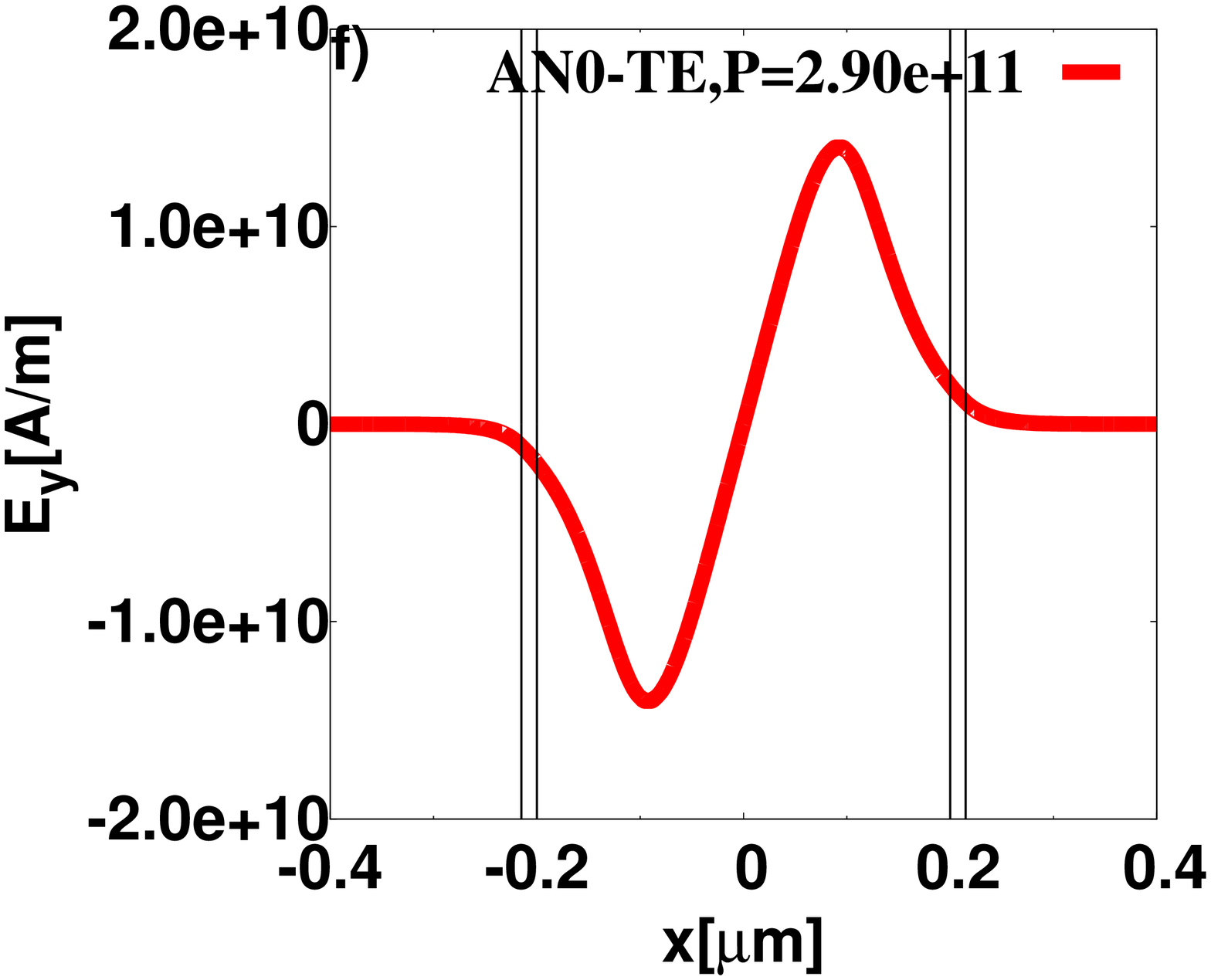}}
\caption{Nonlinear field profiles $E_y(x)$ for the improved NPSW in the TE case. The top green curves and the bottom red curves for AS1-TE and AN0-TE, respectively, with $d_{core}=400$ nm, $\varepsilon_{buf}=2.5^2$, and $d_{buf}=15$ nm. The power is given in W/m.}
\label{fig:TEAsymm-Anti-nonlinear-profile-modal-dbuf15nm-transition}
\end{figure}

In Fig.~\ref{fig:nl_dispersion_curve_TE0}, we present the nonlinear dispersion diagram for the main nonlinear TE modes in the improved NPSW. The global shape of this diagram is similar to the ones obtained for the TM case such that the relative positions of the three main modes are preserved. However, the field profiles are different. The blue, green, and red curves represent the symmetric, asymmetric, and antisymmetric nonlinear TE modes, respectively. The field profiles for the nonlinear symmetric mode S0-TE are given in Fig.~\ref{fig:TEsymmetric-nonlinear-profile-modal-dbuf15nm-transition}(a), (b), and (c) in which the intensity is located in the middle of the waveguide without any plasmonic tails at the interfaces between the linear buffer layers and the metal regions. The antisymmetric field profiles are depicted in Fig.~\ref{fig:TEAsymm-Anti-nonlinear-profile-modal-dbuf15nm-transition}(d), (e), and (f).

Due to the nonlinearity, the nonlinear asymmetric mode (denoted by AS1-TE), bifurcates at a critical value of the power through a Hopf bifurcation from the nonlinear symmetric mode S0-TE (see the green and the blue curves in Fig.~\ref{fig:nl_dispersion_curve_TE0}). The nonlinear asymmetric mode AS1-TE profiles are shown in Fig.~\ref{fig:TEAsymm-Anti-nonlinear-profile-modal-dbuf15nm-transition}(a), (b), and (c) in which the mode is shifted-off the slot center and moves to one of the interfaces (we remind that the asymmetric mode is doubly degenerated). This kind of Hopf bifurcation is similar to the ones obtained for the TM wave both for the improved NPSW (see Fig.~\ref{fig:nl-dispersion-curve}) and for the simple NPSW~\cite{Walasik15a,Walasik15b}. If we compare the bifurcation thresholds for the TE case (see green curve in Fig.~\ref{fig:nl_dispersion_curve_TE0}) with its counterpart in the TM case (shown in Fig.~\ref{fig:nl-dispersion-curve}(c)) using the same parameters, we found that the asymmetric nonlinear mode in the TE case bifurcates from the symmetric nonlinear mode at a higher power than its counterpart in the TM case. Furthermore, in the TM case, the asymmetric mode bifurcates from the symmetric plasmonic mode S0-plas, and turns gradually to the asymmetric solitonic mode AS1-solI as it is shown in Fig.~\ref{fig:nonlinear-profile-modal-dbuf15nm-transition}(b), while in the TE case, the asymmetric mode AS1-TE, bifurcates from the nonlinear symmetric mode S0-TE and does not exhibit any spatial transition (see Fig.~\ref{fig:nl_dispersion_curve_TE0}). At high power, above the bifurcation threshold in the nonlinear dispersion diagram shown in Fig.~\ref{fig:nl_dispersion_curve_TE0}, the nonlinear symmetric core localized mode S0-TE turns gradually to a nonlinear two-humped mode. We call it SII-TE (see the blue and thick dark blue curves in Fig.~\ref{fig:nl_dispersion_curve_TE0}), in which the soliton intensity profile in the core displays two peaks in the middle of the waveguide. In Fig.~\ref{fig:TEsymmetric-nonlinear-profile-modal-dbuf15nm-transition}, we illustrate the modal spatial transition from the nonlinear symmetric mode S0-TE (blue profiles), to the nonlinear symmetric two-humped mode SII-TE (dark blue profiles). It is worth mentioning that such kind of transition between the symmetric modes has already been found in a fully nonlinear dielectric waveguides~\cite{spatial-solitons-Trillo2001} at high power. 

\begin{figure}[htbp]
  \centerline{
    \includegraphics[width=0.49\columnwidth,angle=-0,clip=true,trim= 30 40 50 30]{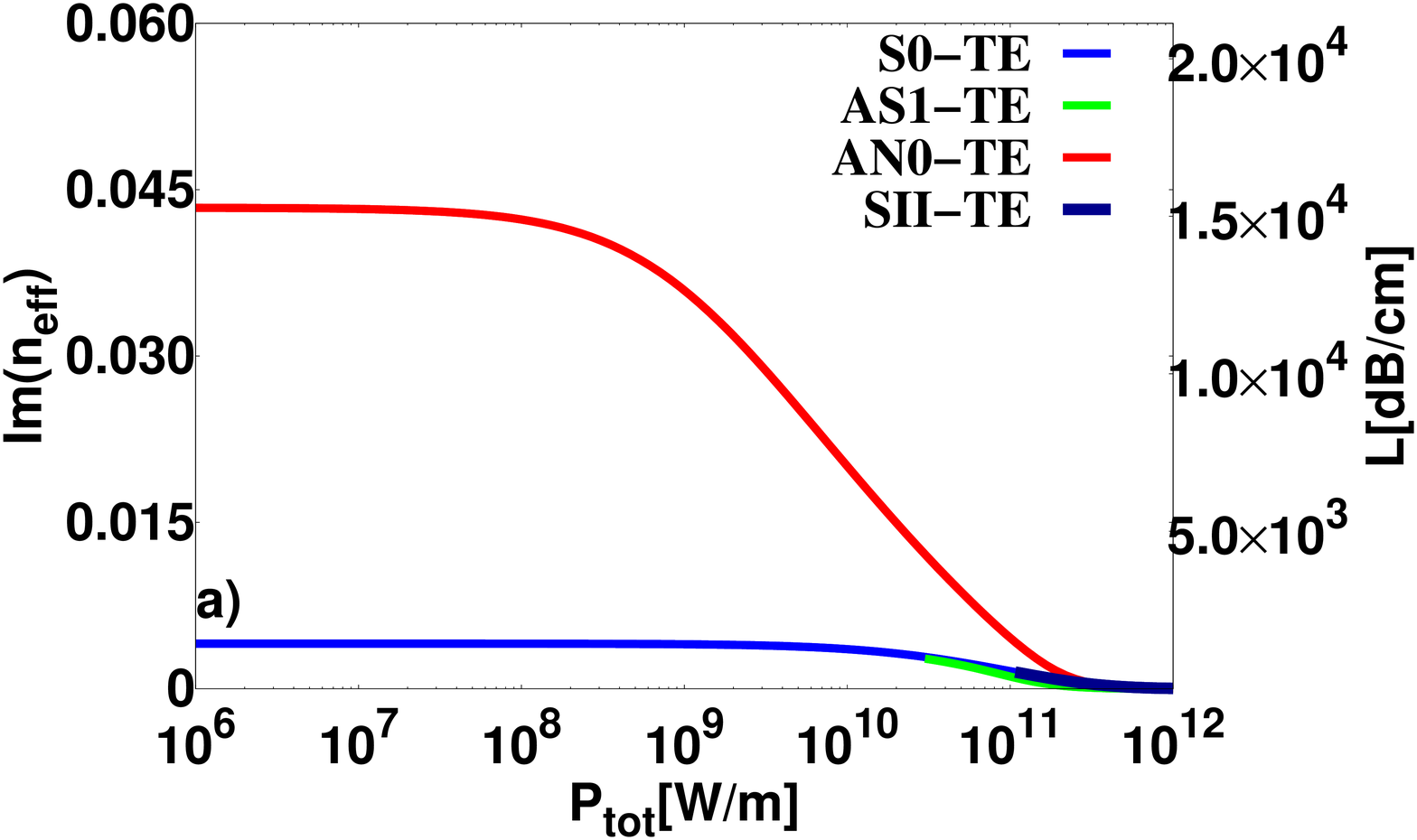}
     \includegraphics[width=0.49\columnwidth,angle=-0,clip=true,trim= 30 40 50 30]{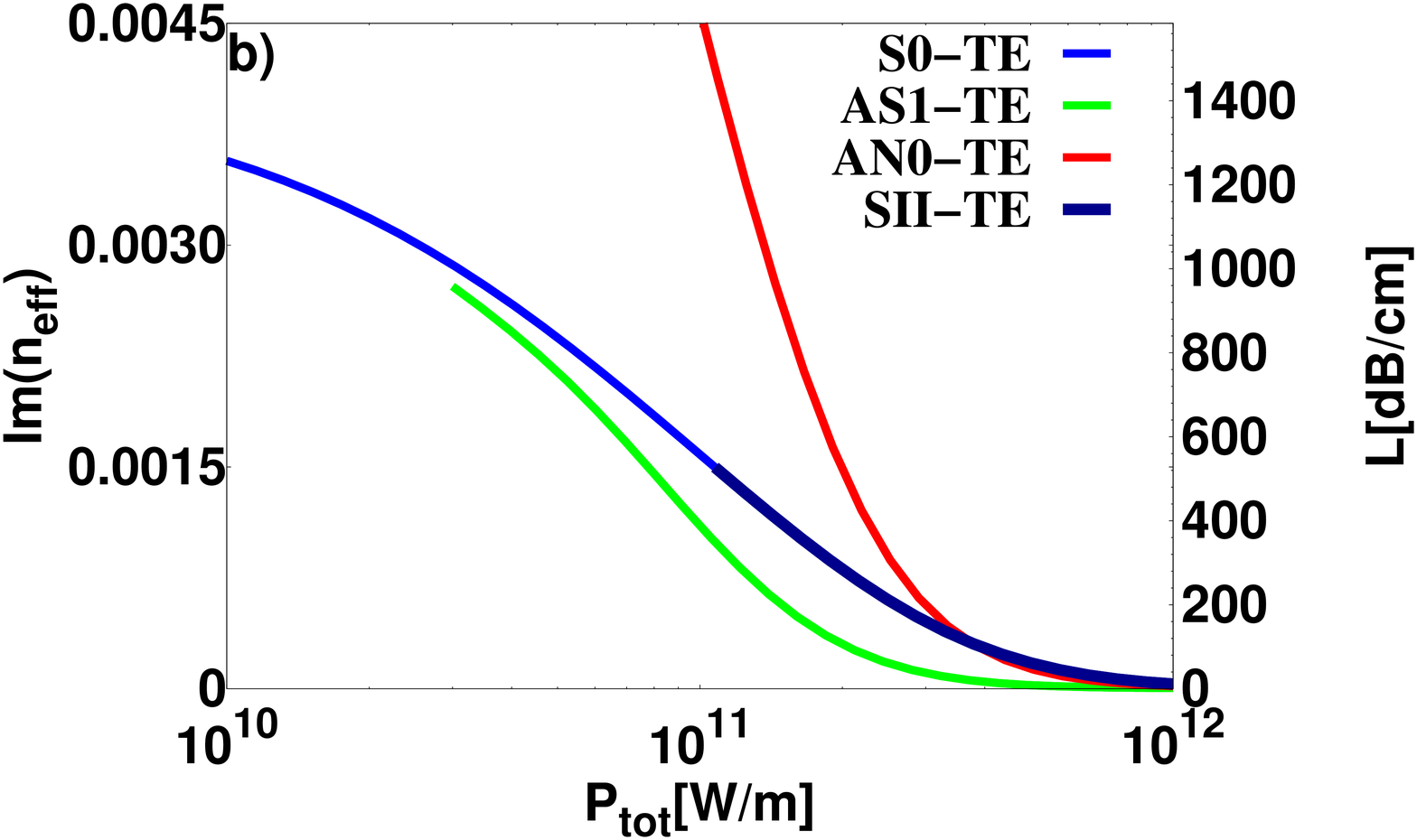}}
  \caption{Im$(n_{eff})$ and the losses for the TE waves in the improved NPSW. The blue, green, and red colors indicate the symmetric, asymmetric and antisymmetric nonlinear TE modes, respectively. $d_{core}=400$ nm, $\varepsilon_{buf}=2.5^2$, and $d_{buf}=15$ nm.}
  \label{fig:losses_TE_case}
 \end{figure}
In Fig.~\ref{fig:losses_TE_case}, we show the imaginary part of the effective indices of the main TE modes with the corresponding values of the losses. One clearly see that, at low power, the antisymmetric mode has the highest value of the losses (see Fig.~\ref{fig:losses_TE_case}(a)) because the peaks are closer to the metal regions as it is shown in the second row of Fig.~\ref{fig:TEAsymm-Anti-nonlinear-profile-modal-dbuf15nm-transition}. Whereas, at high power the losses of the three main modes become of the same order (see Fig.~\ref{fig:losses_TE_case}(b)). In general, the losses decrease with the power increase for all the tested modes. This decrease of the losses with the increase of the power, can be understood from the field profiles in Figs.~\ref{fig:TEsymmetric-nonlinear-profile-modal-dbuf15nm-transition} and~\ref{fig:TEAsymm-Anti-nonlinear-profile-modal-dbuf15nm-transition} in which at high power, all the modes are more localized in the core and move away from the metal regions.
\section{Conclusion}
\label{sec:conclusion}
We have presented a full study of a nonlinear plasmonic slot waveguide improved by the addition of supplementary linear dielectric buffer layers between the nonlinear core of Kerr-type and the semi-infinite metal regions. We have provided a complete description of its main TM and TE modes for both  linear and nonlinear case.  For the linear TM case, in addition to the symmetric linear plasmonic profile obtained in the simple structure (without the buffer layers), our improved structure supports two new types of linear modes; the core localized and the flat modes. These new linear modes are not found in the simple symmetric linear structures whatever the opto-geometric parameters used. For the nonlinear TM case, we have provided the nonlinear phase diagram for the existence and the type of the main TM modes. The buffer layer thickness and the total power are used as parameters such that, for a well-chosen opto-geometric parameters, the main TM modes exhibit spatial modal transitions which can be controlled by the power. We have also shown that this structure allows a decrease of the losses compared to the simple nonlinear slot waveguide and that they decrease with the power for most of the cases. The stability properties of the main TM modes has also been studied numerically using the FDTD method. We have also demonstrated the existence (for some configurations), and studied the properties of the main TE modes for both linear and nonlinear cases. This kind of structure could be fabricated and characterized experimentally due to the realistic parameters chosen to model it.


\acknowledgments
This work has been partially funded by the PhD school of Physics and Materials Science, ED352, and the University of Aix-Marseille. It was not funded by the French National Research Agency.

\end{document}